\documentclass[twocolumn, twocolappendix]{aastex701}
\usepackage{amsmath,amstext}
\usepackage[T1]{fontenc}
\usepackage{multirow}
\usepackage{xcolor}
\usepackage{graphicx}

\newcommand{\hst}{\textit{HST}}

\newcommand{\jwst}{\textit{JWST}}

\newcommand{\hb}{\hbox{H$\beta$}}
\newcommand{\ha}{\hbox{H$\alpha$}}
\newcommand{\hg}{\hbox{H$\gamma$}}
\newcommand{\hd}{\hbox{H$\delta$}}
\newcommand{\oii}{\hbox{[O \textsc{ii}]}}
\newcommand{\oiii}{\hbox{[O \textsc{iii}]}}
\newcommand{\nii}{\hbox{[N \textsc{ii}]}}
\newcommand{\sii}{\hbox{[S \textsc{ii}]}}
\newcommand{\Mgi}{\hbox{Mg \textsc{i}}}
\newcommand{\Mgii}{\hbox{Mg \textsc{ii}}}
\newcommand{\NaD}{\hbox{Na \textsc{d}}}

\newcommand{\cigale}{\hbox{\textsc{cigale}}}
\newcommand{\bagpipes}{\hbox{\textsc{Bagpipes}}}
\newcommand{\eazy}{\hbox{\textsc{Eazy}}}

\newcommand{\ebvg}{$\mathrm{E(B-V)_{gas}}$}
\newcommand{\ebvs}{$\mathrm{E(B-V)_{star}}$}
\newcommand{\sm}{$M_\ast$}

\begin{document}

\title{Quiescent Galaxies at $z > 2$ from CAPERS spectroscopy and their Number densities}


\author[0000-0001-9495-7759]{Lu Shen}
\affiliation{Department of Physics and Astronomy, Texas A\&M University, College Station, TX, 77843-4242 USA}
\affiliation{George P.\ and Cynthia Woods Mitchell Institute for
 Fundamental Physics and Astronomy, Texas A\&M University, College Station, TX, 77843-4242 USA}
\email[show]{lushen@tamu.edu}

\author[0000-0003-3424-3230]{Weida Hu}
\affiliation{Department of Physics and Astronomy, Texas A\&M University, College Station, TX, 77843-4242 USA}
\affiliation{George P.\ and Cynthia Woods Mitchell Institute for
 Fundamental Physics and Astronomy, Texas A\&M University, College Station, TX, 77843-4242 USA}
\email{weidahu@tamu.edu}

\author[0000-0001-7503-8482]{Casey Papovich}
\affiliation{Department of Physics and Astronomy, Texas A\&M University, College Station, TX, 77843-4242 USA}
\affiliation{George P.\ and Cynthia Woods Mitchell Institute for
 Fundamental Physics and Astronomy, Texas A\&M University, College Station, TX, 77843-4242 USA}
\email{papovich@tamu.edu}

\author[0000-0002-6348-1900]{Justin Cole}
\affiliation{Department of Physics and Astronomy, Texas A\&M University, College Station, TX, 77843-4242 USA}
\affiliation{George P.\ and Cynthia Woods Mitchell Institute for
 Fundamental Physics and Astronomy, Texas A\&M University, College Station, TX, 77843-4242 USA}
\email{jwc68@tamu.edu}

\author[0000-0001-5414-5131]{Mark Dickinson}
\affiliation{NSF NOIRLab, 950 N. Cherry Ave., Tucson, AZ 85719, USA}
 \email{mark.dickinson@noirlab.edu}

\author[0000-0002-7959-8783]{Pablo Arrabal Haro}
\affiliation{Center for Space Sciences and Technology, UMBC, 5523 Research Park Dr, Baltimore, MD 21228 USA }
\affiliation{Astrophysics Science Division, NASA Goddard Space Flight Center, 8800 Greenbelt Rd, Greenbelt, MD 20771, USA}
\email{}

\author[0000-0003-1282-7454]{Anthony J. Taylor}
\email{anthony.taylor@austin.utexas.edu}
\affiliation{Department of Astronomy, The University of Texas at Austin, Austin, TX, USA}
\affiliation{Cosmic Frontier Center, The University of Texas at Austin, Austin, TX, USA}
\email{}

\author[0000-0002-1482-5818]{Adam C. Carnall}
\affiliation{Institute for Astronomy, University of Edinburgh, Royal Observatory, Edinburgh, EH9 3HJ, UK}
 \email{adamc@roe.ac.uk}

\author[0009-0004-1163-0160]{\'Angel Chandro-G\'omez}
\affiliation{International Centre for Radio Astronomy Research (ICRAR), The University of Western Australia, 35 Stirling Highway, Crawley, WA 6009, Australia}
\affiliation{ARC Centre for All-Sky Astrophysics in 3 Dimensions (ASTRO 3D), Australia}
\email{angel.chandrogomez@research.uwa.edu.au}

\author[0000-0001-8551-071X]{Yingjie Cheng}
\affiliation{Department of Astronomy, University of Washington, Seattle, WA 98195, USA}
\affiliation{University of Massachusetts Amherst, 710 North Pleasant Street, Amherst, MA 01003-9305, USA}
\email{}

\author[0000-0001-7151-009X]{Nikko J. Cleri}
\affiliation{Department of Astronomy and Astrophysics, The Pennsylvania State University, University Park, PA 16802, USA}
\affiliation{Institute for Computational and Data Sciences, The Pennsylvania State University, University Park, PA 16802, USA}
\affiliation{Institute for Gravitation and the Cosmos, The Pennsylvania State University, University Park, PA 16802, USA}
 \email{cleri@psu.edu}

\author[0000-0002-1404-5950]{James S. Dunlop}
\email{james.dunlop@ed.ac.uk}
\affiliation{Institute for Astronomy, University of Edinburgh, Royal Observatory, Edinburgh EH9 3HJ, UK}

\author[0009-0000-0272-5468]{Carter Flayhart}
\email{cbf2388@rit.edu}
\affiliation{Laboratory for Multiwavelength Astrophysics, School of Physics and Astronomy, Rochester Institute of Technology, 84 Lomb Memorial Drive, Rochester, NY 14623, USA}

\author[0000-0001-8519-1130]{Steven L. Finkelstein}
\affiliation{Department of Astronomy, The University of Texas at Austin, Austin, TX, USA}
\affiliation{Cosmic Frontier Center, The University of Texas at Austin, Austin, TX, USA}
\email{stevenf@astro.as.utexas.edu}

\author[0000-0002-7831-8751]{Mauro Giavalisco}
\affiliation{University of Massachusetts Amherst, 710 North Pleasant Street, Amherst, MA 01003-9305, USA}
\email{}

\author[]{Michaela Hirschmann}
\affiliation{Institute of Physics, Laboratory of Galaxy Evolution, Ecole Polytechnique Fédérale de Lausanne (EPFL), Observatoire de Sauverny, 1290 Versoix, Switzerland}
\email{michaela.hirschmann@epfl.ch}

\author[0000-0002-0786-7307]{Eric F. Jiménez-Andrade}
\affiliation{Instituto de Radioastronomía y Astrofísica, Universidad Nacional Autónoma de México, Antigua Carretera a Pátzcuaro \# 8701,\\ Ex-Hda. San José de la Huerta, Morelia, Michoacán, México C.P. 58089}
\email{}

\author[0000-0001-9187-3605]{Jeyhan S. Kartaltepe}
\affiliation{Laboratory for Multiwavelength Astrophysics, School of Physics and Astronomy, Rochester Institute of Technology, 84 Lomb Memorial Drive, Rochester, NY 14623, USA}
\email{}

\author[0000-0003-3021-8564]{Claudia D.P. Lagos}
\affiliation{International Centre for Radio Astronomy Research (ICRAR), The University of Western Australia, 35 Stirling Highway, Crawley, WA 6009, Australia}
\affiliation{Cosmic Dawn Center (DAWN), Denmark}
\email{claudia.lagos@uwa.edu.au}

\author[0000-0003-0486-5178]{Ho-Hin Leung}
\affiliation{ Institute for Astronomy, University of Edinburgh, Royal Observatory, Edinburgh EH9 3HJ, UK}
\email{ hleung2@roe.ac.uk }

\author[0000-0002-7530-8857]{Arianna S. Long}
\affiliation{Department of Astronomy, The University of Washington, Seattle, WA USA}
\email{aslong@uw.edu}

\author[0000-0003-1581-7825]{Ray A. Lucas}
\affiliation{Space Telescope Science Institute, 3700 San Martin Drive, Baltimore, MD 21218, USA}
\email{lucas@stsci.edu}

\author[0000-0001-7089-7325]{Eric J. Murphy}
\affiliation{National Radio Astronomy Observatory, 520 Edgemont Road, Charlottesville, VA 22903, USA}
\email{}

\author[0000-0002-8951-4408]{Lorenzo Napolitano}
\affiliation{INAF – Osservatorio Astronomico di Roma, via Frascati 33, 00078, Monteporzio Catone, Italy}
\affiliation{Dipartimento di Fisica, Università di Roma Sapienza, Città Universitaria di Roma - Sapienza, Piazzale Aldo Moro, 2, 00185, Roma, Italy}
\email{lorenzo.napolitano@inaf.it}

\author[0000-0003-4528-5639]{Pablo G. P\'erez-Gonz\'alez}
\affiliation{Centro de Astrobiolog\'{\i}a (CAB), CSIC-INTA, Ctra. de Ajalvir km 4, Torrej\'on de Ardoz, E-28850, Madrid, Spain}
\email{}

\author[0000-0002-6386-7299]{Raymond C.\ Simons}
\affiliation{Department of Engineering and Physics, Providence College, 1 Cunningham Sq, Providence, RI 02918 USA}
\email{raycsimons@gmail.com}

\author[0000-0001-5642-752X]{Struan D Stevenson} 
\affiliation{Institute for Astronomy, University of Edinburgh, Royal Observatory, Edinburgh EH9 3HJ, UK}
\email{}

\author[0000-0001-8728-2984]{Elizabeth Taylor}
\affiliation{Institute for Astronomy, University of Edinburgh, Royal Observatory, Edinburgh EH9 3HJ, UK}
\email{etaylo2@roe.ac.uk }
 
\author[0000-0002-9373-3865]{Xin Wang}
\affiliation{School of Astronomy and Space Science, University of Chinese Academy of Sciences (UCAS), Beijing 100049, China}
\affiliation{National Astronomical Observatories, Chinese Academy of Sciences, Beijing 100101, China}
\affiliation{Institute for Frontiers in Astronomy and Astrophysics, Beijing Normal University, Beijing 102206, China}
\email{xwang@ucas.ac.cn}

\author[0000-0003-3466-035X]{{L. Y. Aaron} {Yung}}
\affiliation{Space Telescope Science Institute, 3700 San Martin Drive, Baltimore, MD 21218, USA}
\email{l.y.aaronyung@gmail.com}

\author[0000-0002-7051-1100]{Jorge A. Zavala}
\affiliation{University of Massachusetts Amherst, 
710 North Pleasant Street, Amherst, MA 01003-9305, USA}
\email{jzavala@umass.edu}

\begin{abstract}

Recent \jwst\ observations have revealed an overabundance of quiescent galaxies at high redshift. 
Photometric methods, such as photometry-derived sSFRs, $UVJ$ colors, and synthetic $ugi_s$ colors, are commonly used to identify quiescent galaxies, but their completeness and purity remain poorly constrained.
Here we present a spectroscopically selected sample of 19 quiescent galaxies at $2 \leq z \leq 4$ from the CANDELS-Area Prism Epoch of Reionization Survey (CAPERS). 
They are selected with $M_\ast > 10^{9.5}\,M_\odot$ and a specific star formation rate below 20\% of the inverse cosmic age. 
These galaxies formed 50\% of their stellar mass at $z \sim 3.0$--4.6 and quenched rapidly on timescales of $\sim0.1$--0.2 Gyr. 
These quenching timescales are comparable to those of other high-redshift QGs but shorter than those at lower redshift.
Using the spectroscopic sample as a benchmark, we assess commonly adopted photometric QG selection methods. We find completeness of 63--89\% and purities of only 40--63\%, with strong emission-line galaxies constituting the dominant source of contamination. 
Applying completeness and purity corrections calibrated from the spectroscopic sample, we measure the number density of quiescent galaxies at $2 \leq z \leq 5$. 
The resulting number densities exceed the predictions of most current galaxy-formation simulations at $2<z<4$, with the largest discrepancy at $M_\ast \geq 10^{10.5}\,M_\odot$. 
This discrepancy suggests that current cosmological models may not assemble and quench massive galaxies sufficiently early or efficiently.

\end{abstract}


\section{Introduction} 

Massive galaxies provide crucial constraints for testing $\Lambda$ cold dark matter ($\Lambda$CDM)-like cosmological models. 
In standard galaxy formation models, galaxies assemble hierarchically through star formation and a series of mergers \citep{Somerville2015}. 
Their stellar mass growth is governed by the growth rate of dark matter halos ($dM_\mathrm{halo}/dt$), the cosmic baryon fraction ($f_b$), and the efficiency of converting baryons into stars ($\epsilon$), such that $dM_\ast/dt = \epsilon f_b dM_\mathrm{halo}/dt$.
This simple relation allows us to predict the maximum stellar masses and number densities of galaxies as a function of redshift \citep{Behroozi2018, Boylan-Kolchin2023}. 
However, the recent observations with the James Webb Space Telescope (JWST) unveiled a large number of massive galaxies ($M_\ast\gtrsim 10^{11}\ M_\odot$) already in place at $\sim1$ Gyr after the Big Bang, lying at the very edge of these limits \citep{Labbe2023,Xiao2024,Barrufet2025}. 

More intriguingly, JWST spectroscopy shows that many of these massive galaxies had already become quiescent for several hundred million years \citep{Carnall2023,Glazebrook2024,deGraaff2025,Nanayakkara2024}. 
The advantage of studying quiescent galaxies is that their stellar masses can be estimated more accurately than those of star-forming galaxies, as their absorption features provide robust constraints on stellar-population ages, star-formation histories, and, consequently, on stellar masses. 
Moreover, quiescent galaxies offer additional constraints on the physical mechanisms responsible for the rapid cessation of star formation and the subsequent maintenance of quiescence \citep{Hartley2023, DeLucia2024, Lagos2024, Weller2025, Lagos2025, Vani2025, Chandro-Gomez2026a, Chandro-Gomez2026b}. 

Recent JWST observations have consistently found a higher abundance of massive quiescent systems ($M_\ast > 10^{10}\ M_\odot$) at $z\gtrsim3$ than those predicted by the state-of-the-art analytic models and cosmological simulations \citep{Valentino2023, Carnall2023, Carnall2023b, Carnall2024, Glazebrook2024, Baker2025a, Baker2025b, Nanayakkara2024, Nanayakkara2025, Stevenson2025}. 
Such discrepancies were already hinted at in pre-JWST studies \citep[e.g.,][]{Straatman2014, Forrest2020}.
This suggests that massive quiescent galaxies may have experienced both an exceptionally rapid buildup of stellar mass \citep[e.g., ][]{Dekel2023, Ferrara2023} and highly effective feedback processes capable of quenching star formation on short timescales from active galactic nuclei \citep[AGN; e.g.,][]{Fabian2012}, or driven by galaxy interactions \citep[e.g.,][]{Hopkins2008, Suess2025,Hu2025}.

Over the past decades, massive quiescent galaxies have been successfully identified at $z<2$ using classical photometric rest-frame color selections, such as $UVJ$  selection criteria \citep[e.g.,][]{Williams2009, Muzzin2013, Schreiber2015} and $NUVrJ$/$NUVrK$\citep[e.g.,][]{Ilbert2013, Arnouts2013}. 
At higher redshifts, photometric redshift uncertainties increase significantly due to sparse photometric sampling, which can also introduce biases in the derived stellar masses \citep[e.g.,][]{Forrest2024}. 
To mitigate this, surveys designed to split the $J$, $H$, and $K_s$ bands to better sample the Balmer/4000 \AA\ break of galaxies at $1 < z < 3$ in the NEWFIRM Medium-Band Survey \citep{Whitaker2011} and the FourStar galaxy evolution survey \citep[ZFOURGE, ][]{Straatman2016}, and at $z>4$ in the FENIKS survey \citep{Esdaile2021}. 
These surveys led to the identification of a large population of quiescent galaxies \citep{Whitaker2012, Straatman2014}, some of which have been confirmed spectroscopically \citep{Glazebrook2017, Antwi-Danso2025}. 

More recently, deep JWST NIRCam and MIRI imaging enable rest-frame optical and near-IR sampling of high-redshift galaxies, substantially improving the robustness of the photometric redshift and galaxy properties estimates. 
However, even with JWST NIRCam photometry, rest-frame $J$-band fluxes at $z \gtrsim 3$ remain less accurate because they require extrapolation beyond the observed wavelength range, which can lead to misclassification of galaxies \citep{Antwi-Danso2023}. 
Secondly, rest-frame color selections such as $UVJ$ can be significantly affected by strong nebular emission lines, particularly \hb+\oiii\ can boost the rest-frame $V$-band flux and bias color-based classifications. 
Moreover, 20–30\% of sources located within the $UVJ$ quiescent region could have substantial ongoing star formation \citep{Forrest2020b}, and $UVJ$ color-selection also tends to miss the recent quiescent or post-starburst galaxies, which seem to become increasingly prevalent at $z > 3$ \citep[e.g.,][]{Belli2019, Baker2025a}. 

As an alternative, new color-based selections have been proposed \citep{Antwi-Danso2023, Long2024}. In addition, many recent studies have adopted specific star formation rate (sSFR) thresholds to identify quiescent galaxies, as sSFR provides a more direct indicator of star formation activity \citep[e.g.,][]{Gallazzi2014, Pacifici2016, Carnall2023b, Carnall2024, Baker2025a}.  These sSFR methods are improved by using spectroscopy that includes coverage of SFR-sensitive emission lines such as the Balmer lines (\ha\ and \hb).  

In this paper, we present a spectroscopically confirmed sample of quiescent galaxies at $2 < z < 5$ selected primarily based on their low sSFRs derived from joint spectral and photometric SED fitting. The spectral data are drawn from CAPERS (GO-6368, PI M. Dickinson), a JWST Cycle 3 Treasury Program designed to obtain deep NIRSpec/prism spectroscopy for galaxies in three legacy CANDELS fields \citep{Grogin2011,Koekemoer2011}: the Extended Groth Strip (EGS), the Ultra-deep Survey (UDS), and the Cosmic Evolution Survey (COSMOS). 
Using this spectroscopically confirmed sample, we reassess the completeness and purity of various photometry-based quiescent-galaxy selection methods at high redshift.
We then derive the number density of quiescent galaxies using the full photometric sample, applying completeness and purity corrections based on the spectroscopic sample. 

The outline for this paper is as follows. 
In Section \ref{sec:data}, we describe the spectroscopy and photometry datasets 
In Section \ref{sec:methods}, we describe the SED fitting methods used to derive stellar population properties from spectroscopic and broadband photometric data, and emission line and spectral break measurements. 
In Section \ref{sec:sample}, we present the selection of both photometric quiescent galaxies and spectroscopically confirmed quiescent galaxies (``QGs'' hereafter). 
In Section \ref{sec:results}, we present the properties of the QGs and compare them with those selected through various photometry-based selection methods. 
In Section \ref{sec:discussion}, we present the number density of quiescent galaxies at $2 \leq z \leq 5$, and compare them with observational measurements and predictions from seven galaxy formation simulations. We also compare the formation and quenching times of quiescent galaxies across different cosmic epochs based on the literature studies. 
Finally, we summarize our findings in Section \ref{sec:summary}. 
Throughout this paper, all magnitudes are presented in the AB system \citep{Oke1983, Fukugita1996}. We adopt a standard $\Lambda$–cold dark matter ($\Lambda$CDM) cosmology with $H_0$ = 70 km s$^{-1}$, $\Omega_{\Lambda,0}$ = 0.70, and $\Omega_\mathrm{M,0}$ = 0.30. 

\section{Data} \label{sec:data}

In this Section, we describe the spectra and photometry data used.

\subsection{CAPERS spectra} \label{sec:data_spec}

CAPERS is a JWST Cycle 3 Treasury Program designed to obtain deep NIRSpec prism spectra of galaxies selected from EGS, UDS, and COSMOS. 
CAPERS targeted 7 pointings in each of the three fields using the NIRSpec disperser/filter configuration of PRISM/CLEAR (R $\sim100$), which provides wavelength coverage from 0.6 to 5.3 $\mu$m. 
{Targets were selected using a homogeneous priority scheme across the three fields. Each source was assigned a base weight defined as an exponential function of its photometric redshift. For sources fainter than the estimated continuum-detection threshold of 28 AB mag, this weight was further attenuated according to source brightness. The reference magnitude was defined as the brightest available measurement among the F277W, F356W, and F444W NIRCam bands. }

For each pointing, CAPERS designed three NIRSpec MSA configurations with the lower-priority objects being replaceable. 
Therefore, this setup enables high-priority objects to be observed multiple times to achieve a high signal-to-noise ratio (S/N), while observing a large sample of lower-priority objects to maximize scientific yield. 
Each MSA configuration is carried out using a three-shutter slitlet with nodding between the three shutters.
The effective exposure time of each MSA configuration is 5,690 s, and the total effective exposure time of each pointing is 17,079 s.
The NIRSpec data were reduced using the JWST Calibration Pipeline\footnote{\url{https://github.com/spacetelescope/jwst}} \citep[][]{bushouse2025} with several customized steps similar to \citet{ArrabalHaro2023}. 
{All spectra data were reduced using pipeline v1.17.1 and Calibration Reference Data System (CRDS) context pmap 1350.}
The 1D spectra are then extracted using an optimal extraction \citep{horne1986}. 
A detailed description of target selection and data reduction will be presented in a forthcoming survey paper. 

\begin{figure*}
    \centering
    \includegraphics[width=\textwidth]{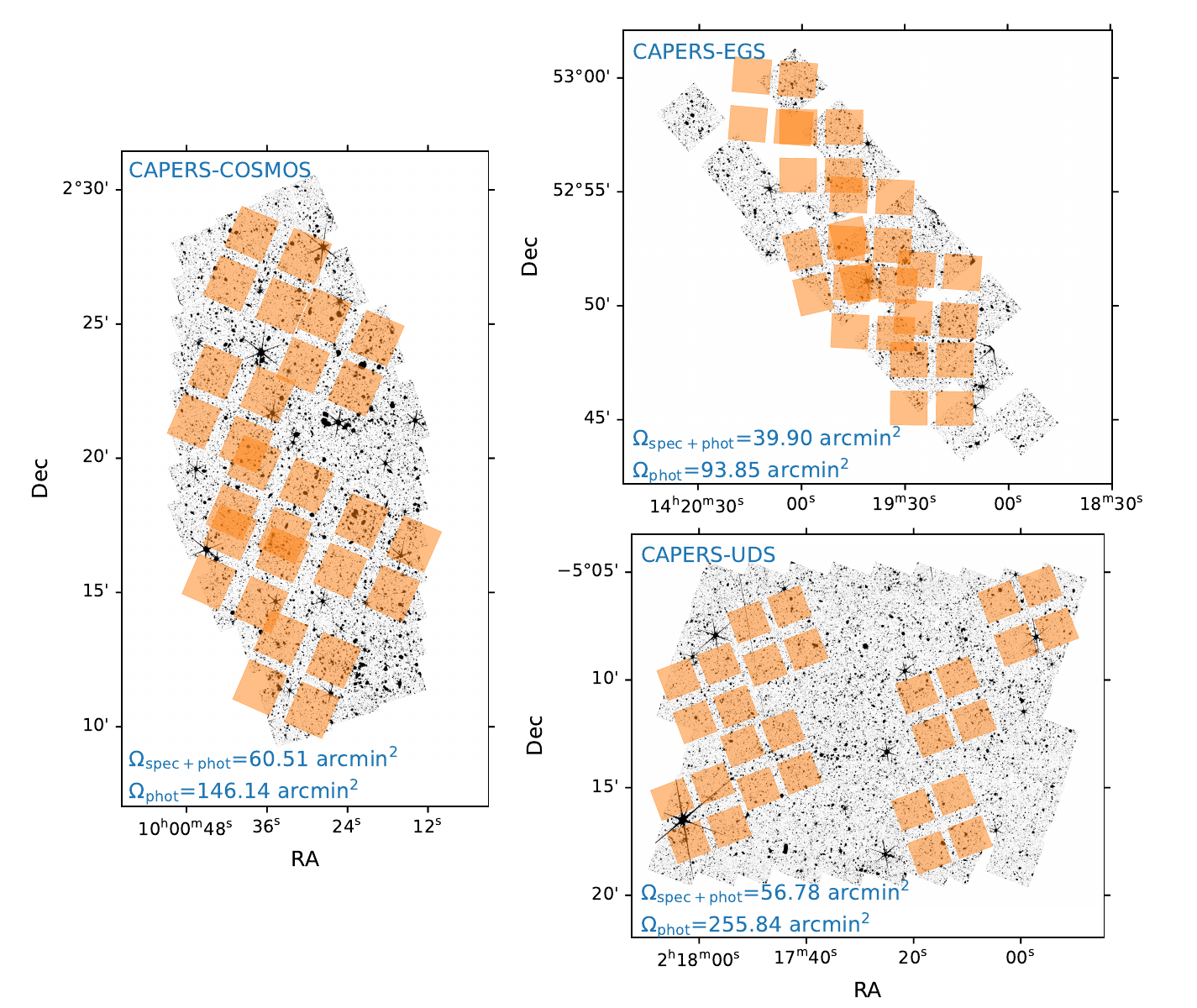}
    \caption{The current CAPERS spectroscopic survey footprints (orange) overlaid on the \jwst/NIRCam images (F277W+F356W) from PRIMER (COSMOS and UDS fields) and CEERS (EGS field). These are the MSA pointings in the current study. An additional three MSA pointings are planned in the UDS fields. The overlapping survey areas covered by both the CAPERS spectroscopy and photometric catalogs, as well as the photometry survey areas, are indicated in each panel. The total survey areas of CAPERS spectroscopy covered in this study and the full photometry are 138.61 and 495.84 arcmin$^2$, respectively. } 
    \label{fig:capers_area}
\end{figure*}

For this study, we used CAPERS {observations comprising seven pointings in each of the UDS, EGS, and COSMOS fields (see Figure \ref{fig:capers_area})}. 
The total number of sources with spectra to date is {8169}.  

The spectroscopic redshifts of galaxies are measured using a variety of software, including a modified version of Marz \citep{Hinton2016} \citep[see details in][]{Hu2024}, a modified version of \textsc{MSAEXP} \citep{Brammer2022}, \bagpipes\ \citep{Carnall2019}, \cigale\ \citep{Burgarella2025}. 
The results are then combined and vetted by the CAPERS team. 
In total, {4820} galaxies have secure spectroscopic redshifts up to $z \sim 11$ with high-quality spectra. 
Because our SED fitting requires NIRCam photometry, some galaxies selected from the CANDLES HST catalog \citep{Stefanon2017} are excluded, which restricts the sample to the {3672} galaxies within the NIRCam footprint from CEERS and PRIMER (see Figure~\ref{fig:capers_area}).

{The effective exposure times of the spectra for our QGs range from $3.79$ to $11.38$ ks, with a median of $5.69$ ks. We characterize the spectral quality using the S/N per spectral pixel near rest-frame $3800\,\text{\AA}$, blueward of the 4000-$\text{\AA}$ break. Among the 17 QGs with coverage at this wavelength, the median S/N per pixel is 38.6, with a 16th–84th percentile range of 22.3–66.8 and a full range of 9.7–92.0. The other two QGs (UDS-35916 and EGS-12953) lack spectral coverage at rest-frame $3800\,\text{\AA}$.}

\subsection{Optical/NIR Imaging and Photometry} \label{sec:data_phot}

We utilized a comprehensive set of deep imaging taken with \hst\ and \jwst\ available in the COSMOS, EGS, and UDS fields. 
This includes \hst\ ACS F435W, F606W, and F814W from the latest reductions processed as part of the Cosmic Assembly Near-IR Deep Extragalactic Legacy Survey (CANDELS, \citealp{Koekemoer2011, Grogin2011}). 
For \jwst, we include NIRCam imaging in the F090W, F115W, F150W, F200W, F277W, F356W, F410M, and F444W bands. These data were obtained from the Cosmic Evolution Early Release Science survey \citep[CEERS;][]{Finkelstein2025} for the EGS field, and from the Public Release IMaging for Extragalactic Research survey (PRIMER; Dunlop et al., in prep) for the COSMOS and UDS fields. 

We adopt the photometric catalogs as detailed in \citet{Finkelstein2024} as parts of the Uniform Near-Infrared CatalOg from Robust imagiNg \citep[UNICORN;][Finkelstein et al.\ in prep.]{}. 
{We briefly summarize the relevant procedures here. All \hst/ACS bands and all \jwst/NIRCam bands bluer than F277W are matched to the F277W PSF. For bands with larger PSFs, correction factors are determined by convolving the F277W image to the PSF of each band and measuring the ratio of the flux in the native F277W image to that in the convolved image. Photometry was performed with Source Extractor \citep{Bertin1996} using a weighted sum of the  F277W and F356W as the detection image. Fluxes are measured within Kron apertures using a Kron factor of 1.1 and a minimum radius of 1.6. 
The PSF corrections for bands with PSFs larger than that of F277W, together with aperture corrections to total flux, are then applied to the measured fluxes.}
The corresponding flux uncertainties are empirically derived and include the same aperture corrections. 
Photometric redshifts were estimated with \textsc{eazy}, following the methodology in \citet{Finkelstein2023}, including templates from \citet{Larson2023}. 

Rest-frame colors are also obtained from \textsc{eazy} using photometric redshift or spectroscopic redshift if they exist from CAPERS. 

\section{Methods} \label{sec:methods}

In this section, we describe the methods used to obtain stellar population properties of galaxies in our sample. 
Specifically, we perform three stages of SED fitting: (1) an initial fit using \cigale\ to remove low-mass galaxies, (2) \bagpipes\ photometry-only fitting to derive the properties of photometrically selected galaxies, and (3) joint spectroscopic–photometric \bagpipes\ fitting to obtain the properties of galaxies with CAPERS spectroscopy.
We primarily adopt the measurements from \bagpipes\ for the following analysis, as \bagpipes\ allows for joint fitting of photometric and spectroscopic data. 
The configurations of SED fittings are described in Sections \ref{sec:cigale}, \ref{sec:bagpipes-phot}, and \ref{sec:bagpipes-spec}, while the details of sample selection will be presented in Section \ref{sec:sample}.
We then present the measurement of emission line fluxes in Section \ref{sec:spec_fitting}.

\subsection{CIGALE} \label{sec:cigale} 

This study is based on a photometric quiescent galaxy sample selected from the full photometric catalogs of CEERS, PRIMER-UDS, and PRIMER-COSMOS fields (see Section \ref{sec:sample_phot}). 
Performing SED fitting with \bagpipes\ to the full photometric catalog requires substantial computational resources.
To expedite the selection process, we utilize the SED fitting Code Investigating GALaxy Emission (\cigale) \citep{Boquien2019, Yang2020} for a preselection to remove the abundant low-mass galaxies.
We adopt simplified models to enable a fast and approximate stellar mass estimate for the sample.

We adopt the photometric redshifts or spectroscopic redshifts from CAPERS when available.
We adopt a delayed star formation history (SFH) allowing $\tau$ and stellar age to vary from 0.05--10~Gyr and 0.01--10~Gyr, respectively. 
We assumed a \citet{Chabrier2003} IMF and the stellar population synthesis models presented by \citet{Bruzual2003} with metallicity {vary between 0.004, 0.008 and 0.02 (0.4~$Z_\odot$, 0.2~$Z_\odot$, and $Z_\odot$, respectively)}. 
We include nebular emission using templates of \citet{Inoue2011}. We allow the ionization parameter $\log(U)$ to vary between $-3$ to $-1$, the gas metallicity ($Z_\mathrm{gas}$) to vary between 0.002, 0.005 and 0.02, 
and a fixed electron density of 100~cm$^{-3}$. 
{We adopt the \citet{Calzetti2000} attenuation law} for attenuating the stellar continuum, and the \citet{Cardelli1989} extinction law with $R_V=3.1$ for attenuating the emission lines. 
We allow the dust attenuation in emission lines from nebular regions \ebvg\ to vary from 0 to {1.8}, and a fixed dust attenuation ratio between emission lines and stellar continuum (\ebvs/\ebvg $= 0.44$).  

Because the primary purpose of this SED fitting is to exclude low-mass galaxies,  we only adopt the stellar mass from the \cigale\ SED fitting (hereafter $M_\mathrm{*,cigale}$). 

\subsection{BAGPIPES with photometric-only data} \label{sec:bagpipes-phot}

We use the Python package \bagpipes\ \citep{Carnall2018, Carnall2019} to fit the massive galaxies selected from the photometric sample. 
We employ the default Stellar Population Synthesis models from \citet{Bruzual2003}.  
For the star formation history parameterization, we use the Gaussian Process model from \texttt{dense\_basis} \citep{Iyer2019}, where the star formation history is split into 4 dynamically adjusted time bins, and during each time bin, $25\%$ of the total stellar mass is formed.
The stellar metallicity is allowed to vary between 0.00355 and 3.5~$Z_\odot$.  
We adopt the Calzetti dust attenuation law \citep{Calzetti2000} with $A_V$ ranging from 0.0 to 8.0~mag.  
We include nebular emission with the metallicity of the gas equal to that of the stellar populations, and an ionization parameter, $\log U$, in the range $-4$ to $-1$.

We adopt the photometric redshifts or CAPERS spectroscopic redshifts when available. 
{For every galaxy, regardless of whether the adopted redshift is photometric or spectroscopic, we allow the redshift to vary within} $\Delta z = \pm$0.25 to account for the redshift uncertainties. 
For galaxies with CAPERS spectroscopic redshift, the broader redshift prior ($\Delta z= \pm 0.25$) ensures that the \bagpipes\ results largely do not rely on the spectroscopic redshift. 

From \textsc{Bagpipes}, we adopt the posterior median values of stellar mass ($M_{*}$), SFR averaged over the past 10~Myr ($\mathrm{SFR_{10}}$), SFR averaged over the past 100~Myr ($\mathrm{SFR_{100}}$), and dust attenuation ($A_V$). 
{To preserve the covariance between stellar mass and SFR, we calculate $\mathrm{sSFR_{100}}=\mathrm{SFR_{100}}/{M_*}$ for each posterior sample and adopt the median of the resulting $\mathrm{sSFR_{100}}$ posterior.} 
We also adopt the formation time ($t_{50}$) and quenching time ($t_{90}$), defined as the {lookback time from the observed epoch at which the} galaxy had formed 50 and 90 per cent of its total formed stellar mass, without accounting for mass loss due to stellar evolution. 
Correspondingly, we define $z(t_{50})$ and $z(t_{90})$ as the redshifts at which these mass fractions were reached. 
The uncertainties on all derived parameters correspond to the 16th and 84th percentiles of the posterior distributions of SED fitting. 
To distinguish between results derived from photometry-only and joint spectroscopy and photometry fitting (as described in Section \ref{sec:bagpipes-spec}), we use subscript notation for the photometry-only case (e.g., $M_\mathrm{*,p}$ denotes stellar mass estimated using only photometry data). Results without this subscript refer to the full spectral fitting. 

\subsection{BAGPIPES with spectroscopy data} \label{sec:bagpipes-spec}

For galaxies with CAPERS spectra, we perform joint spectroscopic–photometric fitting using \bagpipes.
We adopt the same \bagpipes\ configuration described in \ref{sec:bagpipes-phot}. 
To incorporate NIRSpec spectroscopy, 
we adopt a multiplicative scaling factor (assumed to be a second-order Chebyshev polynomial) in \bagpipes\ to estimate the wavelength-dependent flux calibration to the broadband photometry. 
We allow the zeroth order to vary from 0.1 -- 10, while the first and second orders are allowed to vary from $-1$ to 1. 
To account for the variable resolution of the NIRSpec spectrum, we utilize the resolution model for the prism spectra from the JWST User Documentation\footnote{\url{https://jwst-docs.stsci.edu/jwst-near-infrared-spectrograph/nirspec-instrumentation/nirspec-dispersers-and-filters}}.
We adopt the vetted spectroscopic redshifts from CAPERS and allow a small varying range of $\Delta z = \pm$0.05 to account for uncertainties from the low-resolution spectra. 

\begin{figure*}
    \centering
    \includegraphics[width=\textwidth]{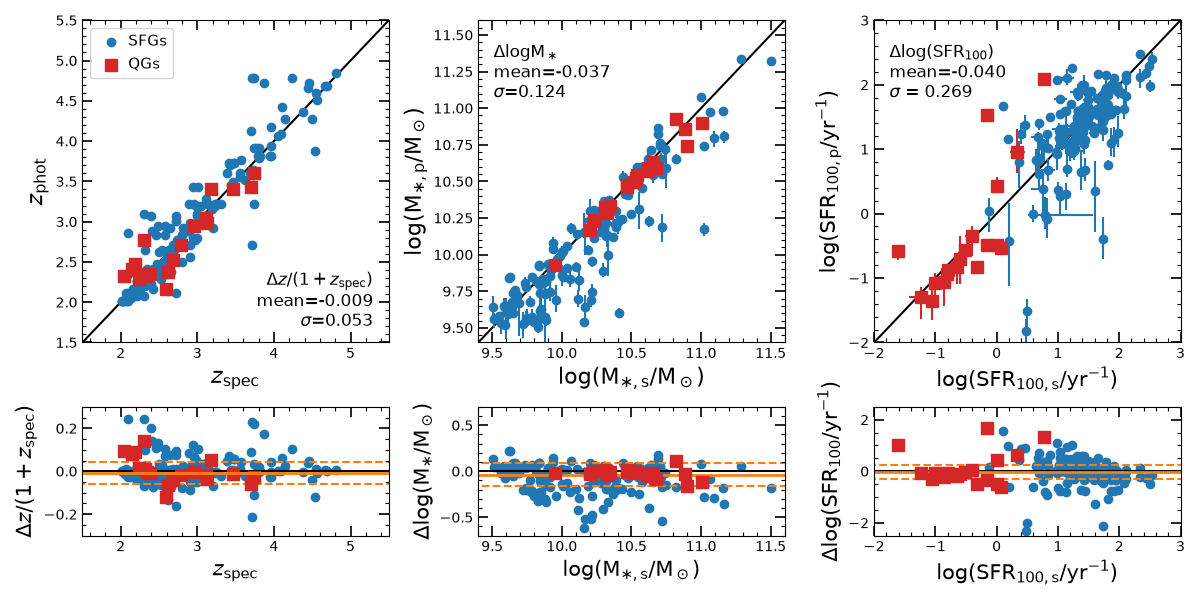}
    \caption{\textit{Top left:} spectroscopic versus photometric redshifts for SFGs (blue dots) and QGs (red squares). The mean and $1\sigma$ scatter derived from fitting a Gaussian to $\Delta z/1+z_\mathrm{spec}$ are indicated in the bottom-right corner. \textit{Top middle and right:} Stellar masses and $\mathrm{sSFR_{100}}$ derived from full spectral fitting compared to those from photometric-only data. The mean and $1\sigma$ scatter derived from fitting a Gaussian to $\Delta \log {M_\ast}$ and $\Delta \log \mathrm{SFR_{100}}$ are indicated in the top-left corner. \textit{Bottom:} spectroscopic redshift (\textit{left}), stellar mass (\textit{middle}), and $\mathrm{SFR_{100}}$ (\textit{right}) versus their difference relative to those derived from photometric-only data. The black line shows zero difference, and the orange lines show the mean and $\pm~1\sigma$ range for the combined sample of SFGs and QGs. } 
    \label{fig:delta_z_mass_sfr}
\end{figure*} 

Because the spectra provide valuable information, including emission lines, absorption features, and spectral breaks, we treat the results from the full spectral fits as the more reliable reference.
We quantify the systematic offsets and scatter in the parameters derived from photometry-only fits by comparing them to those obtained from the full spectral fits, as shown in Figure~\ref{fig:delta_z_mass_sfr}. 
We separate the spectroscopic sample into QGs and SFGs, as described in Section \ref{sec:sample_spec}. 
In details, we fit Gaussian functions to the distributions of $(z_\mathrm{phot} - z_\mathrm{spec})/(1+z_\mathrm{spec})$, $\log(\mathrm{M_{*,p}})-\log(\mathrm{M_{*,s}})$, and $\log(\mathrm{SFR_{100,p}})-\log(\mathrm{SFR_{100,s}})$ and adopt the best-fit means and standard deviations as the offsets and scatters. 
The resulting offsets and scatters are {$-0.009$} and {$0.053$} for $\Delta z/(1+z_\mathrm{spec})$, 
{$-0.037$} dex and {$0.124$} dex for $\Delta \log(M_\ast)$, 
and {$-0.040$} dex and {$0.269$} dex for $\Delta \log(\mathrm{SFR_{100}})$ (see Figure~\ref{fig:delta_z_mass_sfr}). 
For the QGs, the median offsets are {$\Delta z/(1+z_\mathrm{spec}) = -0.008$, $\Delta \log(M_\ast) = -0.021$ dex, and $\Delta \log(\mathrm{SFR_{100}}) = -0.082$ dex}, all within the scatter derived from the spectroscopic sample. 
We noticed that the spectroscopic and photometric stellar masses agree well for the QGs, but the SFGs show a significant tail of objects with $M_{*,p} < M_{*,s}$.  
On the other hand, the SFRs comparison shows significantly larger scatter. 
{In particular, three QGs (UDS-6050, COSMOS-43671, and EGS-12661) show significantly higher $\mathrm{SFR_p}$ than spectroscopically derived values by more than 1 dex. 
For COSMOS-43671, the photometric fluxes are $\sim3$--$4$ times higher than the corrected spectrum at rest-frame $2000$--$3000~\text{\AA}$, due to the contamination from a nearby low-redshift interloper (at $z\sim0.8$, see Figure~\ref{fig:QGs_z2z3_1}). 
UDS-6050 shows the largest absolute offset, $\Delta\mathrm{SFR}=118.6~M_\odot\,\mathrm{yr}^{-1}$. Its photometry-only fit favors a shallower Balmer/D4000 break and stronger nebular emission. The presence of Mg\,{\sc ii} emission suggests AGN activity, which is not included in our SED model and may therefore be misinterpreted as recent star formation. For EGS-12661, the $>1$ dex offset arises because both SFRs are very low, and their absolute difference is only $\sim0.24~M_\odot\,\mathrm{yr}^{-1}$. This discrepancy could be dominated by the uncertainty of the SED fitting and does not affect its classification as a quiescent galaxy. Thus, the origin of the SFR discrepancy may differs among galaxies. Possible causes include photometric blending, unmodeled AGN emission and uncertainties associated with measuring extremely low SFRs. Spatial color gradients and aperture effects may also contribute. }

\subsection{Emission line and Spectral Break Measurements} \label{sec:spec_fitting}

All spectral measurements are performed on spectra after applying the calibration from the full spectral \bagpipes\ results. 
To account for stellar absorption, we subtract the underlying stellar continuum using the best-fit \bagpipes\ model before fitting the emission lines. 
We then simultaneously fit \hb, \oiii, \ha, {\nii}, and \sii\ emission lines. 
In detail, we assume a single Gaussian profile for \hb\ and \ha, emission lines, and two double Gaussian profiles for \oiii, {\nii}, and \sii. 
The central wavelengths of all Gaussian profiles are tied to {that of} \ha. {Doublet ratios of \oiii\ $\lambda$4960/$\lambda$5008 and \nii\ $\lambda$6550/$\lambda$6585 were fixed to 1/2.98 and 1/2.96, respectively, while the \sii\ doublet ratio was allowed to vary. The two components of each doublet were constrained to have the same Gaussian width. Uncertainties were estimated using 100 Monte Carlo realizations. In each realization, the observed spectrum was perturbed according to its flux uncertainties, and a fitted stellar continuum spectrum was randomly drawn from the \bagpipes\ posterior before repeating the emission-line fit. The 16th and 84th percentiles of the resulting line-flux distributions were adopted as the lower and upper uncertainty bounds, respectively.}
For non-detections ($<$3$\sigma$), we estimate the line flux upper limit using the FWHM of the spectrum resolution at the line wavelength. 
The 1$\sigma$ flux uncertainty is calculated by multiplying the spectral errors by $\sqrt{\mathrm{FWHM}}$. We adopt the 3$\sigma$ value as the upper limit. 

In addition, we calculate the 4000 \AA\ break ($D_n4000$), and the Balmer break at 3646 \AA\ ($D_B$). 
The 4000-\AA\ break spectral feature ($D_n4000$) is prominent in galaxies with evolved stellar populations, typically with an age of $>$1–2 Gyr. 
This break originates from the absorption of stellar continuum by ionized metals in the atmospheres of stars (including Ca II H and K), and it increases monotonically with stellar age, with an additional dependency on metallicity \citep{Kauffmann2003a}.
The $D_n4000$ index is defined as the ratio of the median flux in the 4000–4100 \AA\ and 3850–3950 \AA\ windows \citep{Balogh1999}. 

Balmer break at 3646 \AA\ ($D_B$) originates from the absorption of stellar continuum by the hydrogen Balmer series and is strongest in A-type stars. 
It peaks for stellar populations with ages of $\sim0.3$–$1$ Gyr and declines thereafter \citep{Kriek2006}. 
The Balmer break is defined as the ratio of the median flux in the 3800–3950 \AA\ and 3500–3650 \AA\ windows \citep{Kriek2006}. 
The uncertainties on the $D_n4000$ and $D_B$ are estimated from the 16th and 84th percentiles of Monte Carlo sampling, using the 1D spectra uncertainties in the same method as described in the emission line fluxes.

\section{Sample Selection} \label{sec:sample}

 \begin{figure*}
    \centering
    \includegraphics[width=\textwidth]{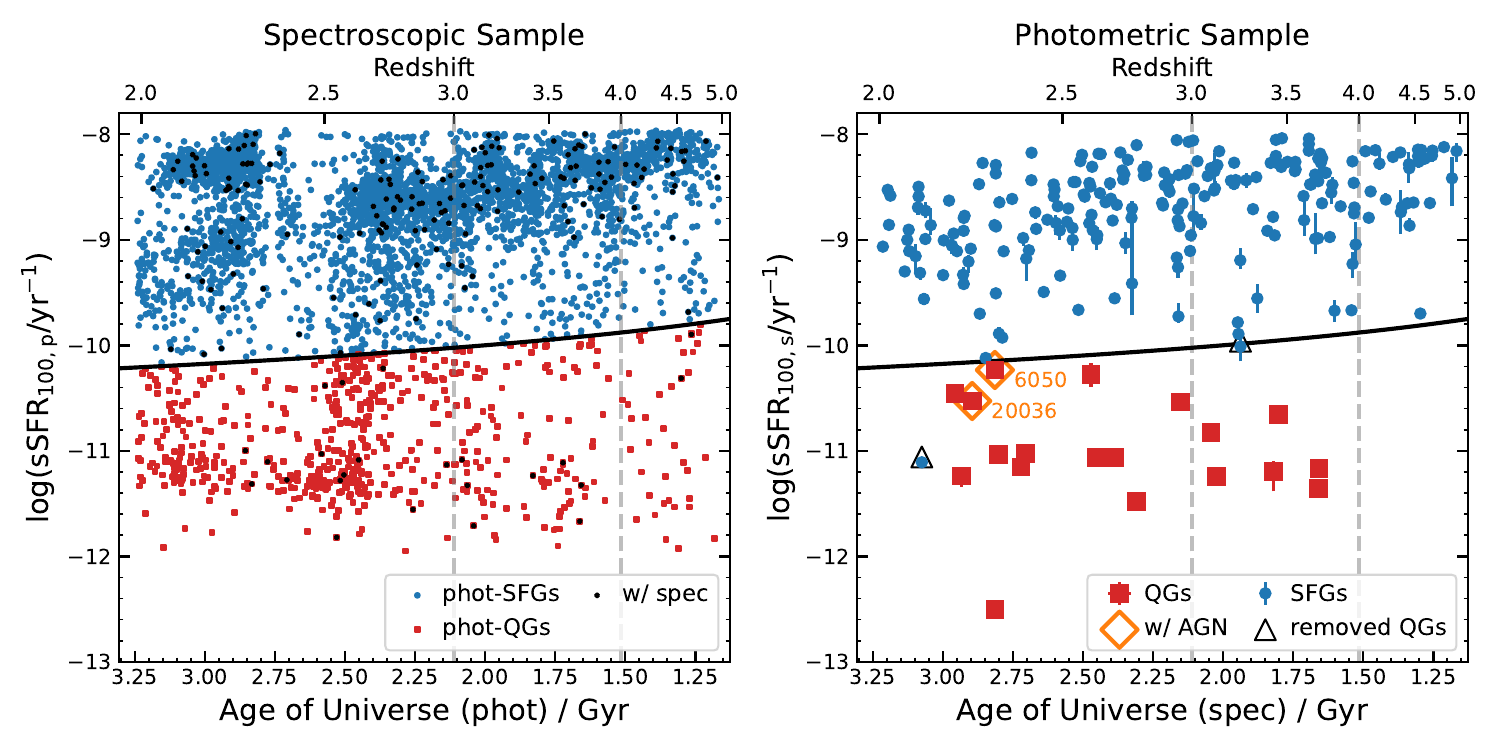}
    \caption{\textit{Left:} Age of the universe at observed redshift versus $\mathrm{sSFR_{100}}$ for photometric quiescent galaxies (red squares) and photometric SFGs (blue dots) selected from the photometric sample at $2 \leq z \leq 5$.  The quiescent selection criteria $\mathrm{sSFR_{100}} \leq 0.20/t_\mathrm{obs}$ is shown as a solid black line. \textit{Right:} The same quantities for QGs (red squares) and SFGs (blue dots) selected from the CAPERS spectroscopic sample at $2 \leq z \leq 5$. Two galaxies (COSMOS-34313 and EGS-10563) removed from the quiescent sample after visual inspection are marked with open black triangles. Two galaxies hosting AGN (UDS-6050 and EGS-20036) are shown as orange open diamonds, with their ID labeled.} 
    \label{fig:QGs_selection}
\end{figure*}

In this section, we describe the criteria for selecting photometric quiescent galaxies from the photometric catalog and spectroscopically confirmed quiescent galaxies from the CAPERS data. 
The properties of the spectroscopically confirmed quiescent galaxies are presented in Section~\ref{sec:results}. 
These two samples are combined in Section \ref{sec:number_density} to determine the number density of massive quiescent galaxies.

\subsection{Selection of photometric quiescent galaxies}\label{sec:sample_phot}

For the photometric sample, we pre-select galaxies with photometric redshifts in the range $1.75 < z_\mathrm{phot} < 5.25$ and require $(z_{68\%~\mathrm{upper}} - z_{68\%~\mathrm{lower}})/(1 + z_\mathrm{phot}) \leq0.5$, where $z_\mathrm{phot}$ is the best-fitting photometric redshift and $z_{68\%~\mathrm{upper}}$ and $z_{68\%~\mathrm{lower}}$ are the 68\% confidence values on the redshift from \eazy. 
The latter criterion ensures that we only include galaxies with relatively well-constrained photometric redshifts. 
{We also remove galaxies with missing NIRCam coverage in two or more of the F090W, F115W, F150W, and F200W filters, as these bands are critical for constraining quiescent features such as the Balmer/4000~\AA\ break in galaxies at $z \sim 2$–$5$. }

Additionally, we apply a \cigale\ mass limit of $M_\mathrm{\ast, cigale} \geq 10^{9} M_\odot$ to further reduce the sample size. 
This threshold lies $0.5$ dex below the final mass limit of $M_\ast\geq10^{9.5}\,M_\odot$, providing a buffer for uncertainties and systematic differences in stellar mass. A comparison between the \cigale-derived masses and those obtained from the joint spectroscopic+photometric \bagpipes\ fits yields an offset of $-0.17$ dex and a scatter of $0.14$ dex, both smaller than the $0.5$-dex buffer. The initial cut is therefore unlikely to exclude galaxies above the final mass threshold. 
The resulting preselected photometric sample contains {10652} galaxies.

This preselected photometric sample is then fitted with \bagpipes\ using photometry-only data, as described in Section \ref{sec:bagpipes-phot}. 
We select galaxies with $M_\mathrm{\ast, p} \geq 10^{9.5} M_\odot$ and at $2 < z_\mathrm{bagpipes, p} < 5$ as the photometric galaxy sample. 
This mass limit is adopted because no spectroscopically confirmed quiescent galaxies are found below this threshold.

We then select quiescent galaxies using an sSFR–$t(z)$ criterion, which has been widely used for quiescent galaxy selections at high redshift \citep[$z>2$, ][]{Gallazzi2014, Pacifici2016, Carnall2023b, Carnall2024, Baker2025a}: 
\begin{equation} \label{eq:ssfr-tz}
    \mathrm{sSFR_{100}} \leq \frac{0.2}{t_\mathrm{obs}(z)}, 
\end{equation}
where $\mathrm{sSFR_{100}}$ is the SFR averaged over the past 100 Myrs divided by the total stellar mass formed, and $t_\mathrm{obs}$ is the age of the Universe at the redshift of the galaxy. 
The $\mathrm{sSFR_{100}}$ and redshift are the 50th percentile of posterior sampling from \bagpipes\ with photometry-only data. 

To account for the \bagpipes\ fitting uncertainties and systematic differences between parameters derived from the full spectral fitting and those from photometry-only data fitting, we incorporate these uncertainties using a Monte Carlo approach, and then apply the selection criteria on redshift, stellar mass, and sSFR. 
The \bagpipes\ uncertainties are relatively small, with median values of 0.04 in redshift, 0.04 dex in stellar mass, and 0.07 dex in $\mathrm{SFR_{100}}$. 
The systematic offsets and scatters between parameters derived from the full spectral fitting and those from photometry-only data fitting are described in Section \ref{sec:bagpipes-spec}. 
For each galaxy, redshift and stellar mass are drawn from a Gaussian distribution centered on photometric estimates shifted by the mean offset derived from the spectroscopic comparison, and the quadrature sum of the \bagpipes\ uncertainties and the scatter derived from the spectroscopic comparison as the standard deviation. 
For SFR, only the \bagpipes\ uncertainties and scatter are considered in the Monte Carlo sampling. 
The offset on SFR derived from the spectroscopic comparison is not included here, because the large deviations cannot to be fully captured by Monte Carlo sampling. 
Instead, we correct the number of photometric quiescent galaxies using the true-positive, true-negative, and false-positive rates of quiescent versus star-forming classification measured from the spectroscopic sample (see Section~\ref{sec:number_density}). 
We then apply the redshift, stellar mass, and sSFR-$t(z)$ selection criteria to obtain a mock photometric quiescent sample, and perform $N=1000$ Monte Carlo realizations.  
This procedure results in a mock photometric quiescent sample for each iteration of the Monte Carlo, where the number of quiescent galaxies changes by $\pm$2\% (16-84\%-tile). 

{We estimate the photometric stellar-mass completeness using the limiting-mass method. We conservatively adopt the F356W 5$\sigma$ depth of 28.3 AB mag for PRIMER-UDS, measured in a $0.3\arcsec$-diameter aperture \citep{Donnan2024}. This is the shallowest F356W depth among the three fields. For each galaxy, we calculate the limiting stellar mass defined as the mass it would have if its observed F356W flux were scaled to the adopted 5$\sigma$ limiting flux while keeping its redshift and mass-to-light ratio fixed. In each redshift interval, we select the faintest 20\% of galaxies in F356W and adopt the 90th percentile of their limiting stellar mass distribution as the 90\% stellar-mass completeness limit. We obtain stellar mass completeness limits of $\log(\mathrm{M_\ast/M_\odot})=7.74$, $8.01$, and $7.94$ at $2<z<3$, $3<z<4$, and $4<z<5$, respectively. These limits lie well below our adopted selection threshold of $\log(\mathrm{M_\ast/M_\odot})=9.5$, indicating that the photometric parent sample is sufficiently mass complete for our analysis. }

The selection criteria and sample size for the photometric quiescent galaxies without MC sampling are summarized in Table~\ref{tab:sample}. 
The distributions of sSFR versus $t(z)$ for photometrically selected quiescent galaxies and SFGs are shown in the {left} panel of Figure \ref{fig:QGs_selection}, where the systematic offset is included, but uncertainties and scatter are not accounted for.

\subsection{Selection of spectroscopic quiescent galaxies} \label{sec:sample_spec}

From the preselected photometric sample ($M_{\ast,\mathrm{cigale}} \geq 10^{9}\,{M_\odot}$), we select galaxies with spectroscopic redshifts in the range $2 \leq z \leq 5$ and perform full spectroscopic \bagpipes\ fitting for these galaxies (see Section \ref{sec:bagpipes-spec}). 
We further restrict the sample using the \bagpipes\ results, selecting galaxies with stellar mass $M_\mathrm{\ast,bagpipes} \geq 10^{9.5} M_\odot$. 
We then select {QGs} using the sSFR–$t(z)$ criterion (equation \ref{eq:ssfr-tz}). 

{We visually inspect all spectra along with the \bagpipes\ fitting results and remove sources from spectroscopic sample that are affected by poor spectral data quality or photometric blending. We identify three galaxies that are misclassified as quiescent due to these issues, including two exhibit artificial negative continuum fluxes caused by poor background subtraction when bright neighboring sources fall within the NIRSpec MSA slit, and the third shows excess blue photometric emission relative to its spectrum because of blending with a neighboring galaxy, resulting in an unreliable SED fit.} 
We also identified two galaxies (COSMOS-34313 and EGS-10563) for which \bagpipes\ prefers low-sSFR models, but whose spectra show significant dust content with $A_V=2.3$ and 2.0, more consistent with dusty SFG spectra. In addition, they show a discrepancy in spectroscopic-to-photometric calibration blueward of 4000\AA\ break. 
Using photometry-only data fitting, these galaxies are not classified as quiescent. Including mid- and far-infrared photometry, we further confirm that these are dusty SFGs (also see more discussion in Section~\ref{sec:compare_selection}). 
{We therefore reclassify these two objects as SFGs}.

In total, we identify {19} QGs from the sSFR–$t(z)$ selection. 
The selection criteria and sample size for the spectroscopic sample and spectroscopic quiescent galaxies are summarized in Table~\ref{tab:sample}. 
The distributions of sSFR versus $t(z)$ for QGs and SFGs in the CAPERS spectroscopic sample are shown in the {right} panel of Figure \ref{fig:QGs_selection}. 
The image cutouts, CAPERS spectra, photometry, best-fit SEDs, and recovered star formation histories are shown in Figure~\ref{fig:QGs_z3z4} for the four QGs at $z>3$. 
The remaining QGs at $2 < z < 3$ are shown in the appendix (Figure \ref{fig:QGs_z2z3_1} and \ref{fig:QGs_z2z3_2}).

{The adopted SFH models can affect measured properties, particularly the SFR, and therefore the classification of quiescent galaxies. We assess this effect by rerunning \bagpipes\ with double-power-law and delayed-$\tau$ models.  
Of the 19 QGs selected using the fiducial non-parametric SFH, 13 and 10 are also classified as quiescent using the double-power-law and delayed-$\tau$ models, respectively (see Appendix~\ref{app:diff_SFHs} and Figure~\ref{fig:ssfr_varingSFH}). 
The galaxies whose classifications change are primarily those with fiducial $\log(\mathrm{sSFR}/\mathrm{yr}^{-1})\sim-11$ to $-10$, near the quiescence boundary. 
For the 19 fiducial QGs, the median reduced $\chi^2$ values are $3.99$, $3.93$, and $8.14$ for the fiducial non-parametric, double-power-law, and delayed-$\tau$ models, respectively, with corresponding 16th–84th percentile ranges of 1.96–10.72, 2.22–15.91, and 2.27–22.43. 
Although the non-parametric and double-power-law models have comparable median values, the non-parametric model has a narrower reduced-$\chi^2$ distribution and better reproduces the observed continua and absorption features via visual inspection. 
The two parametric models represent a single, smoothly varying episode of star formation and are less flexible in reproducing the complex SFHs favored for some galaxies, potentially biasing their inferred recent SFRs. 
Our fiducial non-parametric SFH is also motivated by predictions from several cosmological simulations and semi-analytic models, including \textsc{Shark}, \textsc{Galform}, and IllustrisTNG, which indicate that massive quiescent galaxies at these redshifts may experience multiple episodes of star formation \citep{Lagos2024}. 
We therefore retain the non-parametric SFH as our fiducial model while acknowledging the resulting model dependence of individual quiescent classifications.}

\begin{deluxetable}{lc}
\tablecaption{Summary of Photometric and Spectroscopic Quiescent Sample Selection \label{tab:sample}}
\tablehead{
\colhead{Sample Selection Criteria} &
\colhead{Num. of Gals.} \\[-16pt]
\colhead{(1)} &
\colhead{(2)}
}
\startdata
\textbf{Preselected photometric}: & \\ 
\multirow{2}{*}{$1.75 \leq z_\mathrm{phot, eazy} \leq 5.25$ } & 113832 \\
& (39594/29091/45147) \\
\multirow{2}{*}{$M_\mathrm{\ast, cigale} \geq 10^{9} M_\odot$} & 10652 \\
& (3001/2033/5618) \\
\hline
\textbf{Photometric galaxies}: & 4488\\ 
$2 \leq z_\mathrm{phot} \leq 5$ \& $M_\mathrm{\ast, p} \geq 10^{9.5} M_\odot$  &  (1229/835/2424)\\
\hline
\textbf{Photometric quiescent galaxies}: & 603 \\ 
$\mathrm{sSFR_{100, p}} \leq 0.20/t_\mathrm{obs}$ & (151/111/341)\\
\hline
\textbf{Spectroscopic sample}: & 181 \\
$2 \leq z_\mathrm{spec} \leq 5$ \&  $M_{\ast, s} \geq 10^{9.5} M_\odot$ & (80/63/38) \\
\hline
\textbf{Spectroscopic QGs}: & 19 \\
$\mathrm{sSFR_{100, s}} \leq 0.20/t_\mathrm{obs}$+visual check & (5/5/9) 
\enddata
\tablecomments{(1) Selection criteria. Sample names used throughout this paper are shown in bold. (2) Number of galaxies satisfying each criterion. Values in parentheses indicate the number of galaxies in COSMOS, EGS, and UDS, respectively.}
\end{deluxetable}

\begin{figure*}
\centering
\includegraphics[width=0.32\textwidth]{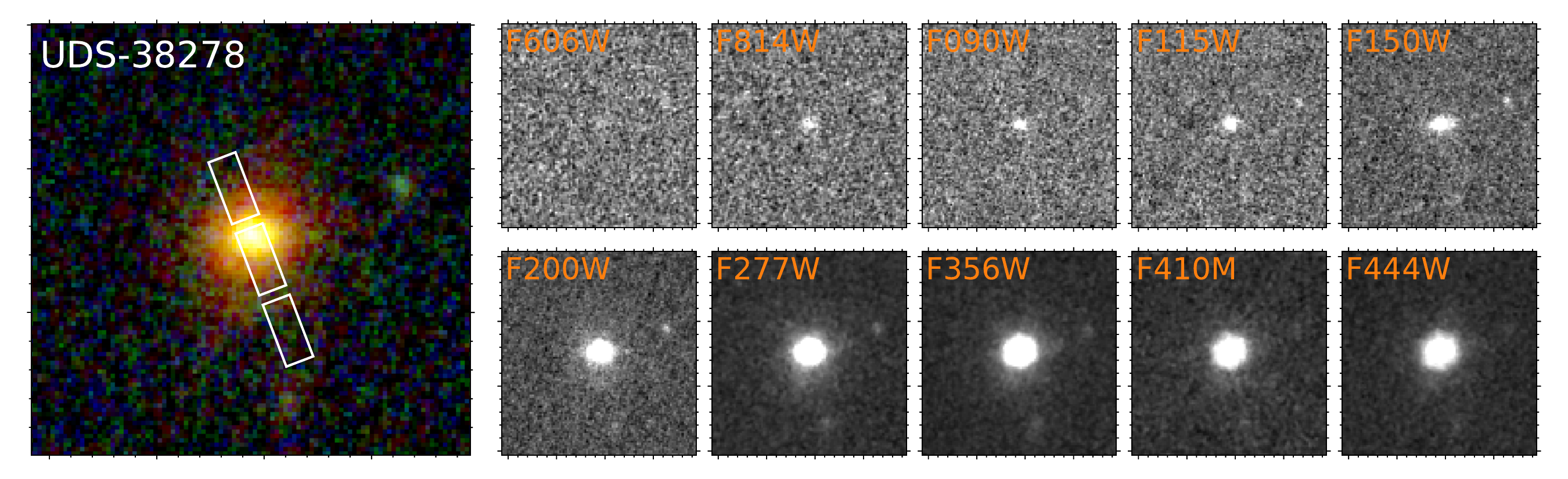}
\includegraphics[width=0.32\textwidth]{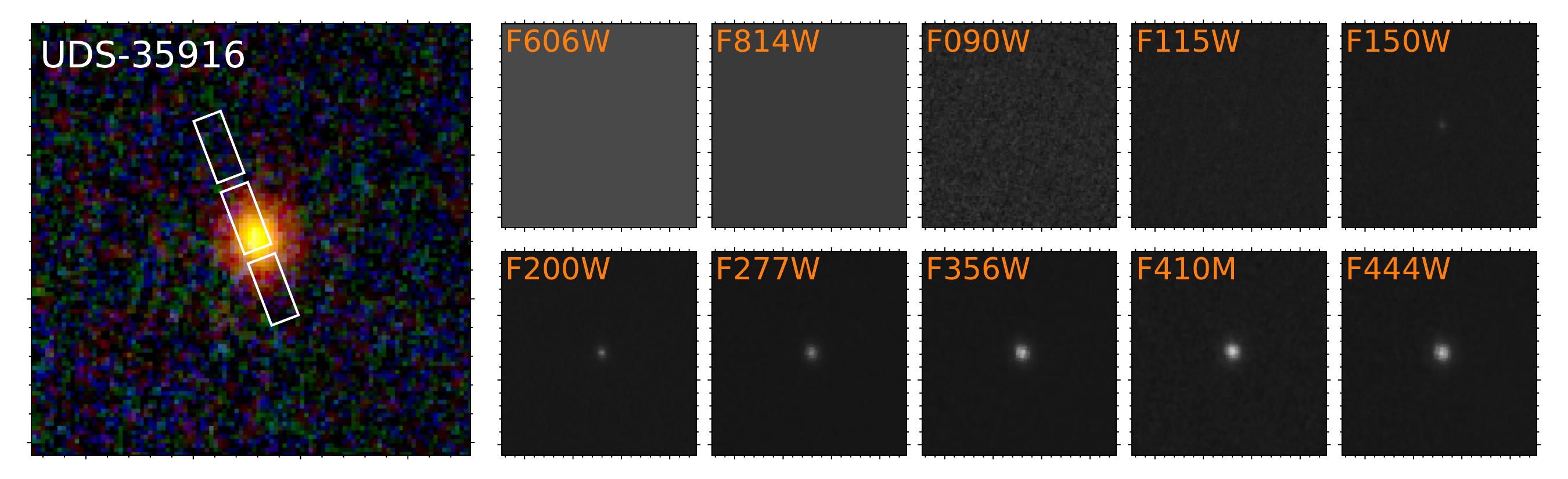}  
\includegraphics[width=0.32\textwidth]{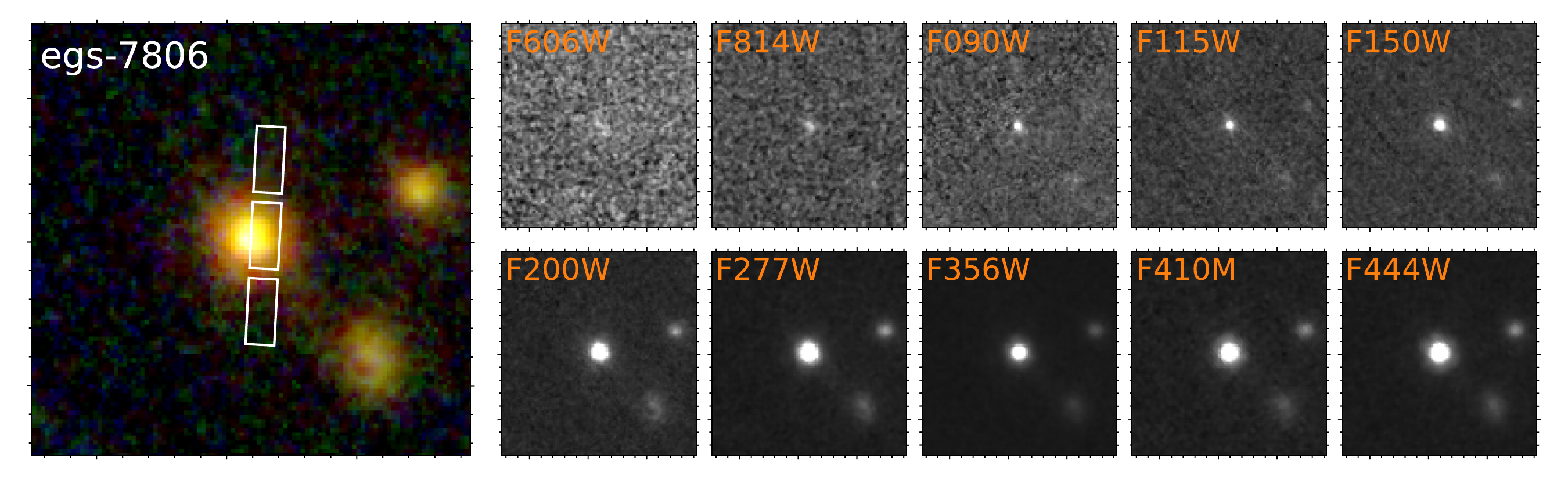} \\
\includegraphics[width=0.32\textwidth]{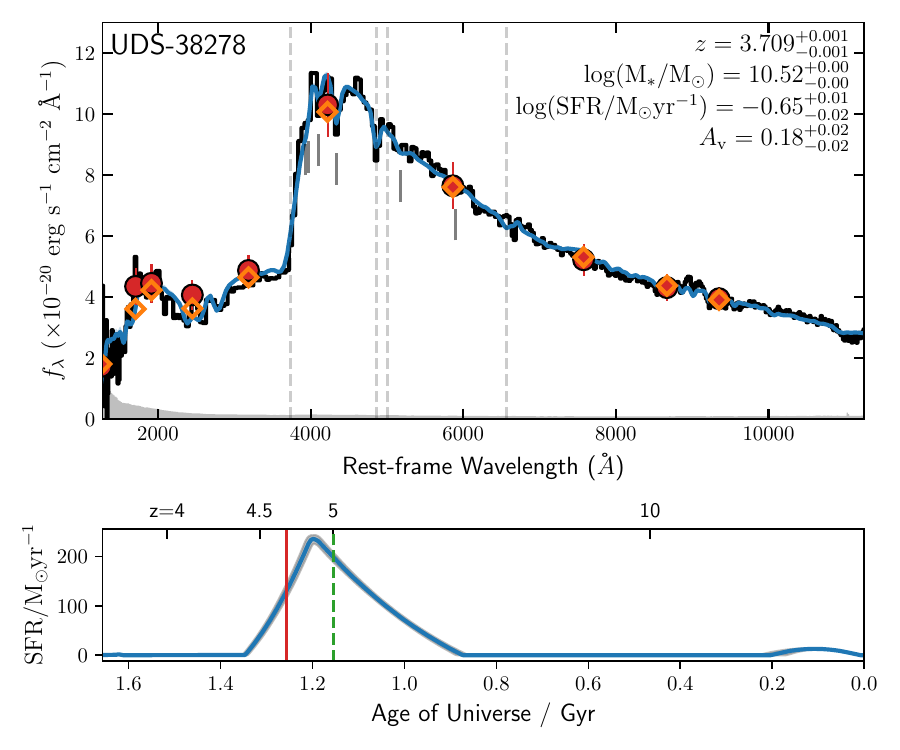}
\includegraphics[width=0.32\textwidth]{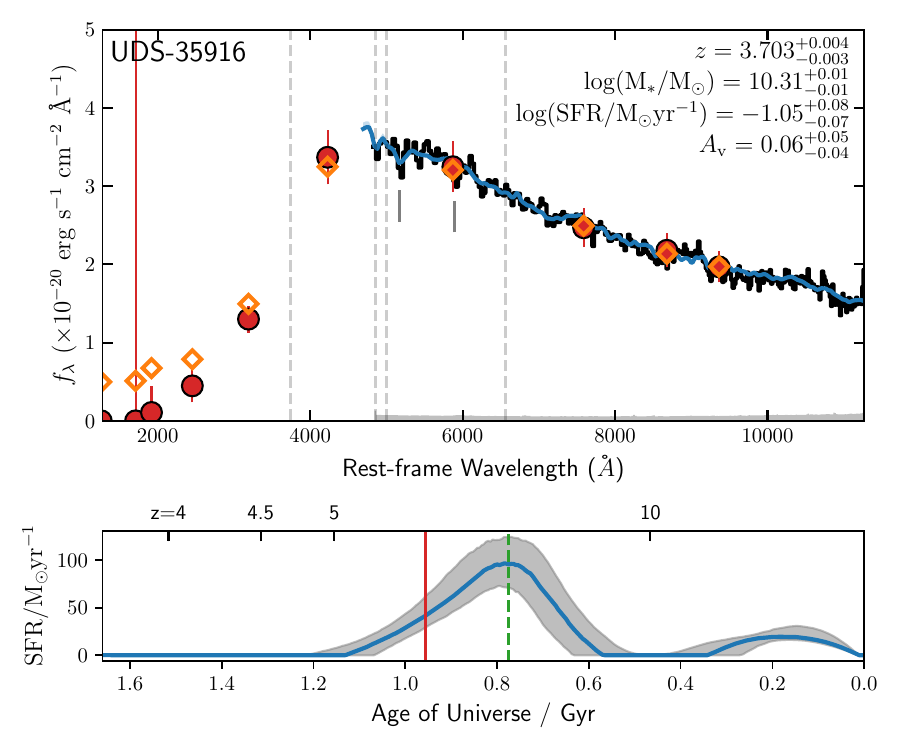}  
\includegraphics[width=0.32\textwidth]{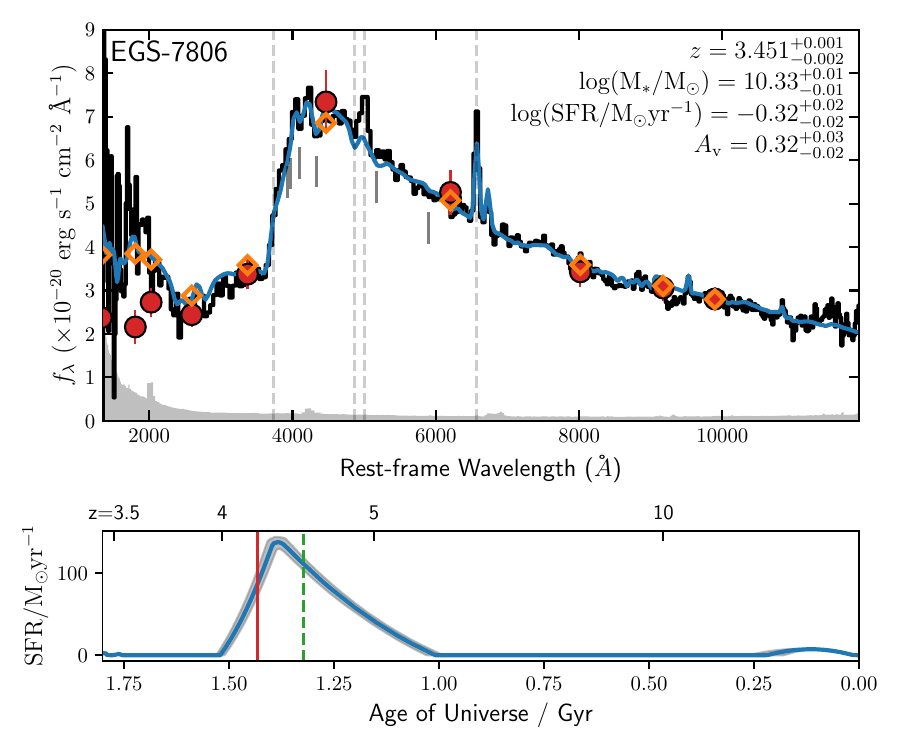} \\
\includegraphics[width=0.32\textwidth]{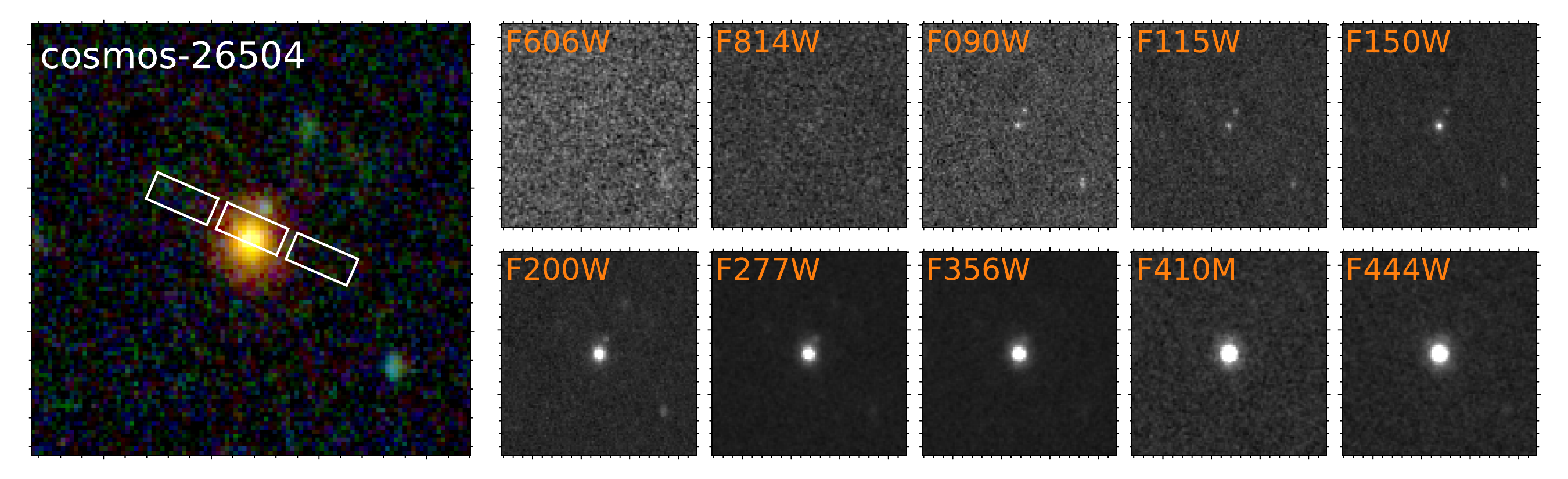}
\includegraphics[width=0.32\textwidth]{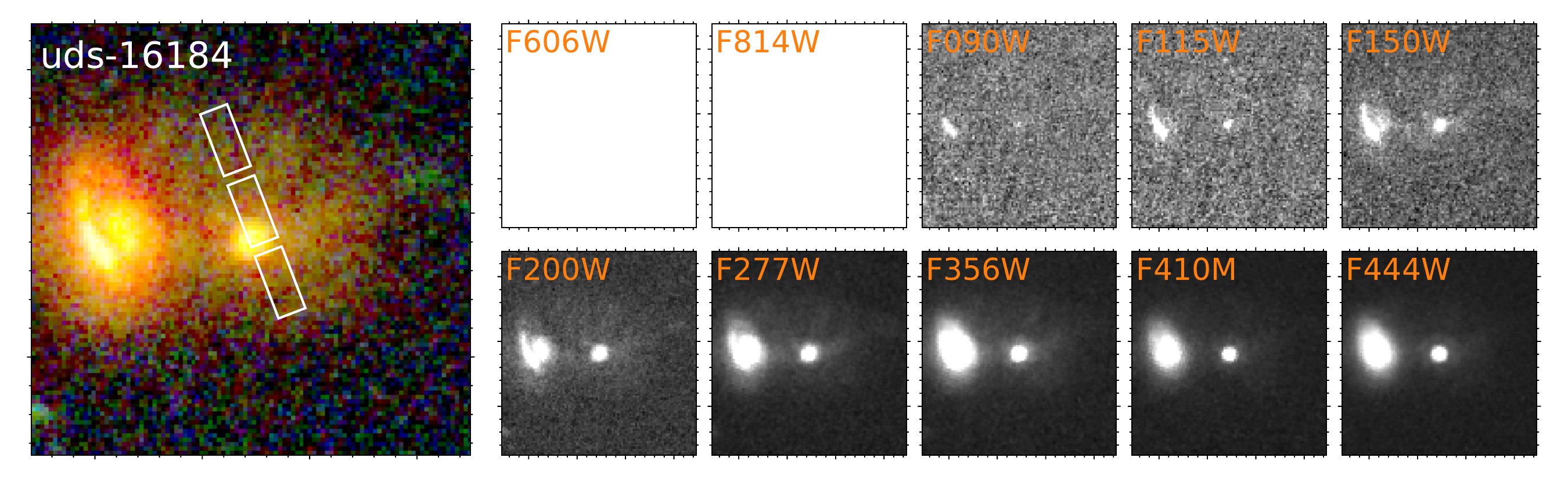}  
\includegraphics[width=0.32\textwidth]{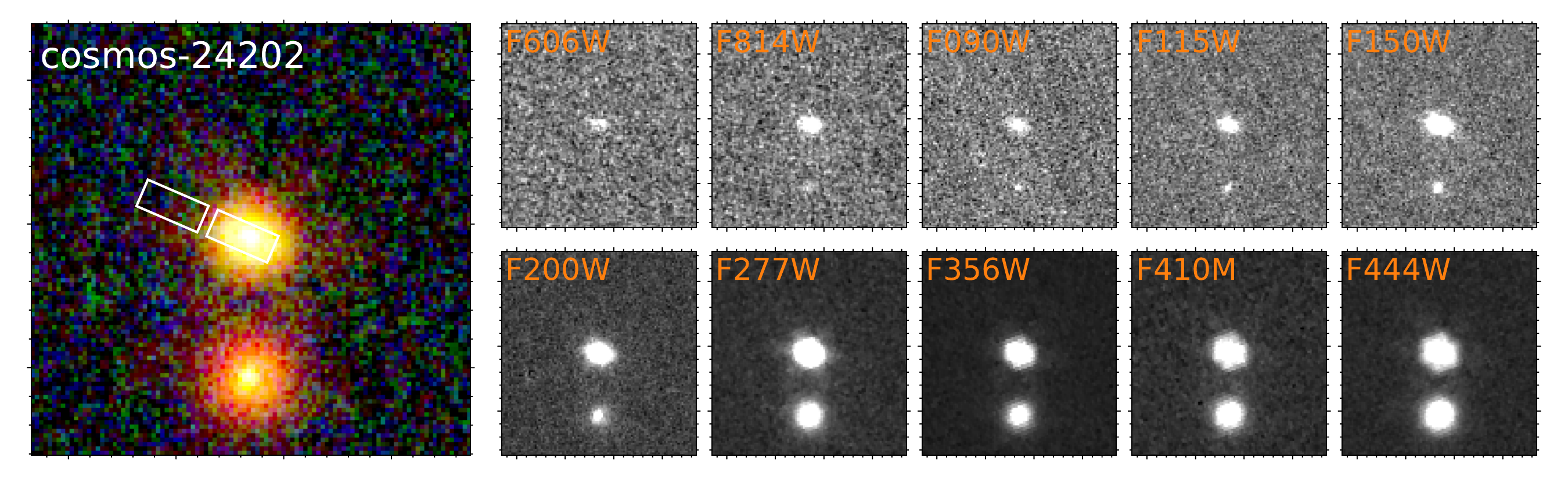} \\
\includegraphics[width=0.32\textwidth]{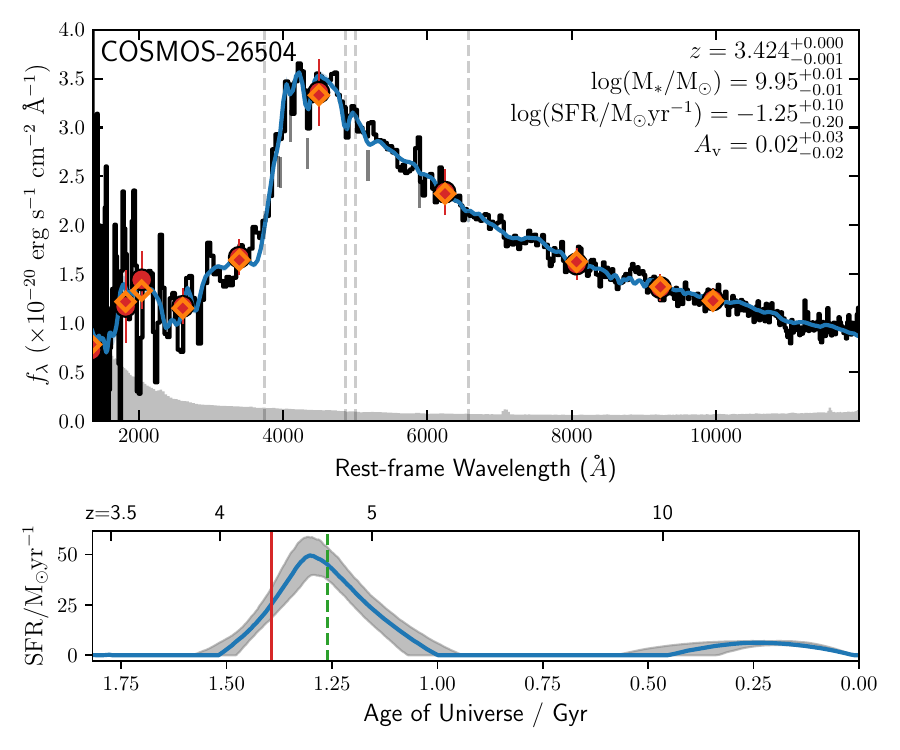}
\includegraphics[width=0.32\textwidth]{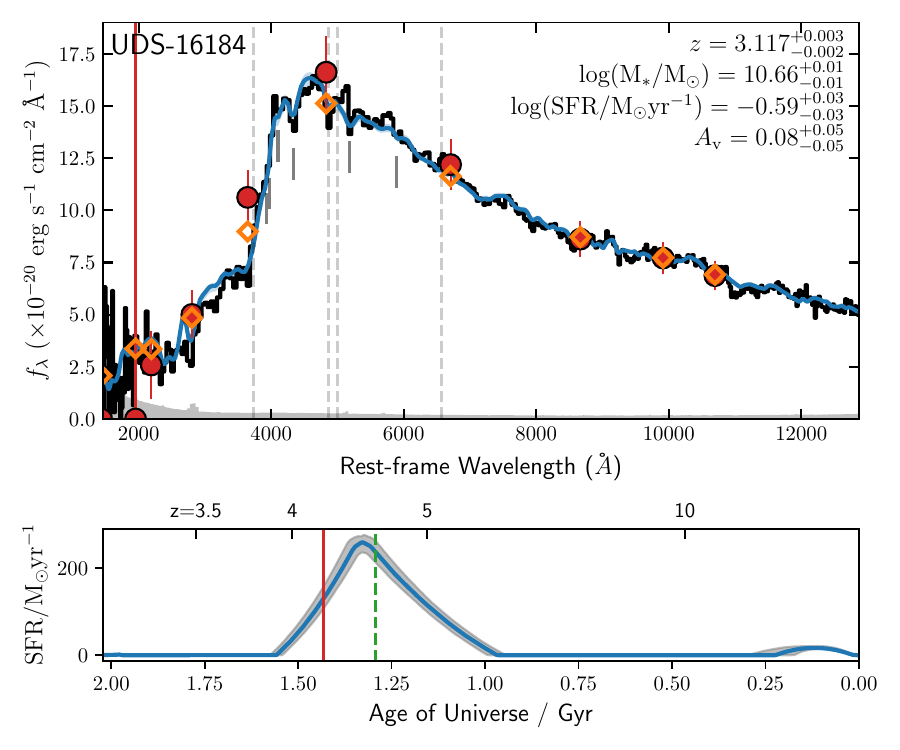}
\includegraphics[width=0.32\textwidth]{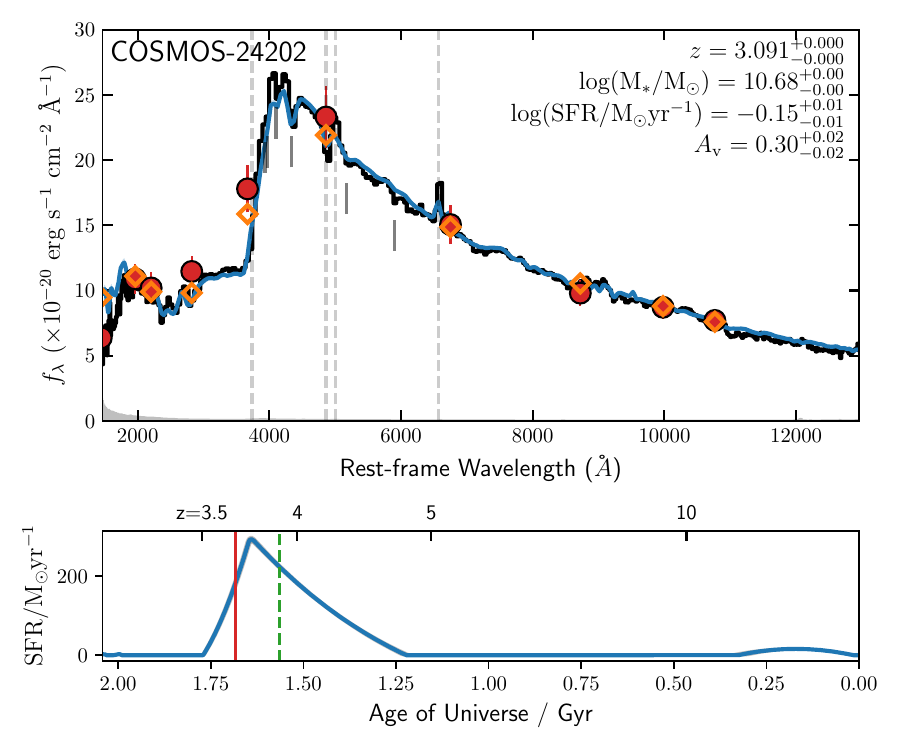} \\
\caption{Spectroscopically-confirmed quiescent galaxies at $3 \leq z_\mathrm{spec} \leq 4$ selected with $\mathrm{sSFR}\leq 0.2/t(z)$. For each galaxy, the \textit{top} panel shows the $3^{\prime\prime} \times 3 ^{\prime\prime}$ false color image using NIRCam F090W, F200W, and F444W, along with individual cutouts from \hst/ACS F606W, F814W, and \jwst/NIRCam F090W, F115W, F150W, F200W, F277W, F356W, F410W, and F444W. The positions of the open NIRSpec MSA shutters (for the first nod position) are shown as white boxes in the false color image. The \textit{middle} panel shows the full \bagpipes\ spectral fitting results. Spectroscopic data flux-calibrated to match the photometry are shown in black, with NIRCam photometry shown as red points. The posterior-median fitted \bagpipes\ models are shown with blue lines. Four emission lines (\oii, \hb, \oiii, and \ha) are indicated as vertical dashed lines. Possible absorption features are indicated as short grey lines (Ca H+K, \hd, \hg, \Mgi, \NaD, from short to long wavelength). The \textit{bottom} panel shows the median of star formation history from \bagpipes\ (blue line), with the 16th--84th percentile range indicated by the shaded region. The $t_{50}$ and $t_{90}$ are indicated with green dashed and red solid lines.  }
\label{fig:QGs_z3z4}
\end{figure*}

\section{Spectroscopically-confirmed Quiescent galaxies} \label{sec:results}

In this section, we first present properties of the spectroscopic quiescent galaxies identified in CAPERS (Section \ref{sec:spec_properties}). 
CAPERS target selection gives higher weight to sources with higher photometric redshifts, but generally does not give special priority to brighter or redder galaxies or quiescent galaxy candidates, unlike some other NIRSpec programs such as EXCELS {\citep{Carnall2024}} or RUBIES {\citep{Zhang2025}}. 
Thus, our QG sample is largely unaffected by selection biases.
Then, in Section~\ref{sec:compare_selection}, we compare the properties of the CAPERS quiescent galaxies to various selection methods, including the primary sSFR–$t(z)$, the sSFR -- \sm, photometry-only sSFR$_\mathrm{p}$–$t(z)$, $UVJ$, and synthetic $ugi$ color criteria.  The goal of this analysis is to investigate the completeness and purity of each selection method.  

\begin{figure}
    \centering
    \includegraphics[width=\columnwidth]{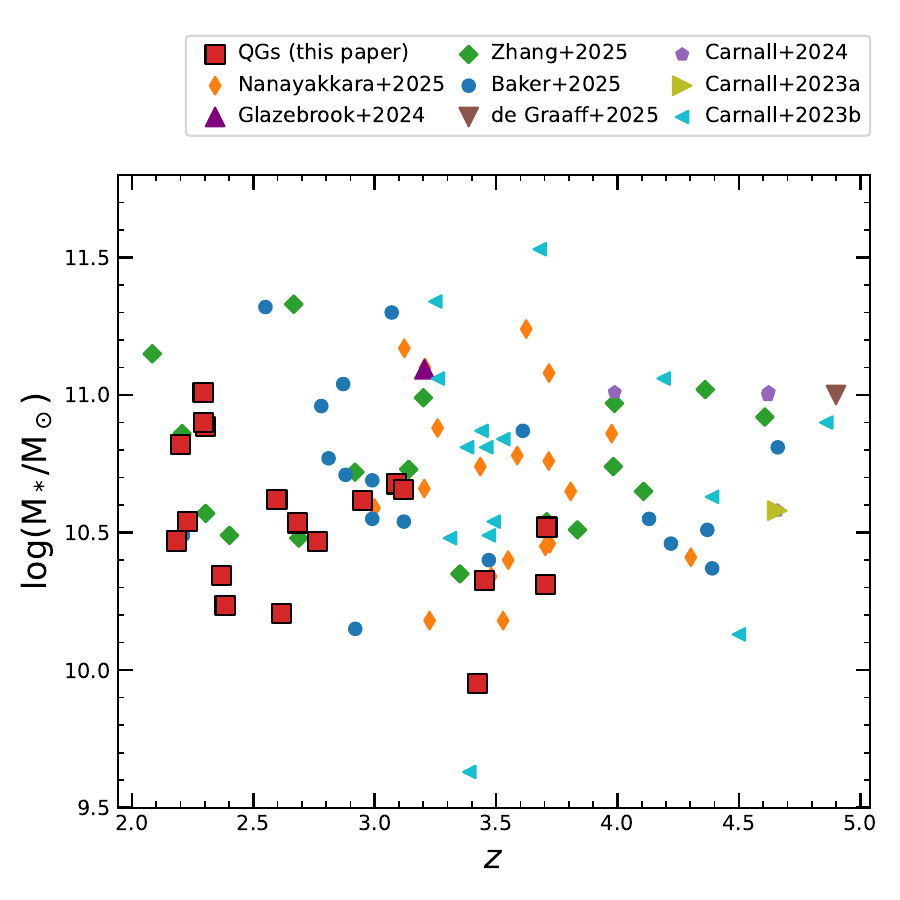}
    \caption{Redshift versus stellar mass for our QGs and quiescent galaxies at $z \sim 2$–5 from recent \jwst\ spectroscopic studies \citep{Carnall2023, Carnall2023b, Carnall2024,  Glazebrook2024, Baker2025a, Nanayakkara2025, Zhang2025, deGraaff2025}. } 
    \label{fig:mass-z}
\end{figure} 

\subsection{Properties of Quiescent Galaxies} \label{sec:spec_properties}

Figure \ref{fig:mass-z} shows stellar mass versus redshift for our spectroscopic quiescent galaxies, compared to quiescent galaxies at $z \sim 2$–5 from recent \jwst\ spectroscopic studies \citep{Carnall2023, Carnall2023b, Carnall2024,  Glazebrook2024, Baker2025a, Nanayakkara2025, Zhang2025, deGraaff2025}. 
Our quiescent galaxies span the redshift range {$z\sim 2$-4}. Although we search up to $z\sim5$, none are found at {$z > 4$} in the CAPERS sample. 
The absence of quiescent galaxies with {$z > 4$} in CAPERS most likely reflects the rarity of that population. 

The median stellar mass of our QGs is $10^{10.54} {M_\odot}$, with a minimum to maximum range of $10^{9.95}$–$10^{11.00} {M_\odot}$. 
For comparison, the median stellar mass of quiescent galaxies identified in recent \jwst\ spectroscopic studies is $\log(M_\ast / M_\odot) \sim 10.74$ at $z \sim 2$–5 and {remains the same at $z \sim 2$–4}. 
Our spectroscopic quiescent galaxies occupy systematically lower stellar masses than these literature samples, both at fixed redshift and across the $z \sim 2$–5 range. Again, this is probably a consequence of the target selection for other surveys that prioritized red, bright galaxies for spectroscopy. 
We do not find additional quiescent galaxies when lowering the stellar mass limit to $\log(M_\ast) > 10^{9} M_\odot$ using the sSFR-$t(z)$ criterion.

\begin{figure}
    \centering
    \includegraphics[width=\columnwidth]{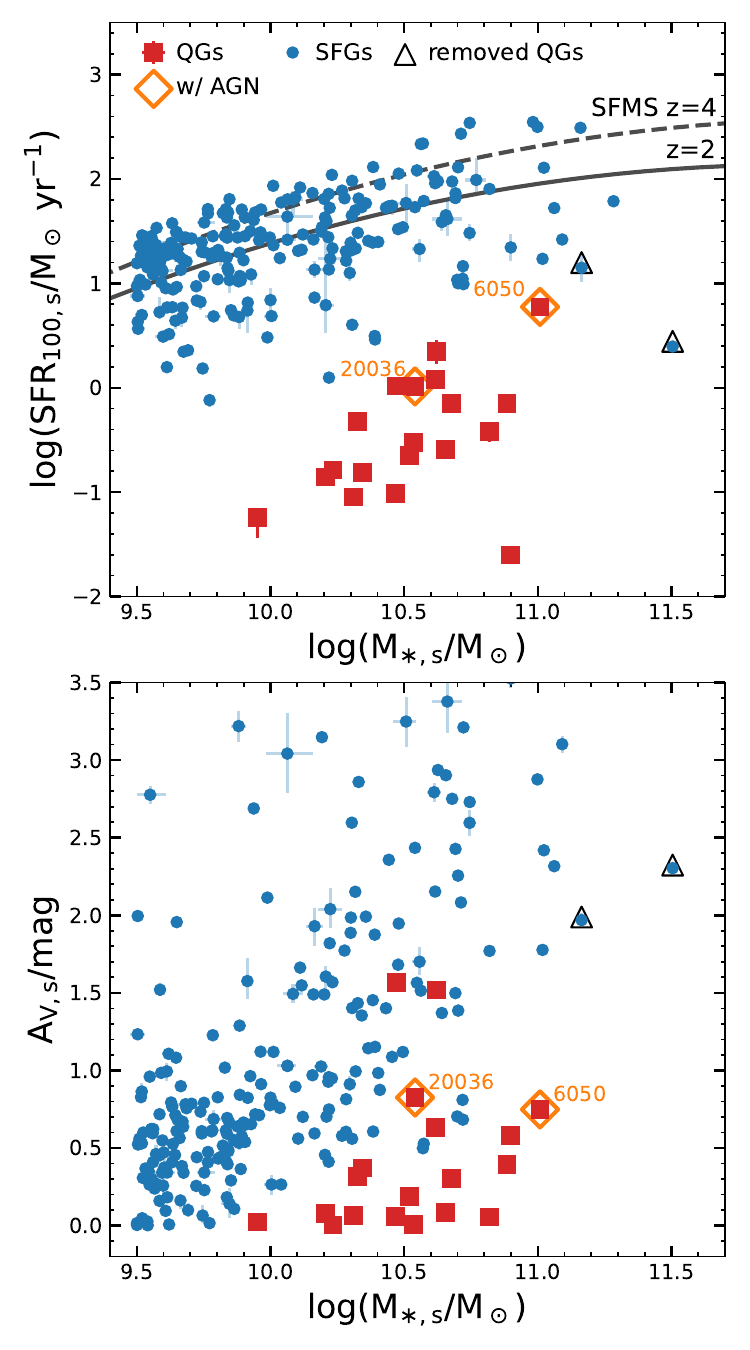}
    \caption{Stellar mass versus $\mathrm{SFR_{100}}$ (\textit{top}), dust attenuation (\textit{bottom}) for QGs (red squares) and SFGs (blue dots) selected from the CAPERS spectroscopic sample at $2 \leq z \leq 5$. AGN are shown as orange open diamonds, with their ID labeled. Two dusty SFGs identified by sSFR-$t(z)$ selection are marked with open black triangles. 
    The star formation main sequences (SFMS) at $z=2$ and $z=4$ from \citet{Popesso2023} are shown as grey solid and dashed lines, respectively.   } 
    \label{fig:spec_q_props1}
\end{figure}

In the top panel of Figure \ref{fig:spec_q_props1}, we present the stellar mass and $\mathrm{SFR_{100}}$ of these QGs, along with SFGs and the star-formation main sequence (SFMS) at $z\sim2-4$ \citep{Popesso2023}. 
QGs lie more than $\gtrsim$1 dex below SFGs and SFMS. 
The median $\mathrm{SFR_{100}}$ and $\log(\mathrm{sSFR_{100}/yr^{-1}})$ of QGs are $0.30~\mathrm{M_\odot~yr^{-1}}$ and $-11.06 \mathrm{yr^{-1}}$, respectively.

Quiescent galaxies are generally dust-poor. In the bottom panel of Figure \ref{fig:spec_q_props1}, we show the dust attenuation as a function of stellar mass for QGs and SFGs at $z\sim2-5$. 
Our QGs have a median $A_V$ of 0.30 mag, with a 16th–84th percentile range of {0.05–0.76} mag. 
These median values are significantly lower than those of SFGs at fixed stellar mass. 
For SFGs at $\log(\mathrm{M_\ast/M_\odot})\geq10$, the median $A_V$ is 1.50 mag, with a 16th–84th percentile range of 0.75–2.74 mag.

\begin{figure*}
    \centering
    \includegraphics[width=\textwidth]{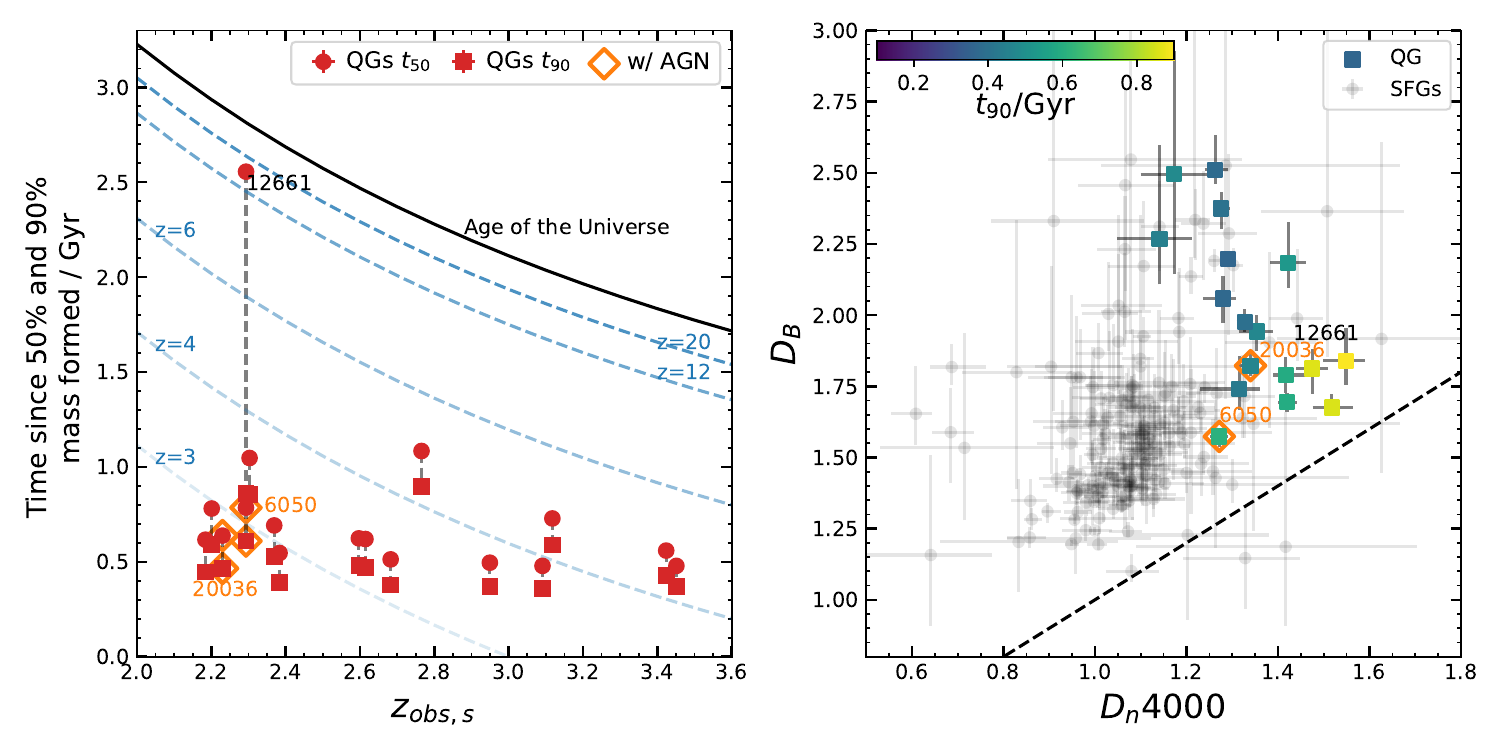}
    \caption{\textit{Left:} Observed redshift versus formation time ($t_{50}$, circles) and quenching time ($t_{90}$, squares) for QGs. The black line indicates the age of the Universe. The blue dashed lines indicate the form/quench at $z=3$ to 20.  \textit{Right:} Balmer break ($D_B$) versus 4000 \AA\ break ($D_n4000$) for QGs (squares color-coded by quenching time) and SFGs (grey dots). Our quiescent galaxies formed half of their stellar masses by {0.50-0.91} Gyr ago, and quenched {0.38-0.72} Gyr ago. } 
    \label{fig:spec_q_props2}
\end{figure*}

\subsubsection{SFHs, formation and quenching times} \label{sec:sfh}

The star formation histories of our QGs are shown in Figures~\ref{fig:QGs_z3z4} for QGs $z\sim3-4$, in Figures~\ref{fig:QGs_z2z3_1} and \ref{fig:QGs_z2z3_2} for those at $z\sim2-3$. 
The majority of galaxies underwent a recent major burst of star formation with peak SFRs $\gtrsim 100~\mathrm{M_\odot~yr^{-1}}$ between $\sim1$-2.5 Gyr after the Big Bang, followed by a rapid decline. 

Figure~\ref{fig:spec_q_props2} shows the formation and quenching times ($t_{50}$ and $t_{90}$, respectively) as a function of the observed redshift of these quiescent galaxies. 
Our QGs formed between {0.50-0.91} Gyr prior to the observed epoch (16th-84th percentile), with a median formation time of 0.62 Gyr, and quenched between {0.38-0.72} Gyr prior to the observed epoch, with a median quenching time of 0.47 Gyr. 
These times correspond to a median (16th-84th percentile) formation and quenching redshifts of $z = 3.5$ ($z \sim 3.0 - 4.8$) and $z = 3.2$ ($z \sim 2.8 - 4.2$), respectively. 
The majority of our QGs are recently formed and quenched within the past $\lesssim1$ Gyr. 
The resulting time difference between formation and quenching is short, with a median of 0.16 Gyr and a 16th-84th percentile range of {0.13-0.19} Gyr, indicating a rapid shutdown of star formation. 
We will further compare the formation and quenching times of quiescent galaxies across different redshifts in Section \ref{sec:formation_times}.

One quiescent galaxy EGS-12661 at $z=2.3$ is a notable outlier from our quiescent galaxy sample.  
According to our full spectral \bagpipes\ fitting, it appears to have formed 50\% of its stellar mass as early as $z \sim 15$ and quenched by $z \sim 3.2$. The star formation history suggests an initial burst of star formation at $z \sim 15$, followed by a secondary, relatively younger episode at $z \sim 3.5$. 
The presence of this early, major star-formation episode is striking, but it may depend on the choice of models and SFH parameterization. {The double-power-law model yields $t_{50}=1.00~\mathrm{Gyr}$ and $t_{90}=0.98~\mathrm{Gyr}$, and the delayed-$\tau$ model yields $t_{50}=0.98~\mathrm{Gyr}$ and $t_{90}=0.77~\mathrm{Gyr}$. 
These values remain toward the upper end of the $t_{50}$ and $t_{90}$ distributions of our QG sample, supporting an early formation history regardless of the adopted SFH model. Higher-resolution, higher-S/N spectroscopy will help place stronger constraints on the stellar ages. }

\subsubsection{Spectra features: Balmer/4000\AA\ breaks} \label{sec:breaks}

The spectroscopic data allow us to quantify the Balmer/4000-\AA\ breaks in our quiescent galaxies. All of our quiescent galaxies show these features in their spectra (see Figure~\ref{fig:QGs_z3z4}, \ref{fig:QGs_z2z3_1}, and \ref{fig:QGs_z2z3_2}). 
We attempt to measure these breaks separately. 
The 4000-\AA\ break and Balmer break are defined in Section~\ref{sec:spec_fitting}. 
We show the $D_B$ versus $D_n4000$ for QGs and SFGs in the right panel of Figure \ref{fig:spec_q_props2}.  
The QGs occupy the region of high $D_B$ and high $D_n4000$, consistent with evolved stellar populations. 
The median $D_n4000$ of QGs is {1.33}, with a 16th-84th percentile range of 1.27-1.45, while the median $D_B$ is {1.94}, with a corresponding range of {1.72–2.31}. 
In comparison, SFGs with $\log(M_\ast/M_\odot) > 10$ show significantly weaker break strengths, with a median $D_n4000$ of 1.10, with 16th–84th percentile of 1.01–1.27 and a median $D_B$ of 1.65, with 16th–84th percentile of 1.37–2.03. 
Overall, QGs show stronger Balmer and 4000 \AA\ break features than SFGs with similar stellar masses. 

When color-coded by $t_{90}$, the QGs show a clear correspondence between color and $D_n4000$. 
In contrast, QGs with larger quenching times ($t_{90} \gtrsim 0.8$ Gyr) tend to show relatively lower $D_B$ values, consistent with the Balmer break peaking for stellar populations with ages of $\sim0.3$–1 Gyr and declining thereafter. 
This trend suggests that the spectral break features for most QGs are dominated by strong Balmer breaks, while older QGs are characterized by an emerging 4000 \AA\ break.

\subsubsection{Spectra features: Emission Lines} \label{sec:emissionlines}

The majority of QGs (13 out of 19) in our spectroscopic sample have \ha+\nii\ detected with $>3\sigma$.  
Elevated \nii/\ha\ has been found in massive quiescent galaxies at $z>4$ \citep{Carnall2024, Perez-Gonzalez2025} in recent JWST observations and at $z\sim2$ \citep{Belli2017, Newman2018}. 
Higher \nii/\ha\ ratios are typically associated with AGN and shocks \citep{Kewley2006, Zhu2025}. 
Indeed, {one} of them are confirmed AGN hosts, as described in Section \ref{sec:agn}.  
Alternatively, ionization from post–Asymptotic Giant Branch (post-AGB) stars can also contribute to elevated \nii/\ha\ ratios in quiescent galaxies. 
\citet{Belfiore2018}, using MaNGA integral field spectroscopy, showed that the LINER region of the BPT diagram is associated with high $D_n4000$, low EW(H$\delta$), and thus old stellar populations. 
Moreover, they found that the majority of spaxels in quiescent galaxies occupy this LINER region. 
However, with our limited resolution of the prism spectra, we cannot separate the \nii\ contribution from \ha\ or distinguish between these ionization mechanisms.

\subsubsection{The presence of AGN} \label{sec:agn}

\begin{figure}
    \centering
    \includegraphics[width=\columnwidth]{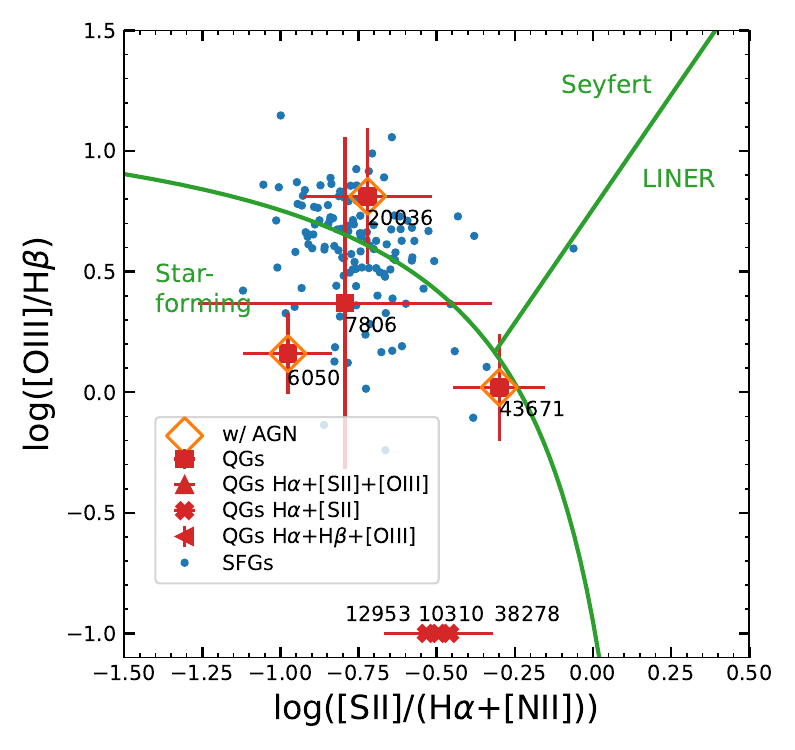}
    \caption{\sii/(\ha+\nii) versus \oiii/\hb\ diagram for the presence of AGN in QGs (red squares) and SFGs (blue dots) for comparison with all four emission lines detected at $\ge3\sigma$. The classification lines are adopted from the \oiii/\hb\ versus \sii/\ha\ diagram of \citet{Kewley2006}. Quiescent galaxies are labeled with their ID. 
    Upper triangles show lower limits in \oiii/\hb\ for sources without \hb\ detections, and left-pointing triangles show upper limits in \sii/\ha\ for sources without \sii\ detections. Cross symbols at the bottom denote quiescent galaxies lacking both \oiii\ and \hb\ detections. Note that \ha\ and \nii\ are blended in the spectra and are fitted with a single Gaussian; thus, the \sii/\ha\ ratio could be higher than measured values. {One} AGN EGS-20036 is identified based on the BPT diagram. } 
    \label{fig:spec_q_props4}
\end{figure}

AGN are known to play a role in quenching star formation. To investigate whether AGN are present in our quiescent galaxy sample, we consider three diagnostics: X-ray detection, radio detection, and the optical BPT diagram.

For the X-ray analysis, we use X-ray catalogs from \citet{Nandra2015} for the EGS field, \citet{Kocevski2018} for the UDS field, and \citet{Marchesi2016} for the COSMOS field. 
The hard-band (2-10 keV) limiting fluxes are $2.5\times10^{-16}$, $6.5\times10^{-16}$, and $2.1\times10^{-15}$ erg~s$^{-1}$~cm$^{-2}$, corresponding to X-ray luminosities of $1.1 \times10^{44}$, $2.9\times10^{44}$, $9.3 \times10^{44}$ erg~s$^{-1}$ at $z=2$ and $2.0\times10^{45}$, $5.2\times10^{45}$, $1.7 \times10^{46}$ erg~s$^{-1}$ at $z=5$, assuming a photon index of $\Gamma=1.4$. 
We cross-match quiescent galaxy coordinates with the X-ray sources using a 1\arcsec\ matching radius, which results in two matches: UDS-6050 and EGS-20036. 
The angular separations between the X-ray detections and the optical coordinates are $0\farcs33$ and $0\farcs25$, respectively. 
Their X-ray luminosity are $3.2\times 10^{44}$ and $5.6\times 10^{43}$ erg~s$^{-1}$. 
We note that these X-ray observations are sensitive to $L_\mathrm{X} \gtrsim 10^{44}$ erg s$^{-1}$ at $z=2$–5, and can detect only luminous unobscured quasars. 
These data cannot exclude the presence of lower-luminosity or heavily obscured AGN.

For radio data, we use a 3~GHz catalog for the EGS field (Jim\'{e}nez-Andrade et al.\ in prep.;  VLA programs 21B-292 and 22A-400, PI: Mark Dickinson) with a modal root mean square (rms) noise of 0.79~$\mu$Jy beam$^{-1}$, the 3~GHz COSMOS catalog from \citet{Smolcic2017}, which reaches a median rms of 2.3~$\mu$Jy~beam$^{-1}$, and a 1.4~GHz radio catalog for the UDS field from \citet{Simpson2006} with an rms of 20 $\mu$Jy beam$^{-1}$. 
For EGS, these data allow radio detection down to radio luminosity of $L_\mathrm{1.4GHz} = 10^{23.2}~\mathrm{W~Hz^{-1}}$ at $z\sim2$ and $L_\mathrm{1.4GHz} = 10^{24.1}~\mathrm{W~Hz^{-1}}$ at $z\sim5$, considering a 5$\sigma$ detection and assuming a spectral index of $\alpha=-0.8$.  
For COSMOS and EGS, typical radio AGN with $L_{1.4\mathrm{GHz}} \gtrsim 10^{24}~\mathrm{W~Hz^{-1}}$ are detectable at $z\sim2$–3, while only the most luminous AGN can be detected at higher redshifts. 
For UDS, the data are shallower by roughly an order of magnitude, so only the most luminous AGN can be detected across the redshift range. 
We cross-match the quiescent galaxy coordinates with the radio catalogs using a 1\arcsec\ radius, but none of our quiescent galaxies are detected in the radio.

For the optical BPT diagram \citep{Veilleux1987}, we adopt the classification scheme based on the \sii/\ha\ versus \oiii/\hb\ diagram from \citet{Kewley2006}. Due to the limited resolution of the prism spectra, \ha\ and \nii\ are blended, which prevents the \nii/\ha\ versus \oiii/\hb\ classification. 
For the same reason, \sii/\ha\ could be underestimated. Correcting for this effect would shift galaxies rightward in the \sii/\ha\ versus \oiii/\hb\ diagram, toward the AGN region.  
\citet{Carnall2024} also found that, among four quiescent galaxies at $z\sim3-5$, two show \nii\ emission significantly stronger than \ha, suggesting that, at least in some galaxies, the \nii\ might dominate our measured \ha+\nii\ flux. 
In addition, we caution that this diagnostic, particularly at low spectral resolutions where \nii\ and \ha\ are blended, becomes increasingly unreliable at higher redshifts \citep{Backhaus2022, Cleri2025, Cleri2023}.

{Four} QGs have all four emission lines detected at $>3\sigma$ significance, and {three} QGs have $>3\sigma$ detections in both \ha\ and \sii\ emission lines. 
These QGs are shown in Figure \ref{fig:spec_q_props4}, along with SFGs that have all four emission lines detected at $>3\sigma$ significance.
Among the QGs, EGS-20036 lies in the Seyfert region of the BPT diagram, {while the uncertainties for COSMOS-43671 and EGS-7806 extend into the Seyfert region. }
We note that UDS-6050 lies in the star-forming region of the BPT diagram, despite being identified as an AGN based on its X-ray detection. 
Its spectra show broader \ha\ and \hb\ components that are not fitted by our single-Gaussian profiles, possibly indicating the presence of a broad-line region. 
In addition, it has a clear \Mgii\ emission line at 2798 \AA, which provides additional evidence of nuclear activity. 
This is the only galaxy in our quiescent sample in which the \Mgii\ line is detected. 

Therefore, a total of {two} quiescent galaxies show evidence of AGN activity, corresponding to {11\%}. 
Using similar diagnostics (BPT and X-ray), \citet{Baker2025a} identified 6 out of 18 (33\%) as AGN hosts, and an additional two based on broad \ha\ emission lines, increasing the AGN fraction to 44\%. These fractions are higher than those in our sample. 
Given the small sample sizes and the incompleteness in AGN identification, including shallow X-ray data and blended emission lines, these fractions should be interpreted with caution, and no strong conclusions can be drawn. 

{Meanwhile, AGN continuum emission may bias stellar-population properties inferred using stellar-only SED models. Because the PRISM spectra do not permit a robust decomposition of the AGN and stellar continua, we interpret the derived properties of both AGN-hosting QGs with additional caution. 
This concern is particularly relevant for UDS-6050, whose possible broad H$\alpha$ and H$\beta$ components and Mg,\textsc{ii} emission identify it as a broad-line AGN. Its non-stellar continuum may affect the stellar-only \bagpipes\ fit, rendering its inferred stellar-population properties and quiescent classification uncertain. 
Nevertheless, excluding these AGN hosts does not affect our conclusions regarding the median properties of the QGs.}

In summary, the QGs lie more than 1 dex below the SFMS, have significantly higher $D_n4000$ values than SFGs, with lower dust attenuation. 
Full spectral SED fitting indicates that they formed between $\sim$0.50-0.98 Gyr and quenched 0.37-0.79 Gyr prior to the observed epoch. 
These times correspond to formation and quenching redshifts of $z \sim 3.0 - 4.6$ and $z \sim 2.8 - 4.2$, respectively. 
The majority of them are consistent with a short quenching timescale (0.1-0.2 Gyr). 
The spectral breaks of most QGs are dominated by strong Balmer breaks, rather than 4000 \AA\ break, consistent with their formation times from SED fitting. 
Finally, we identified {two} QGs that show evidence of AGN activity.

\subsection{Comparison between Quiescent Selections} \label{sec:compare_selection}

In this section, we compare the purity and completeness of various quiescent galaxy selection methods, including the primary sSFR–$t(z)$ criterion, the sSFR–$M_\ast$ relation, the $UVJ$ color–color diagram, and the synthetic $ugi_s$ color–color diagram, and examine which galaxies are consistently identified or missed by these methods. 
{The $UVJ$ and $ugi_s$ colors are derived with \eazy. }

\begin{figure*}
    \centering
    \includegraphics[width=\textwidth]{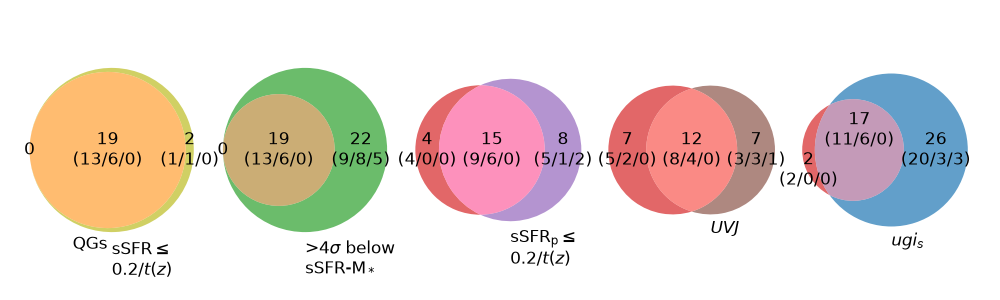}
    \caption{ Venn diagrams showing the number of QGs selected in different quiescent selection methods. Each column compares the QGs (red) with another method: from left to right, sSFR--$t(z)$ selection (yellow), the $>4\sigma$ below the sSFR--$\mathrm{M_\ast}$ relation (green), sSFR$_\mathrm{p}$--$t(z)$ (purple), the $UVJ$ criteria (brown, \citealp{Schreiber2015}), and the synthetic $ugi_s$ diagram (blue, \citealp{Antwi-Danso2023}). {For example, in the leftmost diagram, the numbers in the red-only, overlapping, and yellow-only regions represent QGs not recovered by the sSFR–$t(z)$ criterion, QGs also selected by this criterion, and galaxies satisfying this criterion but not included in the fiducial QG sample, respectively. }The number in parentheses indicates the number of quiescent galaxies in each redshift bin: $2\leq z \leq 3$, $3 < z \leq 4$, and $4 < z \leq 5$. } 
    \label{fig:venn}
\end{figure*}

\begin{figure*}
    \centering
    \includegraphics[width=\textwidth]{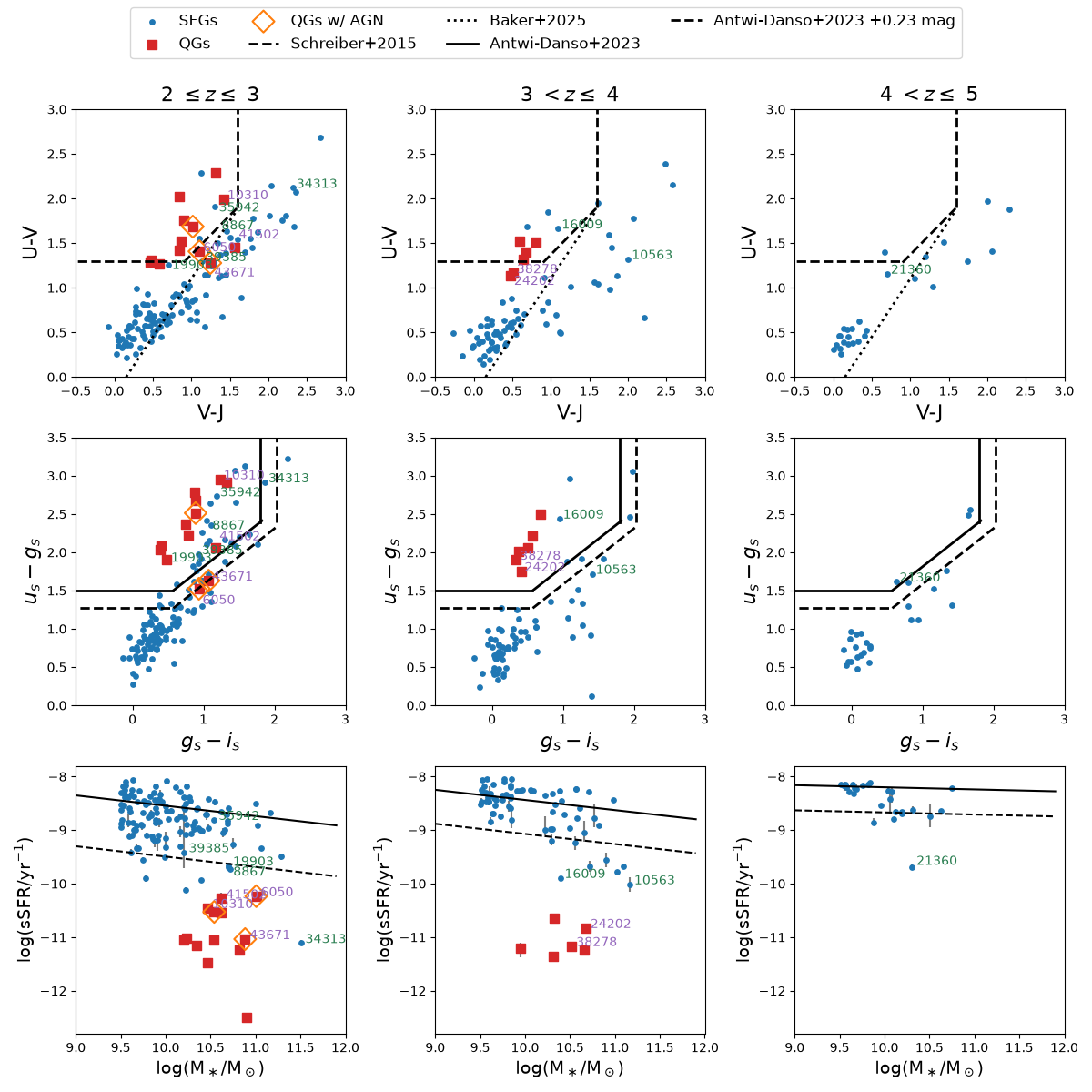}
    \caption{QGs (red squares) and SFGs (blue dots) shown in the $UVJ$ (\textit{top}), $ugi_s$ (\textit{middle}) and sSFR--$\mathrm{M_*}$ (\textit{bottom}) selection diagrams. Each panel spans a different redshift range from left to right: $2 \leq z < 3$, $3 < z \leq 4$, and $4 < z \leq 5$. Galaxies lying 4$\sigma$ below the sSFR–$\mathrm{M_\ast}$ relations are marked by brown open diamonds. In the \textit{top} row, the classification lines from \citet{Schreiber2015} and \citet{Baker2025a} are shown as dashed and dotted black lines, respectively. In the \textit{middle} row, the selection lines from \citet{Antwi-Danso2023} and its extended version (+0.23 mag) are shown as solid and dashed black lines. In the \textit{bottom} row, the fitted sSFR–$\mathrm{M_\ast}$ relation and the $4\sigma$ below relation are shown as solid and dashed black lines, respectively. QGs predominantly lie in the quiescent regions of the $UVJ$ and $ugi_s$ diagrams, and fall below the sSFR--$\ \mathrm{M_\ast}$ relations. {QGs missed by at least one selection method are marked in purple, while contaminants selected by more than one method are marked in green.} } 
    \label{fig:colors_sSFR_M}
\end{figure*}

We compare the numbers of QGs selected by each method, as shown in Figure \ref{fig:venn}. 
In addition, in Figure \ref{fig:colors_sSFR_M}, we show the distributions of QGs and SFGs in the $UVJ$, $ugi_s$, and sSFR–$\mathrm{M_\ast}$ diagrams, separated into three redshift ranges. 
Our QGs predominantly occupy the quiescent regions of the $UVJ$ and $ugi_s$ diagrams and lie more than $4\sigma$ below the sSFR–$\mathrm{M_\ast}$ relations.

All selection methods perform reasonably well in identifying QGs.
By the definition of our QG sample, the sSFR–$t(z)$ criterion identifies all spectroscopically confirmed QGs. 
The sSFR–M$\ast$ criterion similarly achieves complete recovery of the QG sample.
In the case of only photometry data, the sSFR$_\mathrm{p}$–$t(z)$ and two color-selection methods also identified the majority of QGs, with completeness levels of 79\%, 63\%, and 89\% for the sSFR$_\mathrm{p}$–$t(z)$, $UVJ$, and synthetic $ugi_s$ selections, respectively.

Those galaxies missed by the sSFR$_\mathrm{p}$–$t(z)$ and color-selection methods are largely the same across these different methods, indicating that a small subset of objects systematically escapes identification. 
All these selection methods fail to identify COSMOS-43671 and UDS-6050 in the QG sample. 
COSMOS-43671 and UDS-6050 lie close to the boundary of the $UVJ$ and extended $ugi_s$ dividing lines. 
{As discussed in Section~\ref{sec:sample_phot}, the photometry of COSMOS-43671 is affected by blending with a low-redshift interloper, while the photometry-only fit of UDS-6050 may overestimate its SFR because of an unmodeled AGN contribution. As a result, these galaxy are not identified as quiescent by the photometric selection methods.}

In addition to those two galaxies, {two galaxies} in our QGs sample (UDS-10310 and {UDS-41502}) are also missed by the sSFR$_\mathrm{p}$–$t(z)$ method. 
Photometry-only SED fitting suggests that all four galaxies (COSMOS-43671, UDS-6050, UDS-10310, and UDS-41502) are moderately star-forming and dusty, with SFRs of 34, 125, 3, and 9 $\mathrm{M_\odot\ yr^{-1}}$ and $A_V$ values of 1.41, 1.96, 2.76, and 1.59, respectively. 
These fits favor nebular emission along with an underlying older stellar population, combined with significant dust attenuation, to reproduce their strong $D_n4000$ breaks.  

Five additional galaxies in the QG sample are also missed by the traditional $UVJ$ selection. 
They lie just below the $UVJ$ dividing line, within the post-starburst region where galaxies are thought to undergo rapid quenching \citep{Belli2019}. 

While all selection methods are effective at identifying quiescent galaxies, each also includes a fraction of galaxies that are not truly quiescent.
The purities of the QG samples selected by the sSFR–$t(z)$, sSFR–M$\ast$, sSFR$_\mathrm{p}$–$t(z)$, $UVJ$, and $ugi_s$ criteria are {90\%, 46\%, 65\%, 63\%, and 40\%}, respectively.
Below, we examine those galaxies contaminating each selection method. 
Cutouts, SEDs, and SFHs of representative contaminants selected by more than one quiescent galaxy selection criterion are shown in Figure \ref{fig:sed_contaminators}. 

\vspace{0.5em}
{The sSFR--$\mathbf{t(z)}$ selection:}
The sSFR-$t(z)$ selection identified 19 QGs, but also selected two galaxies whose properties are consistent with being dusty, non-quiescent galaxies. These two galaxies are therefore excluded from the QG sample, as briefly described in Section~\ref{sec:sample}. Here, we discuss them in more detail.
Their stellar masses ($\log M_\ast/M_\odot = 11.50$ and 11.16) are higher than those of our confirmed QGs. 
Based on their star formation histories, both galaxies formed and quenched at very early epochs: COSMOS-34313, at $z \sim 2.1$, is formed at $z = 6.6$ (6.4-6.9; 16–84th percentile) and quenched at $z = 5.1$ (4.9 - 5.2), while EGS-10563, at $z = 3.2$, is formed at $z = 9.1$ (8.2-10.4) and quenched at $z = 6.9$ (6.4-7.5). Both galaxies are formed and quenched notably earlier than most QGs in our sample. 
Both spectra exhibit strong dust attenuation, with $A_V = 2.30$ for COSMOS-34313 and $A_V = 1.97$ for EGS-10563, substantially higher than those for our quiescent galaxies (median $A_V = 0.34$). 

Both galaxies show inconsistencies between their spectroscopic and photometric data blueward of 4000 \AA\ break, even with flux calibration applied during the \textsc{Bagpipes} fitting. 
In both cases, the first-order flux calibration polynomial is driven to its lower limit ($-1$), whereas the median value for the full spectroscopic sample is $-0.34$. This behavior indicates significant tension between the spectra and broadband photometry, reflecting spatial mismatches between the integrated photometric apertures and the nuclear regions sampled by the spectra. 
When fitted using photometry alone, neither galaxy would be classified as quiescent based on the same selection criteria, with $\log(\mathrm{sSFR_{100,p}/yr^{-1}}) = -10.08$ for COSMOS-34313 and $-8.70$ for EGS-10563.

In addition, EGS-10563 was previously identified as a submillimeter galaxy (SMG) in the SCUBA-2 survey and is detected at 450~$\mu$m and 850~$\mu$m \citep{Zavala2017}, as well as in several MIRI bands (F770W, F1100W, F1500W, and F2100W) from the MEGA survey \citep{Backhaus2025}. Incorporating these mid- and far-infrared data into the SED fit yields a stellar mass of $\log(M_\ast/M_\odot) = 10.8$ and an SFR of $98~M_\odot~\mathrm{yr^{-1}}$. 
This is consistent with those derived from photometry-only data. These results confirm that EGS-10563 is a dusty, actively star-forming system rather than a truly quiescent galaxy.
COSMOS-34313 is detected in submillimeter/millimeter catalogs from ALMA CHAMPS survey (\citealt{Liu2025,Zavala2026}, Faisst et al. in prep.). 
It is also detected in the COSMOS VLA 3~GHz catalog \citep{Jin2018} with a positional offset of $0\farcs1$ and a 1.4~GHz luminosity of $\log(L_\mathrm{1.4GHz}/\mathrm{W~Hz^{-1}}) = 23.76$, suggesting potential AGN or residual star formation activity. 

Morphologically, both galaxies exhibit disk-like structures. Taken together—their high dust content, moderate ongoing star formation, and disk morphologies—these characteristics strongly indicate that COSMOS-34313 and EGS-10563, at least the overall galaxies, are more consistent with dusty SFGs rather than quiescent galaxies. 
However, we do not exclude the possibility that their central regions, where the spectra were obtained, may be in a more evolved or quiescent phase. 

\vspace{0.5em}
{The sSFR--$M_\ast$ selection:} 
The sSFR--\sm\ selection identified 22 galaxies that are not part of the QG sample. 
Among these contaminants, two are dusty SFGs (COSMOS-34313 and EGS-10563) that are also selected by the sSFR–$t(z)$ criterion and discussed above. 
In addition, we identify five other dusty SFGs with {$A_V > 3$}, and spectral properties similar to these two galaxies; one of them is detected in the JCMT submillimeter survey, and another is detected in radio observations. 
These galaxies exhibit substantial dust attenuation, with $A_V$ values ranging from 3.1 to 3.5. 
{Another four contaminants (including COSMOS-19903 and UDS-8867) show prominent Balmer/4000\AA\ break, and stellar absorption features, indicative of evolved stellar populations. However, their sSFR are $\log(\mathrm{sSFR}/\mathrm{yr}^{-1})\sim-9.6 - -9.8$ do not satisfy our quiescent criterion.} 
The remaining 11 galaxies show nebular emission lines indicative of ongoing star formation or AGN activity. Four of these are identified as AGN hosts based on BPT diagnostics. One of them, EGS-16009 at $z=3.22$, is also detected in X-rays and is misclassified as quiescent by sSFR$_\mathrm{p}$--$\mathbf{t(z)}$ and all the color-based criteria considered here.

\vspace{0.5em}
{The sSFR$_\mathrm{p}$--$\mathbf{t(z)}$ selection:} 
Using only photometry data to derive the sSFR, we identify eight galaxies that are not part of the QG.  
Five are at $2 < z < 3$, one at $3 < z < 4$, and two at $4 < z < 5$. 
Three of these galaxies (COSMOS-19903, COSMOS-39385, and UDS-8867) show prominent Balmer/4000\AA break, and stellar absorption features, indicative of evolved stellar populations. Their sSFR are $\log(\mathrm{sSFR}/\mathrm{yr}^{-1})\sim-9.7$ to $-9.4$. 
The rest of them show strong \ha\ and \oiii\ emission lines, with median rest-frame equivalent widths of EW(\ha) = 252 \AA\ and EW(\oiii) = 211 \AA. 
Four of these five galaxies are confirmed AGN hosts located in the Seyfert region of the BPT diagram.

\vspace{0.5em}
{The $\mathbf{UVJ}$ color selection:} 
The $UVJ$ color selection identified 12 QGs, but also identified {seven} galaxies whose spectra and morphological characteristics are consistent with AGN and/or little red dots.  
Three are at $2 < z < 3$, three at $3 < z < 4$, and one at $4 < z < 5$. 
Similar to those misclassified by the sSFR$_\mathrm{p}$--$t(z)$ method, all show strong \ha\ and \oiii\ emission lines, with median rest-frame equivalent widths of EW(\ha) = 480 \AA\ and EW(\oiii) = 315 \AA. 
They also display very weak 4000 \AA\ breaks, with a median $D_n4000 = 0.9$. Their red $U - V$ colors are therefore driven by strong emission lines rather than by continuum features associated with old stellar populations. 

Morphologically, five out of {seven} are compact and point-like. 
Two of them are confirmed AGN based on the BPT diagram, and both of them are selected by the sSFR$_\mathrm{p}$--$t(z)$ selection. 
Except for EGS-16009, all remaining galaxies at $z > 3$ exhibit “V”-shaped SEDs, consistent with the little red dots \citep[e.g.,][]{Labbe2023}. 
Overall, these $UVJ$ contaminants exhibit spectral and morphological characteristics consistent with AGN and/or little red dots. 

\citet{Baker2025a} introduced an extended $UVJ$ color selection designed to include all 18 massive spectroscopically confirmed quiescent galaxies identified in the JADES survey at $2 \leq z \leq 5$. 
The selection boundaries are shown in the top panels of Figure~\ref{fig:colors_sSFR_M}.
This method achieves a completeness of {89\%} for the QG sample, but yields a relatively low purity {11\%}.  
The majority of contaminants are normal star-forming galaxies. 
Therefore, this color selection alone is not sufficient for identifying quiescent systems; rather, it can be used as a pre-selection technique in combination with other methods, such as the sSFR–$t(z)$ criterion \citep{Baker2025a}.

\vspace{0.5em}
{The $\mathbf{ugi_s}$ color selection:} 
The $ugi_s$ color selection identified {17} QGs and also identified {26} galaxies whose spectral features are more consistent with dusty SFGs and/or AGN. 
Most contaminants (20 of 26) lie at $2<z<3$. At $z>3$, the $ugi_s$ selection has a completeness of $100\%$ and a purity of $50\%$, compared with $67\%$ and $50\%$, respectively, for the $UVJ$ selection. Thus, $ugi_s$ provides higher completeness than $UVJ$ at $z>3$, while their purities are same. Three contaminants (COSMOS-19903, COSMOS-39385, and UDS-8867) exhibit prominent Balmer/4000-$\text{\AA}$ breaks and stellar absorption features indicative of evolved stellar populations; these galaxies are also selected by the sSFR$_{\mathrm p}$–$t(z)$ method. 
Most of the remaining contaminants are SFGs or AGN with strong nebular emission and have a median SFR of $54\,M_\odot\,\mathrm{yr}^{-1}$. Seven lie in the Seyfert region of the BPT diagram.
At $z>3$, the six contaminants are four AGN hosts and two heavily obscured SFGs with $A_V>3$.


\vspace{0.5em}
In summary, the sSFR–$t(z)$ and sSFR–$M_\ast$ selections provide the highest completeness for the QG sample, and the sSFR–$t(z)$ method yields the highest purity. 
The main contaminants in these two methods are dusty SFGs. 
This could arises from discrepancies between the spectra and broadband photometry, as the photometry is extracted from the integrated light of the entire galaxy, whereas the spectra probe only the more compact, central regions. 
For the sSFR–$M_\ast$ method, the selection is further sensitive to the adopted definition of the star formation main sequence. While the main sequence is relatively well established at $2 < z < 3.5$, it becomes increasingly uncertain at higher redshifts.  This could still affect the purity and completeness of quiescent galaxy identification. 

When only photometric data are available, the sSFR$_{\mathrm{p}}$–$t(z)$, $UVJ$, and $ugi_s$ methods yield QG completeness values of {$79\%$, $63\%$, and $89\%$}, respectively, and purity values of {$65\%$, $63\%$, and $40\%$}, respectively. The $ugi_s$ selection provides the highest completeness but the lowest purity, whereas the sSFR$_{\mathrm{p}}$–$t(z)$ selection provides the highest purity. 
Most contaminants in these selections are strong emission-line galaxies, most of which could hosting AGN. 
The presence of strong emission lines can artificially boost broadband fluxes, mimicking a pronounced 4000~\AA\ break and leading to misclassifications in both the $UVJ$ and $ugi_s$ diagrams, as well as in SED-fitting results. 
Although the synthetic $ugi_s$ colors avoid wavelengths dominated by strong nebular emission, this method still suffers from AGN contamination and is additionally more vulnerable to dusty star-forming interlopers.

\section{DISCUSSION} \label{sec:discussion}

\subsection{Number densities of quiescent galaxies} \label{sec:number_density}

In this section, we estimate the number density evolution of quiescent galaxies at $2\leq z \leq 5$ using the photometric quiescent sample and the results from above.
As we described in Section \ref{sec:sample_phot}, we construct the photometric quiescent sample using a Monte Carlo approach. For each galaxy and in each iteration of the Monte Carlo, we sample its redshift, stellar mass, and SFR based on the uncertainties from the \bagpipes\ photometry-only fits, together with the systematic offsets and scatter derived from comparing photometry-only and full spectral fits in the spectroscopic sample. 
We then apply the redshift, stellar mass, and sSFR-$t(z)$ selection criterion to each mock realization of the photometric sample. 
Note that the offset in SFR is not included in the Monte Carlo sampling because this will be dealt with using the ratio of true/false-positives and negatives discussed below. 
The number density is taken as the median of 1000 Monte-Carlo iterations, with the 16th and 84th percentiles as uncertainties. 
In addition, we derived the number density and its uncertainties in each field.

To account for quiescent galaxies that are either missed or incorrectly identified from the photometric selection, we apply a correction factor derived from the true-positive, true-negative, and false-positive rates based on the analysis above.
A true positive (TP) is defined as a galaxy identified as quiescent by both the spectroscopic and photometric SED fitting. 
A true negative (TN) refers to a galaxy classified as a quiescent galaxy based on the spectroscopic SED fitting, but not identified as such using only the photometric SED fitting.
A false positive (FP) refers to a galaxy classified as \textit{not} a quiescent galaxy based on spectroscopic SED fitting, but identified as a quiescent galaxy using the photometric SED fitting.
The correction factor is calculated as: 
\begin{equation} \label{eq:fcorr}
    f_{corr} = \frac{\mathrm{N_{TP} + N_{TN}}}{\mathrm{N_{TP} + N_{FP}}}
\end{equation} 
Uncertainties on $f_{\mathrm{corr}}$ are derived using Poisson statistics following \citet{Gehrels1986}. 
The uncertainties from the Monte Carlo samplings are much smaller than those estimated from Poisson statistics. 
Therefore, we adopt the Poisson uncertainties for $f_{\mathrm{corr}}$.

Finally, the corrected number density of photometrically selected quiescent galaxies is then given by:
\begin{equation} \label{eq:quiescent_density}
\Phi = \frac{N_{\mathrm{quiescent,\ phot}} \times f_{\mathrm{corr}}}{V}, 
\end{equation}
where $N_{\mathrm{quiescent,\ phot}}$ is the number of quiescent galaxies identified from the photometric sample, $f_{\mathrm{corr}}$ is the correction factor, and $V$ is the comoving volume, defined as 
\begin{equation} \label{eq:volume}
    V= \frac{\Omega}{3} (d_{z2}^{3} - d_{z1}^{3}) 
\end{equation}
where $d_{z1}$ and $d_{z2}$ are the comoving distances in Mpc at the lower and upper limits of the redshift range, and $\Omega$ is the survey area in steradians ({$\Omega = 495.84~\mathrm{arcmin}^{2} = 4.20\times10^{-5}~\mathrm{sr}$}). 
{Note that the effective survey area is determined directly from the detection mosaic; regions without photometric coverage, including the gaps between the NIRCam modules, are excluded.}

{Furthermore, we estimate the cosmic variance using five independent realizations of simulated lightcones made with the Santa Cruz semi-analytic model (SAM) for galaxy formation \citep{Somerville2015a, Yung2019, Yung2022, Yung2023}. Each lightcone covers $7200~\mathrm{arcmin}^{2}$, providing a total simulated area of $36{,}000~\mathrm{arcmin}^{2}$, spanning $0<z<10$. Following procedures similar to those described in \citet{Yung2023}, we select simulated galaxies using the same redshift intervals and quiescence criterion, $\mathrm{sSFR}<0.2/t(z)$, adopted for the observed sample. From each lightcone, we draw 250 mock realizations matched to the effective area of our survey, yielding 1,250 realizations in total. In each redshift bin, we calculate the root cosmic variance from the dispersion in the mock galaxy counts, $\sigma_{\mathrm{CV}}=\sigma_N/\langle N\rangle$. For a stellar-mass threshold of $\log(M_*/M_\odot)=9.5$, we obtain $\sigma_{\mathrm{CV}}=0.17$, $0.17$, and $0.37$ at $2<z<3$, $3<z<4$, and $4<z<5$, respectively. 
We also repeat the calculation using thresholds of $\log(M_*/M_\odot)=10.0$ and $10.5$, however, the small numbers of simulated quiescent galaxies above these thresholds, particularly at $z>3$, prevent robust estimates of the cosmic variance. We therefore adopt the estimates obtained with the $\log(M_*/M_\odot)=9.5$ threshold and add the corresponding cosmic variance uncertainties in quadrature in the number densities. }

\begin{figure*}
\centering
\includegraphics[width=\textwidth]{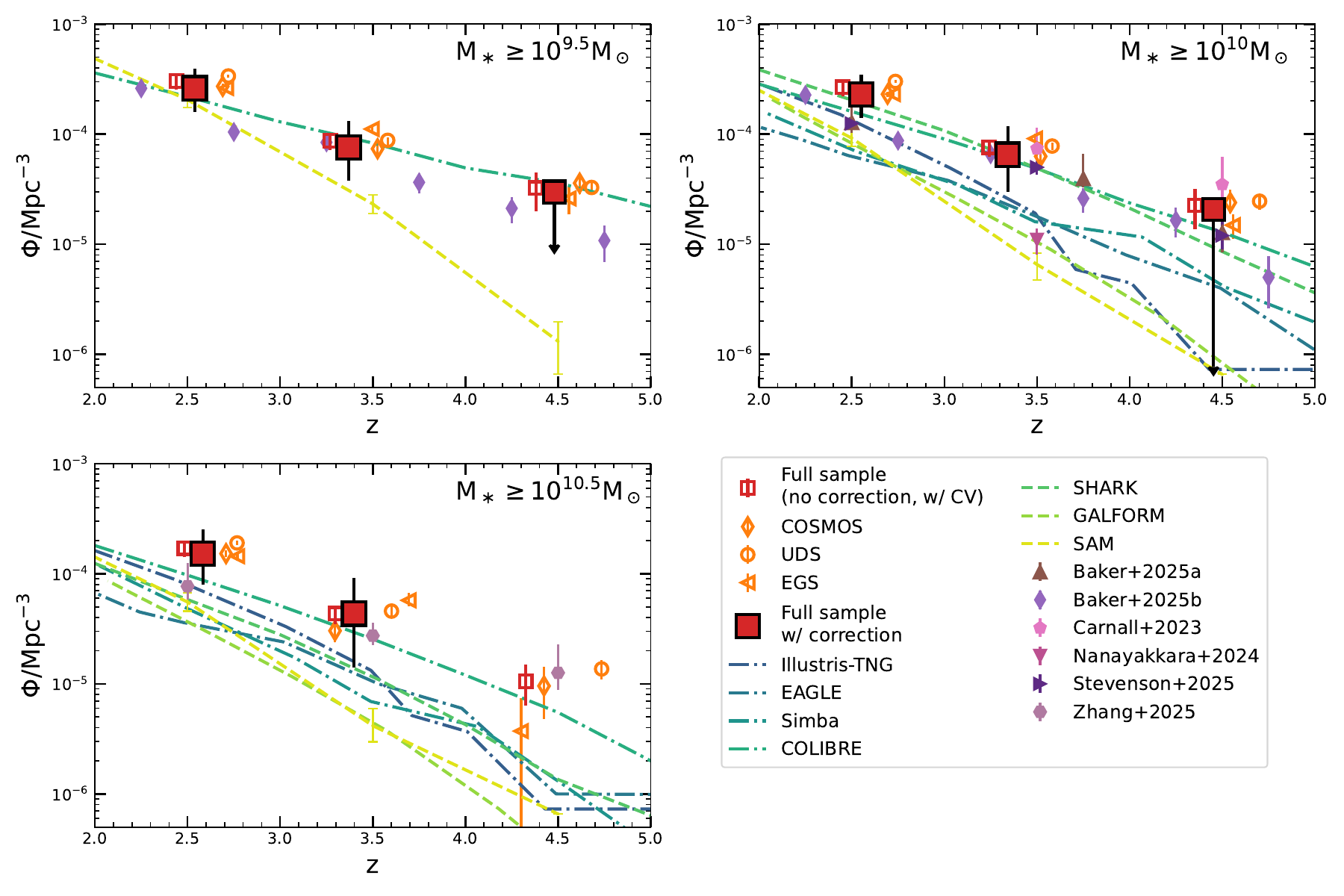} 
\caption{Quiescent galaxy number densities for $\mathrm{M_\ast \geq 10^{9.5} M_\odot}$ (\textit{top left}), $\mathrm{M_\ast \geq 10^{10} M_\odot}$ (\textit{top right}), and $\mathrm{M_\ast \geq 10^{10.5} M_\odot}$ (\textit{bottom left}) are shown as red open squares for the photometric sample, and separately for each fields (COSMOS in orange diamonds; UDS in orange circles, and EGS in orange hexagons), with uncertainties derived from the Monte-Carlo method and {cosmic variance estimated using the SAM lightcones \citep{Yung2023} for the full sample}. Number densities corrected using the spectroscopic sample are shown by black-edged red squares, with uncertainties from the spectroscopic correction. For comparison, observational results from \citep{Baker2025a, Baker2025b, Carnall2023, Nanayakkara2024, Zhang2025, Stevenson2025} are overplotted in the corresponding panel based on their stellar mass cut. Predictions from cosmological simulations (Illustris-TNG300, EAGLE, SIMBA and {COLIBRE}) and semi-analytic models (SHARK,  GAEA, GALFORM and {SAM}) \citep{Lagos2025, Yung2023, Chandro-Gomez2026a} are shown in colored lines in the relevant stellar mass threshold panels. {All these predictions adopt the same sSFR--$t(z)$ threshold as our quiescent sample selection}. }
\label{fig:density}
\end{figure*}

The Poisson errors associated with the correction fraction are significantly larger than those derived from Monte Carlo analysis {and the cosmic variations by factors of $>$4 and $\sim$2, respectively}. Thus, only the Poisson errors are shown for the corrected number densities in Figure~\ref{fig:density}. 
The numbers and number densities of photometrically-selected quiescent galaxies, as well as the numbers of true positive, true negative, and false positive, are summarized in Table \ref{tab:ndens}. 

The number densities for the photometric sample with spectroscopy selection correction, as well as the photometric sample and individual fields without correction, are shown in Figure \ref{fig:density} for samples selected with stellar mass limits of ${M_\ast \geq 10^{9.5} M_\odot}$, ${M_\ast \geq 10^{10} M_\odot}$, and ${M_\ast \geq 10^{10.5} M_\odot}$. 
After applying the $f_\mathrm{corr}$ from equation \ref{eq:fcorr}, the number densities decrease slightly compared to the uncorrected values, but remain consistent within the uncertainties. 
This is because the $f_\mathrm{corr}$ is close to unity, though there are non-negligible numbers of false positives and true negatives in the spectroscopic confirmation process. 
This indicates that the number density estimates based on photometry, at least when the photometry extends to rest-frame wavelengths of $\sim 1 \mu$m, are generally robust. 
At $4 < z < 5$, the corrected number densities for the two lower stellar mass thresholds are upper limits, as no quiescent galaxies have been spectroscopically confirmed in this redshift bin. For the highest stellar mass range, no corrected number density is reported because neither spectroscopically confirmed nor photometrically selected quiescent galaxies have corresponding spectroscopic observations.

We find that the field-to-field variations increase with increasing stellar mass thresholds and redshifts. 
In particular, the differences among fields exceed $3\sigma$ for galaxies with $M_* \geq 10^{10} M_\odot$ at $4 < z < 5$, and for those with $M_* \geq 10^{10.5} M_\odot$ at $3 < z < 4$ and $4 < z < 5$. 
This trend is expected, as the number of quiescent galaxies becomes small at higher redshifts and higher stellar masses, leading to correspondingly larger variations. 
These results indicate that number density measurements at the massive end and high redshift are particular susceptible to large uncertainties when derived from surveys with limited field coverage. 

\subsubsection{Comparing to Other Observational Results}

We compare our results with recent observational studies of quiescent galaxies at $z\sim2-5$ from \citet{Baker2025a, Baker2025b, Carnall2023, Nanayakkara2025, Zhang2025, Stevenson2025} at the corresponding stellar mass threshold. 
\citet{Baker2025b} studied quiescent galaxies ($M_* > 10^{9.5} M_\odot$) selected based on the $UVJ_{baker}$ color–color and sSFR - $t(z)$ criteria using NIRCam photometry data over a combined area of 800 arcmin$^{2}$ from CEERS \citep{Finkelstein2025}, PRIMER (Dunlop et al. in prep), and JADES \citep{Eisenstein2023, Rieke2023} surveys. 
Their selection method is consistent with our photometric quiescent sample; thus, the number densities reported in their study are comparable with our uncorrected values.
We find a slightly higher number density with a difference at 0.1-0.3 dex at $M_* > 10^{9.5} M_\odot$ and at $M_* > 10^{10} M_\odot$, corresponding to {$\sim$0.8–2.5$\sigma$} significance.

There are more studies focus on the $\mathrm{M_\ast \geq 10^{10} M_\odot}$ mass threshold. 
\citet{Baker2025a} derived the quiescent galaxies number density at $\mathrm{M_*} > 10^{10} \mathrm{M_\odot}$ using a photometrically selected sample based on the $UVJ_{baker}$ color–color and sSFR–$t(z)$ criteria, derived from NIRCam $UVJ$ colors of galaxies in the JADES survey (77.1 arcmin$^{2}$). Then they used 18 spectroscopically confirmed quiescent galaxies to calibrate the number densities derived from the photometric sample. 
This approach is consistent with the method used to obtain our corrected number densities, and their results agree with ours within the uncertainties. 
In addition, \citet{Stevenson2025} estimated quiescent number densities using a photometrically selected sample based on sSFR-$t(z)$ in the PRIMER and JADES fields, focusing on galaxies with \hst\ coverage. We find general agreement with their results, although we measure a slightly higher number density in the $2 < z < 3$ bin. 
Two additional studies have focused on spectroscopically confirmed quiescent galaxies. 
\citet{Nanayakkara2025} derived the number density of massive quiescent galaxies at $z\sim3-4.5$ using a sample of 19 quiescent galaxies with $M_* > 3\times 10^{10} M_\odot$ selected from the CEERS.  
\citet{Carnall2023} derived the number density of massive quiescent galaxies at $z\sim3-5$ using a sample of 10 quiescent galaxies with $M_* > 10^{10} M_\odot$ selected from the CEERS. 
The number densities reported by \citet{Carnall2023} are slightly higher than ours, while those from \citet{Nanayakkara2025} are lower. 
Both studies are based purely on spectroscopically confirmed quiescent galaxies; however, neither incorporated corrections for spectroscopic selection functions or incompleteness, which could partly explain the discrepancies with our measurements.

At the highest mass threshold {$M_\ast \geq 10^{10.5} M_\odot$}, \citet{Zhang2025} presented the number density of massive quiescent galaxies using a sample of 17 spectroscopically-confirmed massive ($M_* > 10^{10.3} M_\odot$) quiescent galaxies at $2 < z < 5$ selected from JWST/NIRSpec PRISM spectroscopy from the RUBIES survey \citep{deGraaff2025}. 
To correct for the spectroscopy selection function, they quantified sample incompleteness by accounting for the completeness of target selection, spectral success rate, and quiescent classification efficiency. 
Their results are consistent with ours within the uncertainties, although their number density distribution appears flatter, showing slightly lower values at the lowest redshift bin and slightly higher values at higher redshifts. 
\citet{Zhang2025} adopted a fixed $\mathrm{sSFR < 10^{-10}\ yr^{-1}}$ threshold to identify quiescent galaxies, which, compared to our sSFR–$t(z)$ criterion, would increase the number of quiescent galaxies at $z < 3$ and decrease it at $z > 3$. 
However, our results show the opposite trend, suggesting that the observed difference is unlikely to be driven by the choice of sSFR threshold. 
This discrepancy could reflect residual sample biases as RUBIES prioritizes sources that are in red color (F150W–F444W $>$ 2), bright (F444W $<$ 27 AB mag), or with high photometric redshifts ($z > 6.5$), which may artificially increase the fraction of quiescent galaxies in the spectroscopic sample. 
{Cosmic variance arising from the different survey fields and effective areas could also contribute to the discrepancy between the two measurements. In addition, both studies are affected by small-number statistics at $4<z<5$. Our highest-redshift bin contains a relatively small number ($\sim$15) of photometrically selected quiescent galaxies and no spectroscopically confirmed QGs, and the \citet{Zhang2025} sample contains only four objects in this redshift range.}

\subsubsection{Comparing to Simulations}

Furthermore, we compare our observed number density with predictions from \citet{Lagos2025}, which includes cosmological hydrodynamical simulations: Illustris-TNG300 \citep{Pillepich2018, Springel2018}, EAGLE \citep{Schaye2015, Crain2015}, and SIMBA \citep{Dave2019}, and semi-analytic models: SHARK \citep{Lagos2018, Lagos2024}, and GALFORM \citep{Lacey2016}. 
{In addition, we include predictions from the Santa Cruz SAM \citep{Yung2023} and the COLIBRE cosmological hydrodynamical simulation \citep{Schaye2026}. 
We adopt the same sSFR-based quiescence criterion and the stellar-mass threshold used in this study to select QGs from the simulations. 
We also take account of the measurement uncertainty in stellar masses of QGs by smoothing the stellar mass function with a Gaussian kernel of 0.25 dex \citep{Lagos2025,Chandro-Gomez2026b}, except for Santa Cruz SAM.
We present the theoretical predictions in Figure~\ref{fig:density}.}

{Overall, most models underpredict the observed quiescent-galaxy number densities. 
COLIBRE is the only one that agree with the redshift evolution of observed QG number densities within the uncertainty in all three stellar-mass ranges. 
However, at the highest mass range ($M_* \geq 10^{10.5} M_\odot$), it still offsets by 0.18–0.24 dex.
While the SHARK simulation can reproduce the redshift evolution of observed QG number densities in the mass range $M_* \geq 10^{10} M_\odot$, it fails in the mass ranges $M_* \geq 10^{9.5} M_\odot$ and $M_* \geq 10^{10.5} M_\odot$. This indicates that the SHARK simulation slightly overpredicts the QG number densities in the mass bin $10^{10}M_\odot < M_\ast < 10^{10.5}M_\odot$.}

{The discrepancy is most pronounced at the high-mass end $M_\ast \geq 10^{10.5} M_\odot$, potentially indicating} that massive quiescent galaxies either quench more rapidly, or form more rapidly, or both in the real universe than in current cosmological models. 
AGN feedback is widely considered as one of the primary mechanisms for quenching massive galaxies.
As discussed in \citet{Lagos2025}, {these disparate predictions for massive quiescent galaxies can be linked to differences in their AGN-feedback prescriptions. For example, the black-hole mass dependence for triggering the low-accretion feedback mode in IllustrisTNG and SIMBA makes it difficult for these simulations to produce intermediate-mass quiescent galaxies.}
{By comparison, COLIBRE provides better agreement with our observed number densities, suggesting that it more successfully reproduces the early emergence of massive quiescent galaxies. This agreement may partly reflect its updated prescriptions for black-hole growth and AGN feedback, with super-Eddington accretion potentially playing an important role \citep{Chaikin2026}.}
In addition, other quenching mechanisms, such as environmental effects or stellar feedback, could also contribute to the observed differences. 

On the other hand, insights into galaxy assembly can be gained from the stellar mass function.
\citet{Weaver2023} used the COSMOS2020 catalog to derive the number densities of galaxies from $z = 0.2$ up to $z = 7.5$. When compared to observations, both SHARK and EAGLE underproduce galaxies with $M_\ast > 10^{10.5}~M_\odot$ at $z > 2$, while Illustris-TNG and SIMBA show varying levels of agreement across $z \sim 2$–5. They also found an excess of massive galaxies at $z \sim 3$–5 relative to a Schechter function \citep{Schechter1976}. 
These results suggest that current physical prescriptions in simulations may not be sufficient to assemble massive systems ($M_\ast \sim 10^{10-11.5}~M_\odot$) at $z \sim 3$–5.

\begin{deluxetable}{l|ccccccccc}
\tablenum{2}
\tablecaption{Summary of Quiescent Galaxies from Photometric Sample \label{tab:ndens}}
\tablewidth{0pt}
\tablehead{
\colhead{Redshift} & \colhead{$z_\mathrm{med}$} & \colhead{N$_\mathrm{phot}$}  & \colhead{N$_\mathrm{phot\  QG}$} & \colhead{N$_\mathrm{spec}$} & \colhead{$\mathrm{N_{TP}}$} & \colhead{$\mathrm{N_{TN}}$} & \colhead{$\mathrm{N_{FP}}$} & \colhead{Number Density} & \colhead{Number Density}\\
\colhead{Range} & \colhead{} & \colhead{} & \colhead{} & \colhead{} & \colhead{} & \colhead{} & \colhead{} & \colhead{} & \colhead{with $f_\mathrm{corr}$} \\
\colhead{(1)} & \colhead{(2)} & \colhead{(3)} & \colhead{(4)} & \colhead{(5)} & \colhead{(6)} & \colhead{(7)} & \colhead{(8)} & \colhead{(9)} & \colhead{(10)}} 
\startdata
\hline
\multicolumn{10}{c}{$\log{M_\ast \geq 10^{9.5} M_\odot}$}\\
\hline
$2 < z < 3$ & 2.54 & $3124^{+29}_{-31}$ & $502^{+13}_{-12}$ & 103 & 8 & 4 & 6 & $30.48^{+0.79}_{-0.73} \times 10^{-5}$ & $26.13^{+13.41}_{-10.15} \times 10^{-5}$ \\
$3 < z < 4$ & 3.37 & $1359^{+21}_{-22}$ & $136^{+8}_{-7}$ & 59 & 7 & 0 & 1 & $8.69^{+0.51}_{-0.45} \times 10^{-5}$ & $7.60^{+5.55}_{-3.84} \times 10^{-5}$ \\
$4 < z < 5$ & 4.48 & $504^{+15}_{-14}$ & $46^{+4}_{-4}$ & 21 & 0 & 0 & 2 & $3.24^{+0.28}_{-0.28} \times 10^{-5}$ & $0.00^{+2.99}_{-0.00} \times 10^{-5}$ \\
\hline
\multicolumn{10}{c}{$\log{M_\ast \geq 10^{10} M_\odot}$}\\
\hline
$2 < z < 3$ & 2.55 & $1584^{+22}_{-21}$ & $440^{+12}_{-11}$ & 61 & 8 & 4 & 6 & $26.72^{+0.73}_{-0.67} \times 10^{-5}$ & $22.90^{+11.75}_{-8.89} \times 10^{-5}$ \\
$3 < z < 4$ & 3.34 & $682^{+16}_{-15}$ & $119^{+7}_{-7}$ & 31 & 6 & 0 & 1 & $7.60^{+0.45}_{-0.45} \times 10^{-5}$ & $6.52^{+5.24}_{-3.53} \times 10^{-5}$ \\
$4 < z < 5$ & 4.45 & $203^{+10}_{-10}$ & $32^{+5}_{-4}$ & 9 & 0 & 0 & 2 & $2.26^{+0.35}_{-0.28} \times 10^{-5}$ & $0.00^{+2.08}_{-0.00} \times 10^{-5}$ \\
\hline
\multicolumn{10}{c}{$\log{M_\ast \geq 10^{10.5} M_\odot}$}\\
\hline
$2 < z < 3$ & 2.58 & $681^{+13}_{-14}$ & $281^{+10}_{-10}$ & 26 & 5 & 3 & 4 & $17.06^{+0.61}_{-0.61} \times 10^{-5}$ & $15.17^{+10.20}_{-7.22} \times 10^{-5}$ \\
$3 < z < 4$ & 3.40 & $261^{+11}_{-9}$ & $68^{+6}_{-6}$ & 13 & 4 & 0 & 0 & $4.34^{+0.38}_{-0.38} \times 10^{-5}$ & $4.34^{+4.86}_{-2.94} \times 10^{-5}$ \\
$4 < z < 5$ & 4.43 & $65^{+6}_{-5}$ & $15^{+3}_{-2}$ & 2 & 0 & 0 & 0 & $1.06^{+0.21}_{-0.14} \times 10^{-5}$ & -- \\
\hline
\enddata
\tablecomments{(1)-(2) Redshift bins and the median redshift of the photometric sample. (3)-(4) Median number of photometric and photometrically selected quiescent galaxies in each redshift range and stellar mass ranges using Monte Carlo sampling, with uncertainties given by the 16th–84th percentiles; (5)-(8) Number of spectroscopic galaxies, true positive, ture negative, false positive galaxies (no Monte Carlo applied); (9) Number densities of photometrically selected quiescent galaxies, with uncertainties from the 16th–84th percentiles of the Monte Carlo sampling; (10) Number densities of photometrically-selected quiescent galaxies with $f_\mathrm{corr}$ from equation \ref{eq:fcorr}, with the uncertainties given by the Poisson errors on $f_\mathrm{corr}$. Number densities for $\log\mathrm{M_\ast \geq 10^{9.5} M_\odot}$ and $\log\mathrm{M_\ast \geq 10^{10} M_\odot}$ at $4 < z < 5$ are upper limits. }
\end{deluxetable}

\begin{figure*}
    \centering
    \includegraphics[width=\textwidth]{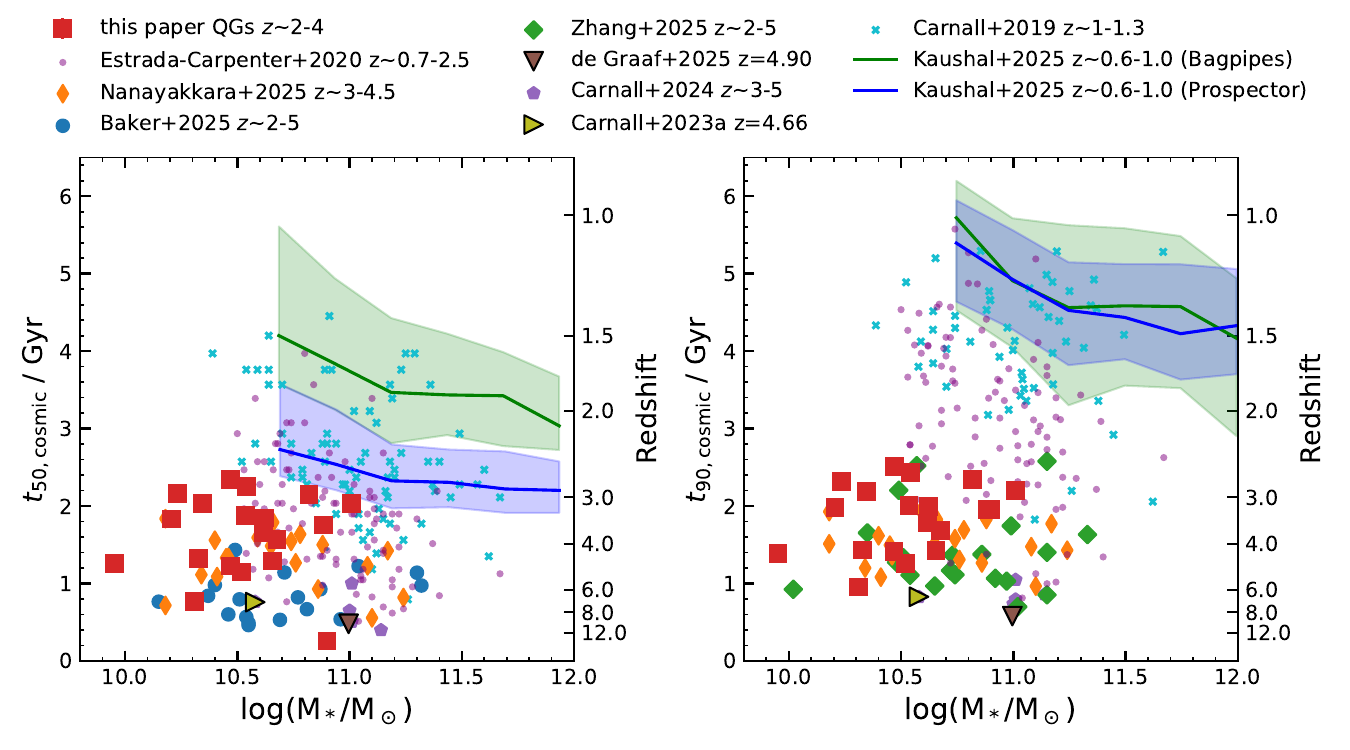}
    \caption{Formation times of 50\% of stellar mass ($t_{50, \mathrm{cosmic}}$, left) and 90\% of stellar mass ($t_{90, \mathrm{cosmic}}$, right) as a function of stellar mass for our spectroscopic QGs (red squares). As comparison, quiescent galaxies at $z \sim 2$–5 from recent \jwst\ spectroscopic studies are shown in colored markers \citep{Carnall2023, Carnall2024, Baker2025a, Nanayakkara2025, Zhang2025, deGraaff2025}, those at $z \sim 0.7$–2.5 from \hst\ grism CLEAR survey as purple dots \citep{Estrada-Carpenter2020}. For lower redshift samples, we show the median trend as lines with the 16th-84th percentile ranges as shaded region: quiescent galaxies at $z \sim 1.0$–1.3 using spectra from VANDELS spectroscopic survey (cyan line, \citealp{Carnall2019}), and those at $z \sim 0.6$–1.0 using spectra from the LEGA-C spectroscopic survey (blue and green lines derived using different SED fitting codes \citealp{Kaushal2024}). Note that \citet{Carnall2019, Carnall2023, Carnall2024}, and \citet{Nanayakkara2025} report time of quenching, defined as the time when sSFR $< 0.1/t_\mathrm{obs}$), instead of $t_{90}$.} 
    \label{fig:formation_times}
\end{figure*}

\subsection{Formation and quenching time of quiescent galaxies} \label{sec:formation_times}

We find our QGs formed 50\% of their stellar mass between 0.50-0.91 Gyr, and assembled 90\% of their stellar mass between 0.38-0.72 Gyr prior to the observed epoch (Figure~\ref{fig:spec_q_props2}).  
These times correspond to formation and quenching redshifts of $z \sim 3.0 - 4.8$ and $z \sim 2.8 - 4.2$, respectively. 
These relatively young ages indicate that the quiescent population at these redshifts has only recently ceased star formation. 
Moreover, the short interval between $t_{50}$ and $t_{90}$ of $\sim0.1$–0.2 Gyr indicates rapid stellar mass assembly followed by efficient quenching. 
We compare the formation times of our quiescent galaxies with those of quiescent systems at other cosmic epochs, as shown in Figure~\ref{fig:formation_times}. 
We defined $t_\mathrm{50, cosmic}$ and $t_\mathrm{90, cosmic}$ as the age of the Universe when galaxies formed 50\% and 90\% of their total stellar mass, respectively.

At high redshift ($z \sim 2$–5), we include individual quiescent galaxies from recent \jwst\ spectroscopic studies, including 18 quiescent galaxies identified from the JADES survey at $z \sim 2$–5 \citep{Baker2025a}, 19 quiescent galaxies at $z \sim 3$–4.5 \citep{Nanayakkara2025}, 20 quiescent galaxies identified from the RUBIES survey at $z \sim 2$–5 \citep{Zhang2025}, a quiescent galaxy at $z=4.66$ \citep{Carnall2023}, and a quiescent galaxy at $z=4.90$ \citep{deGraaff2025}. For all of these samples, formation and quenching times are derived using full spectroscopic SED fitting. 
Overall, these high-redshift quiescent galaxies have formation and quenching times that are generally consistent with our sample. 
Nevertheless, we find that our quiescent galaxies have slightly larger median formation and quenching times, with $t_\mathrm{50, cosmic}=1.76$ Gyr and $t_\mathrm{90, cosmic}=1.95$ Gyr, compared to median values of $t_\mathrm{50, cosmic}=0.98$ Gyr and $t_\mathrm{90, cosmic}=1.38$ Gyr for these literature samples. 
In particular, only one galaxy in our sample formed half of its stellar mass within the first 1 Gyr after the Big Bang, whereas \citet{Baker2025a} find that 14 out of 18 quiescent galaxies and \citet{Nanayakkara2025} find 4 out of 19 quiescent galaxies have $t_{\mathrm{50, cosmic}} < 1$ Gyr. 
{For our QG sample, adopting double-power-law and delayed-$\tau$ SFHs would reduce the median $t_{50,\mathrm{cosmic}}$ by only $0.01$ and $0.09$ Gyr, respectively, while reducing the median $t_{90,\mathrm{cosmic}}$ by $0.03$ Gyr in both cases. These small shifts are insufficient to explain the differences from the literature samples.}
The offsets may instead partly due to the fact that our sample lies at a slightly lower redshift and lower stellar mass than the comparison surveys. 

At intermediate and low redshift, we compare our results to quiescent galaxies from the \hst\ grism CLEAR survey at $z \sim 0.7$–2.5 \citep{Estrada-Carpenter2020}, quiescent galaxies at $z \sim 1.0$–1.3 from the VANDELS spectroscopic survey \citep{Carnall2019}, and quiescent galaxies at $z \sim 0.6$–1.0 from the LEGA-C spectroscopic survey \citep{Kaushal2024}. 
\citet{Estrada-Carpenter2020} used \texttt{Prospector} to fit nonparametric star-formation histories to a sample of $\sim$100 quiescent galaxies spanning $0.7 < z < 2.5$ and stellar masses $10.5 < {\log(M_\ast/M_\odot)} < 12$. 
They found a wide range of formation times, with $t_{50, \mathrm{cosmic}}$ spanning 1.1–2.6 Gyr and $t_{90, \mathrm{cosmic}}$ spanning 2.2–4.1 Gyr. 
\citet{Kaushal2024} estimates $t_{50, \mathrm{cosmic}}$ and $t_{90, \mathrm{cosmic}}$ for $\sim$1000 quiescent galaxies at $z\sim0.6$-1 with $10.5 < {\log(M_\ast/M_\odot)} < 12$. They use two different modeling approaches: a parametric star-formation history implemented with \bagpipes\ and a nonparametric star-formation history modeled with \texttt{Prospector}. The nonparametric SFHs consistently favor earlier stellar mass assembly, yielding lower $t_{50}$ values than the parametric fits, while the inferred $t_{90}$ values remain broadly consistent between the two methods. 
Compared to quiescent galaxies at $z \sim 1$, we find that our quiescent galaxies have lower $t_{50}$ values by $\sim1$ Gyr and much lower $t_{90}$ values by $\sim3.5$ Gyr. This indicates that the dominant quiescent galaxy population at lower redshift formed and quenched at later cosmic times. Alternatively, the difference may suggest that high-redshift QGs may subsequently rejuvenate and temporarily or permanently leave the quiescent population.

Another comparison with these formation times is the quenching timescale, which we define as the interval between $t_{90}$ and $t_{50}$. For our quiescent galaxies, we find short quenching timescales of $0.1$–$0.2$ Gyr. These values are comparable to those reported by \citet{Nanayakkara2025}, who find quenching timescales of $0.03$–$0.35$ Gyr for quiescent galaxies at $z \sim 3$–4.5. 
They are also consistent with results for two quiescent galaxies at $z \sim 4.5$–5, for which quenching timescales are estimated to be $\sim200$ Myr 
\citep{Carnall2023, deGraaff2025}. 
In contrast, studies at lower redshift find substantially longer quenching timescales: \citet{Estrada-Carpenter2020} report values of $0.8$–$1.8$ Gyr, while \citet{Kaushal2024} find quenching timescales of $1.7$–$2.7$ Gyr using nonparametric SFHs with \texttt{Prospector} and $0.2$–$2.3$ Gyr using parametric SFHs with \bagpipes. 
The much shorter quenching timescales observed at $z \gtrsim 2$ suggest that quiescent galaxies at early cosmic times experience more rapid suppression of star formation, potentially driven by intense feedback or rapid gas exhaustion, and these processes are more effective at earlier epochs. 
This is supported by recent cosmological simulations.  Using the Magneticum Pathfinder simulation suite, \citep{Kimmig2025} find that these galaxies are rapidly assembled through an intense burst of star formation and subsequent quenched stellar and AGN feedback, triggered by a particularly isotropic collapse of surrounding gas, which ejects most of the gas far outside the galaxy and quench on timescales of $\lesssim200$ Myr.
{A similar result is found in COLIBRE, where AGN feedback depletes the molecular gas and dust reservoirs of these systems \citep{Chandro-Gomez2026b} on timescales consistent with the observed formation and quenching timescales ($\lesssim600$ Myr in Figure 5; \citealp{Chandro-Gomez2026a}).}

\subsection{Caveat on Aperture Effects}\label{sec:diss_caveat}

{We discuss a caveat regarding aperture effects in this section. Limited by the relatively small slit width of NIRSpec shutters, emission extending beyond the shutter aperture cannot be fully captured by the spectrum, which could potentially bias our studies. However, an object-by-object slit-loss correction is almost impossible, because the fraction of enclosed light relies on source centering, the nodding pattern, barshadow, the wavelength-dependent PSF, galaxy morphology and color gradients, and the spectral background subtraction. To mitigate such an effect, we use a second-order Chebyshev polynomial in the joint spectroscopic+photometric \textsc{Bagpipes} fits to mimic the smooth wavelength-dependent slitloss and calibrate the spectroscopy to the integrated broadband photometry. }

{However, this method assumes the spectral features are uniform across the target, such as the emission line equivalent width and Balmer/4000\AA\ breaks. Spatially resolved studies of quiescent galaxies at $0.6 < z < 2.5$ demonstrate a color gradient in the sample and show that the gradient may depend on stellar-population age \citep{Suess2020, Cheng2024}. In particular, \citet{Suess2020} find that younger quiescent galaxies generally have flatter color gradients, whereas older systems exhibit stronger negative gradients. 
In this scenario, our estimation of stellar mass would be overestimated, while the estimation of SFR would be underestimated. On the other hand, at $0.6 < z < 1.0$, \citet{Cheng2024} find average age and [Mg/Fe] gradients consistent with being flat, with median slopes of 0.007 and -0.008, respectively, but mildly negative [Fe/H] gradients with a median slope of -0.048. These results suggest that aperture-related age biases may be modest on average, at least at lower redshifts, although metallicity gradients and substantial object-to-object variations remain uncertain at high redshift. }

{To further evaluate how the slitloss affects the photometry, we measure the fraction of the light enclosed by the NIRSpec shutter. We first measure the observed half-light radius of QGs in F150W and F444W using the circular curve of growth. The measured radii range from $0.116''$–$0.239''$ and $0.127''$–$0.268''$, with a median of $0.150''$ and $0.168''$ respectively. Compared to the size of MSA shutters ($0.2''\times0.46''$ ), this corresponds to a median enclosed-light fraction of only 42\% and 35\%. This indicates that the systematic introduced by the slitloss could be non-negligible in our study, if the high-redshift quiescent galaxies have already built the color gradient. These aperture effects remain a systematic limitation of our analysis and will be investigated using spatially resolved modeling in future work.}

\section{Summary} \label{sec:summary}

In this paper, we studied a sample of {19 quiescent galaxies at $2 \leq z \leq 4$} with stellar masses ${M_\ast} \geq 10^{9.5}~{M_\odot}$ identified in the JWST NIRSpec/MSA spectroscopy from CANDELS-Area Prism Epoch of Reionization Survey (CAPERS) in the COSMOS, EGS, and UDS fields. 
Quiescent galaxies are defined as systems with specific star formation rates averaged over 100 Myr below 20\% of the inverse cosmic age (sSFR$_{100} \leq 0.2/t(z)$). 
Galaxy properties are derived from full spectral SED fitting of the spectra combined with broadband photometry. 
We have obtained the following findings: 

\begin{itemize} 
    \item Quiescent galaxies lie more than 1 dex below the SFMS, have lower dust attenuation, and show significantly stronger Balmer and 4000 \AA\ breaks compared to those of SFGs with similar stellar mass. 
    {Two} quiescent galaxies show evidence of AGN activity.

    \item Full spectral SED fitting indicates that these galaxies formed 50\% of their stellar mass $0.50$--$0.91$ Gyr before observation (corresponding to $z\sim3.0$--$4.8$) and assembled 90\% of their stellar mass $0.38$--$0.72$ Gyr before observation (corresponding to $z\sim2.8$--$4.2$). The short interval between these two epochs, $\sim0.1$--$0.2$ Gyr, implies rapid stellar-mass assembly followed by efficient quenching. {These quenching timescales are consistent with those of other high-redshift QGs but are substantially shorter than those measured at lower redshift, suggesting more rapid quenching in the early Universe. }

    \item We compare our spectroscopic sample of quiescent galaxies to photometric selection methods: sSFR$\mathrm{p}$–$t(z)$, $UVJ$, and $ugi_s$. 
    {These methods achieve completeness values of 79\%, 63\% and 89\%, respectively, but relatively low purities of 65\%, 63\% and 40\%, respectively.}
    Most contaminants in these photometric selection methods are strong emission-line galaxies, with many of them show evidence of AGN activity. 

    \item We derive the number density of quiescent galaxies at $2 \leq z \leq 5$ using the sSFR–$t(z)$ criterion applied to the full photometric sample. Uncertainties in redshift, stellar mass, and SFR, as well as systematic offsets between spectroscopic and photometric \bagpipes\ fits, are accounted for through Monte Carlo simulations. A correction factor, based on the completeness and purity of the spectroscopically confirmed sample, is applied. The resulting number densities are consistent with other JWST studies. 

    \item Compared to {seven} state-of-the-art cosmological galaxy formation simulations and semi-analytic models, {most models underpredict our observed quiescent-galaxy number densities across all three stellar-mass thresholds at $2<z<4$, while COLIBRE generally provides good agreement. The discrepancy is greatest at $M_\ast\geq10^{10.5}\,M_\odot$, where the models other than COLIBRE underpredict the observations by $0.34$–$0.91$ dex ($\sim$1.0-2.3$\sigma$). } 
    This suggests that massive quiescent galaxies may have either quenched more rapidly, formed more efficiently, or both, in the real universe than predicted by most current cosmological models. 
    
\end{itemize}


\begin{acknowledgments}
We acknowledge the hard work of our colleagues in the CAPERS collaboration and everyone involved in the \jwst\ mission. 
We wish to thank the anonymous referee for a thorough and constructive report that improved the quality and clarity of this work.   
This work benefited from support from the George P. and Cynthia Woods Mitchell Institute for Fundamental Physics and Astronomy at Texas A\&M University. 
CP thanks Marsha and Ralph Schilling for generous support of this research. 
This work is based on observations made with the NASA/ESA/CSA James Webb Space Telescope. The data were obtained from the Mikulski Archive for Space Telescopes at the Space Telescope Science Institute, which is operated by the Association of Universities for Research in Astronomy, Inc., under NASA contract NAS 5-03127 for JWST. 
These observations are associated with \jwst\ programs GO-6368, ERS-1345, and GO-1837. 
Some/all the {\it JWST} data presented in this paper were obtained from the Mikulski Archive for Space Telescopes (MAST) at the Space Telescope Science Institute. 
The CAPERS observations are associated with program 6368 and can be accessed via DOI: 10.17909/0q3p-sp24. 
E.F.-J.A. acknowledges support from UNAM-PAPIIT project  IA104725, and from CONAHCyT Ciencia de Frontera project ID: CF-2023-I- 506. 
HL acknowledges support from a UKRI Frontier Research Grantee Grant (PI Carnall; grant reference EP/Y037065/1).
National Radio Astronomy Observatory is a facility of the National Science Foundation operated under cooperative agreement by Associated Universities, Inc. 

\end{acknowledgments}

\appendix
\restartappendixnumbering

\section{Image, SED and properties for quiescent galaxies}\label{app1}

\begin{figure*}
\centering
\includegraphics[width=0.32\textwidth]{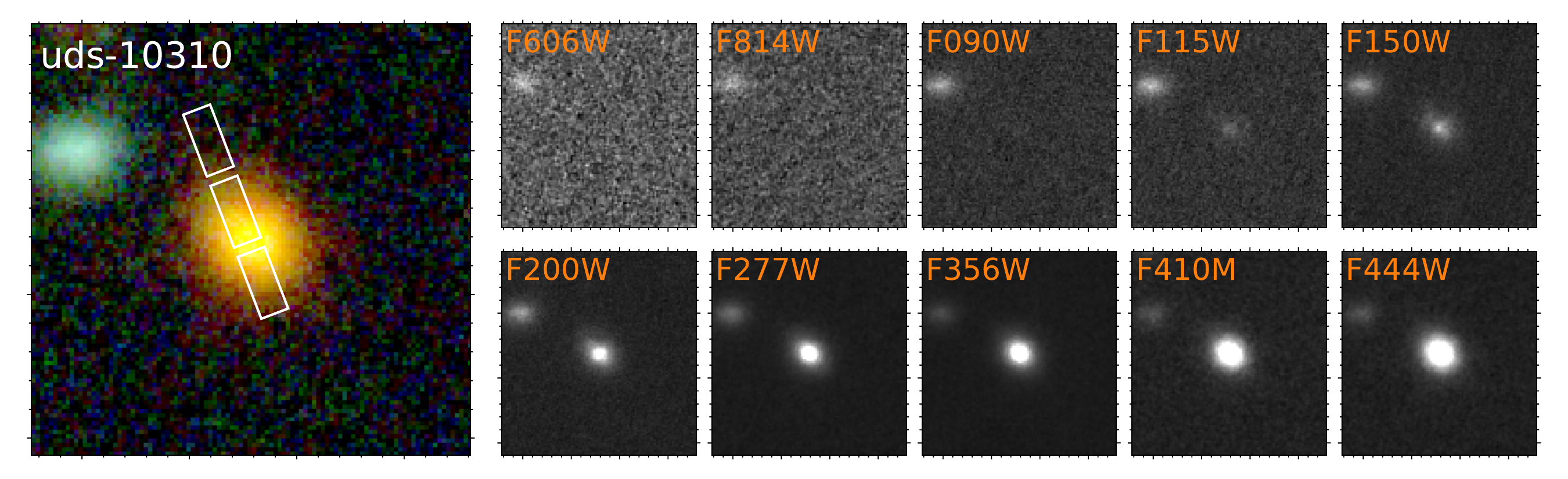}
\includegraphics[width=0.32\textwidth]{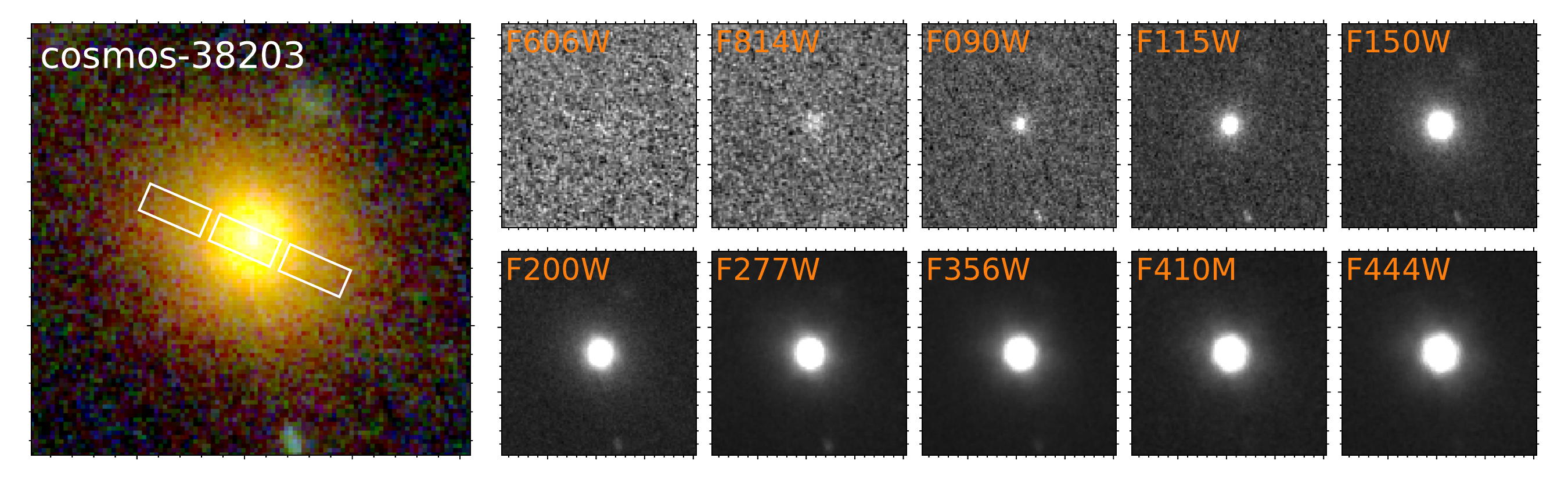} 
\includegraphics[width=0.32\textwidth]{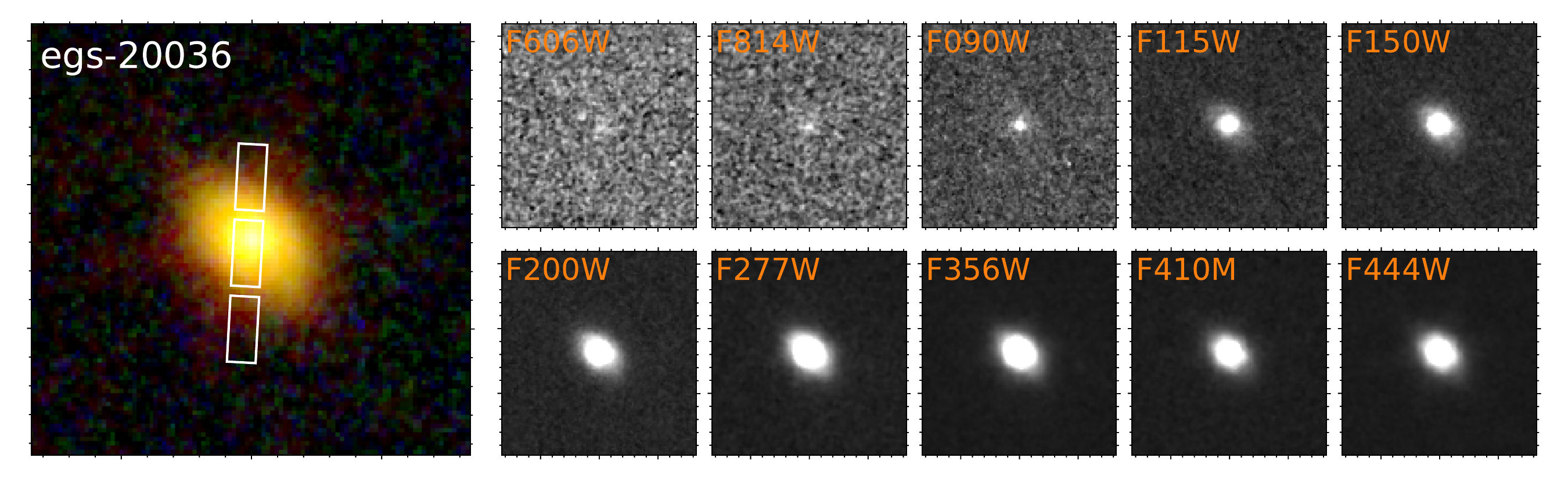} \\
\includegraphics[width=0.32\textwidth]{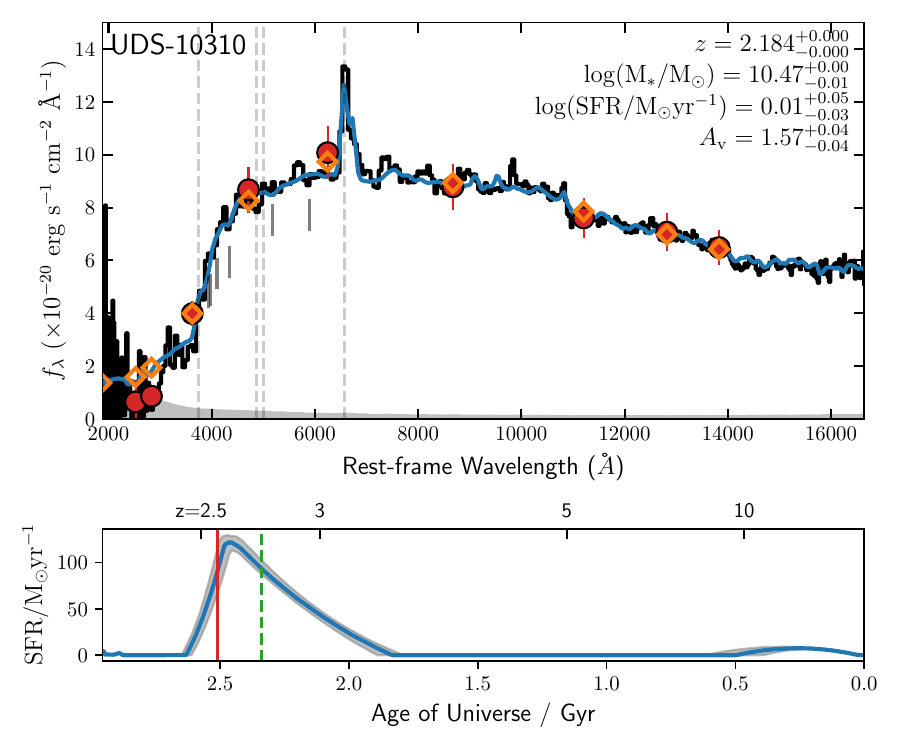} 
\includegraphics[width=0.32\textwidth]{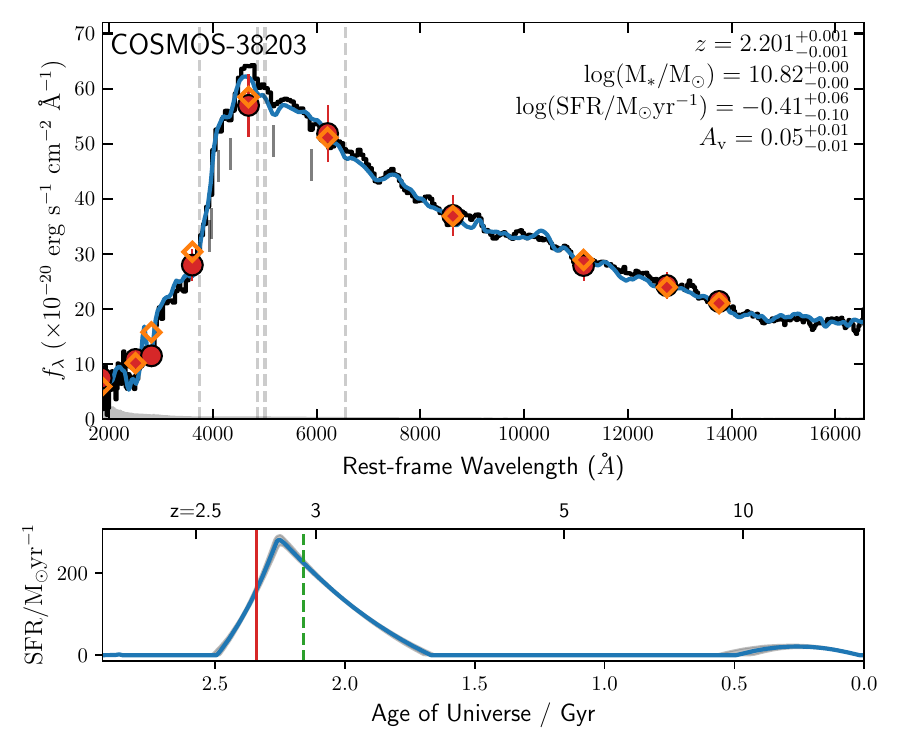} 
\includegraphics[width=0.33\textwidth]{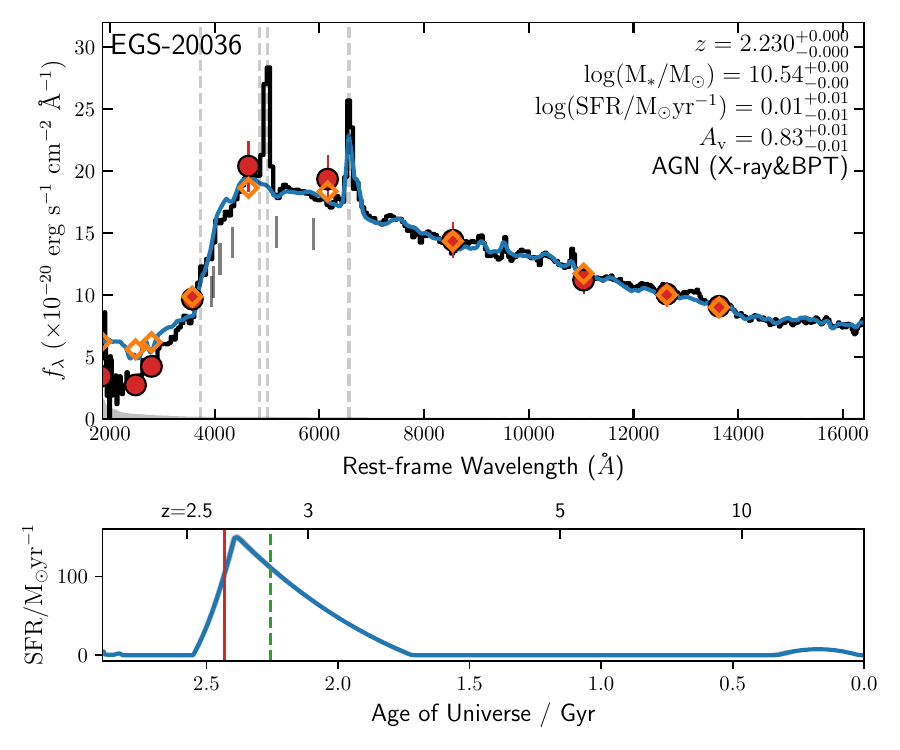} \\
\includegraphics[width=0.33\textwidth]{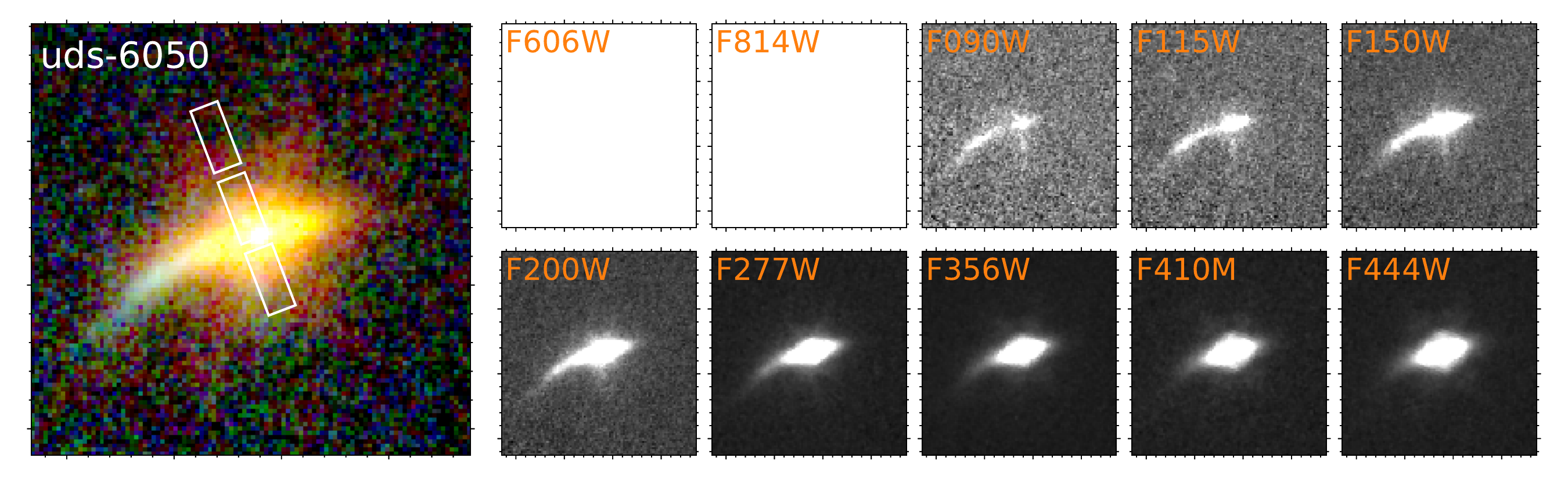} 
\includegraphics[width=0.32\textwidth]{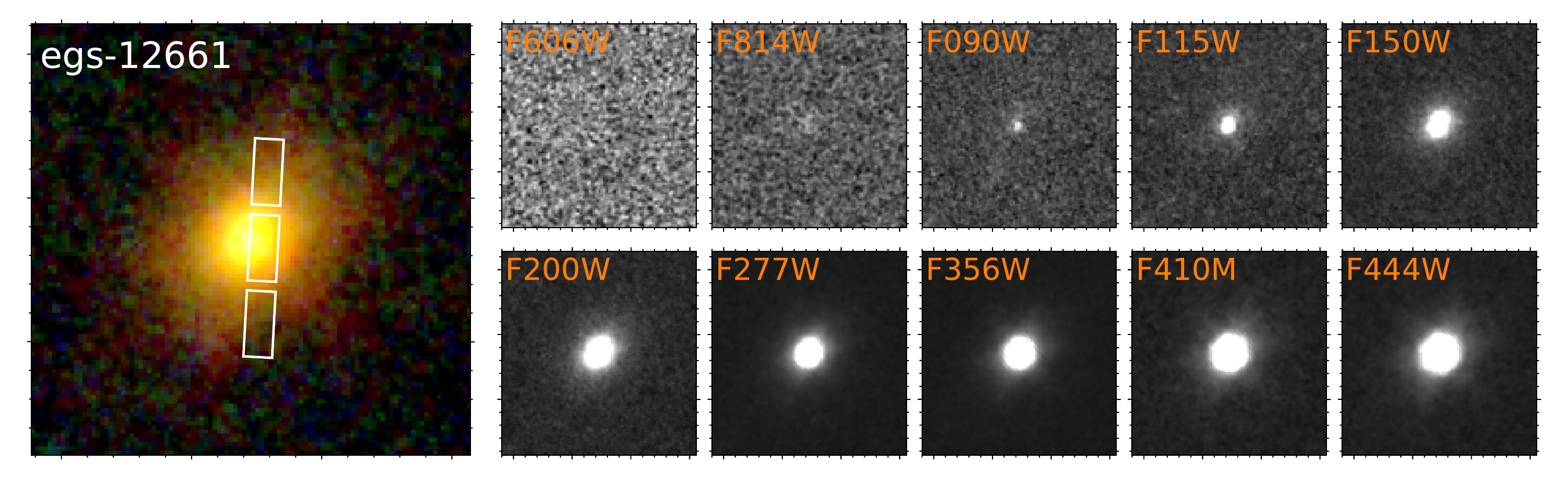} 
\includegraphics[width=0.33\textwidth]{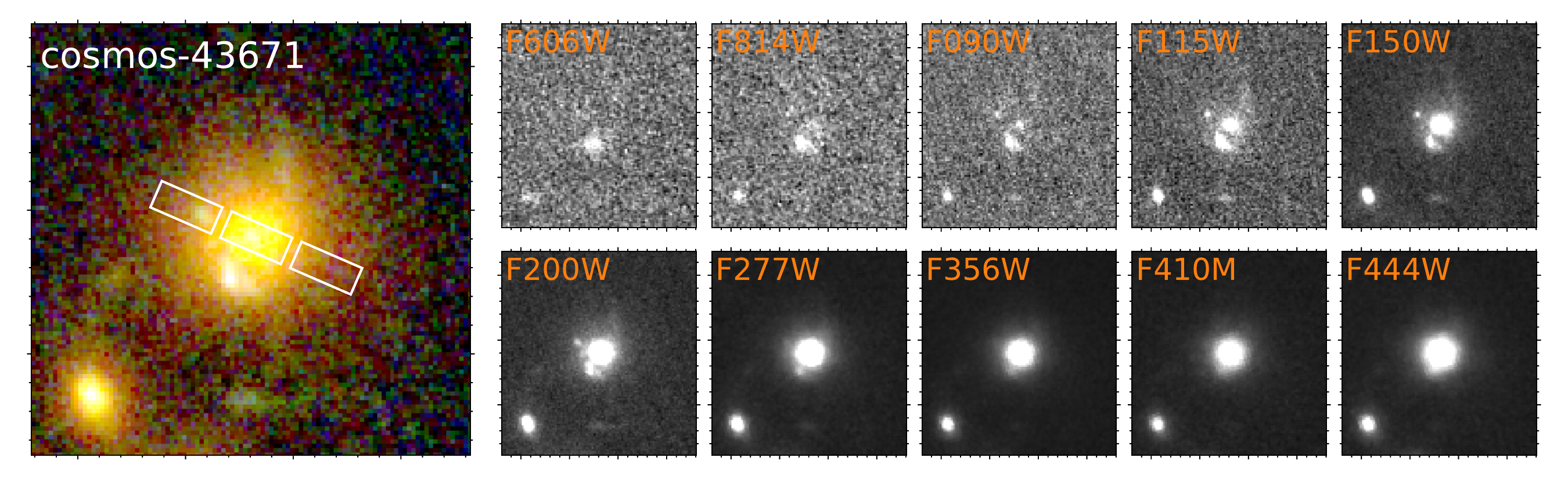} \\
\includegraphics[width=0.33\textwidth]{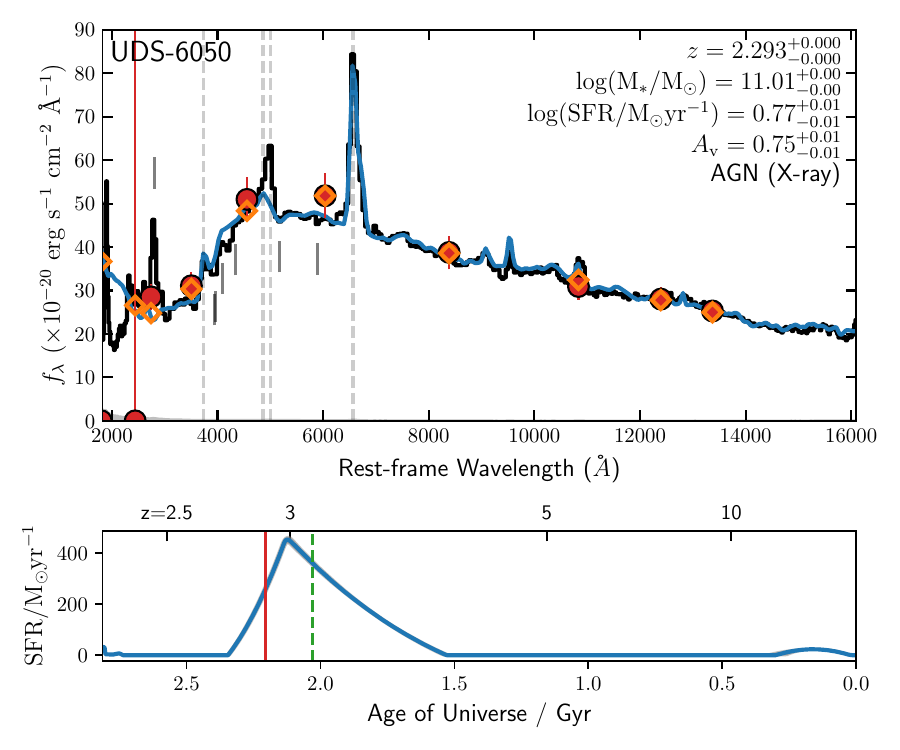} 
\includegraphics[width=0.32\textwidth]{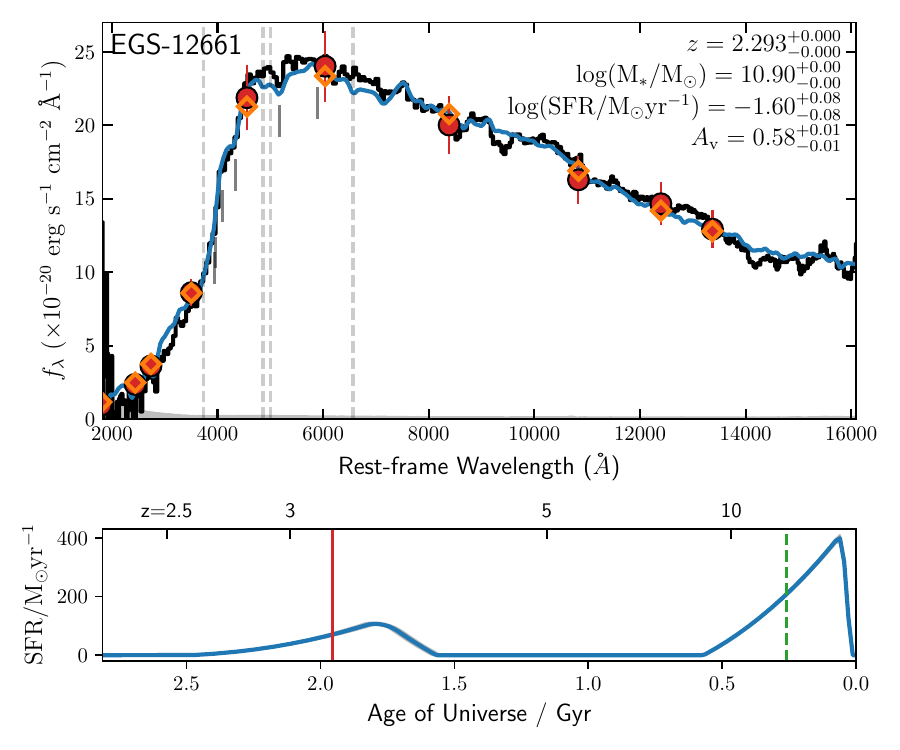} 
\includegraphics[width=0.33\textwidth]{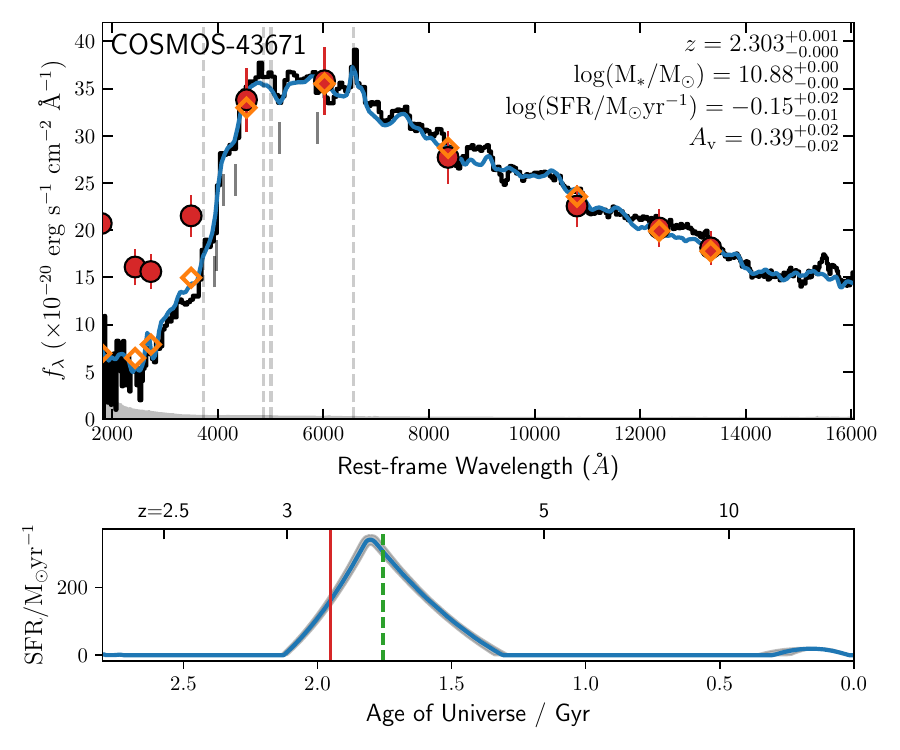} \\
\includegraphics[width=0.32\textwidth]{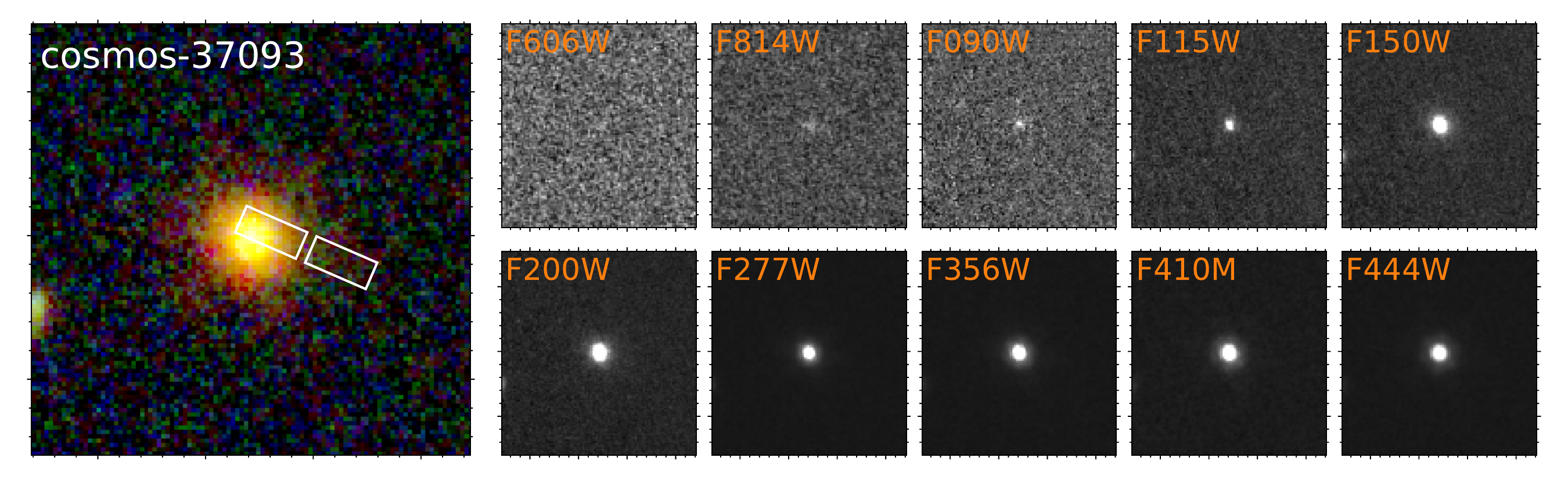} 
\includegraphics[width=0.32\textwidth]{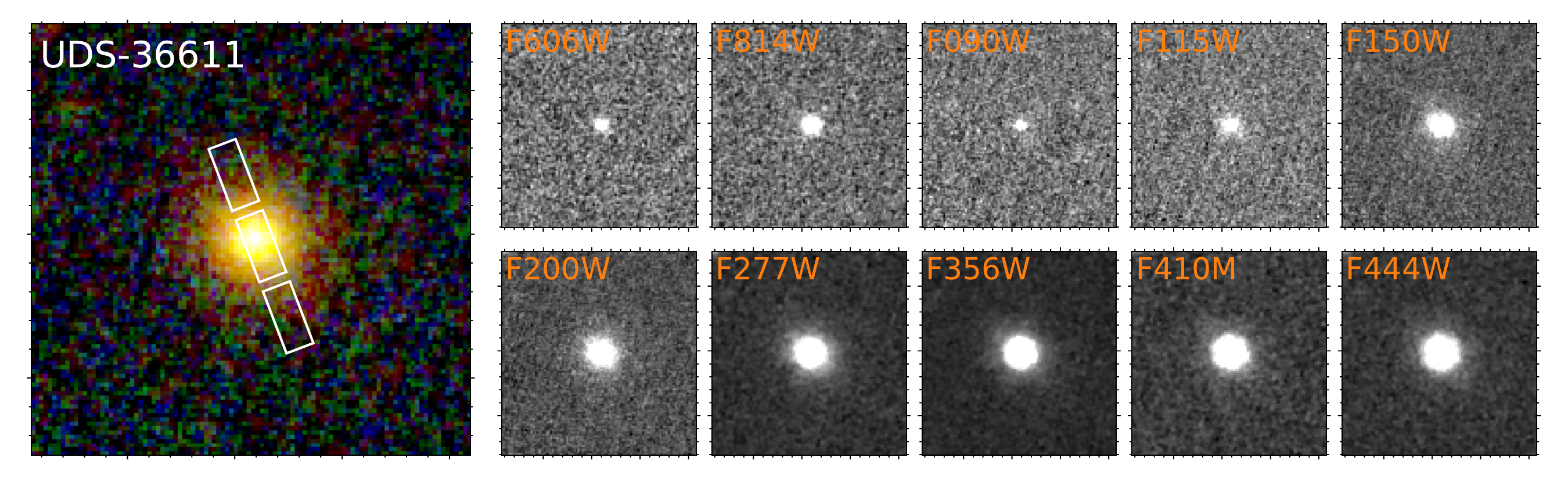} 
\includegraphics[width=0.32\textwidth]{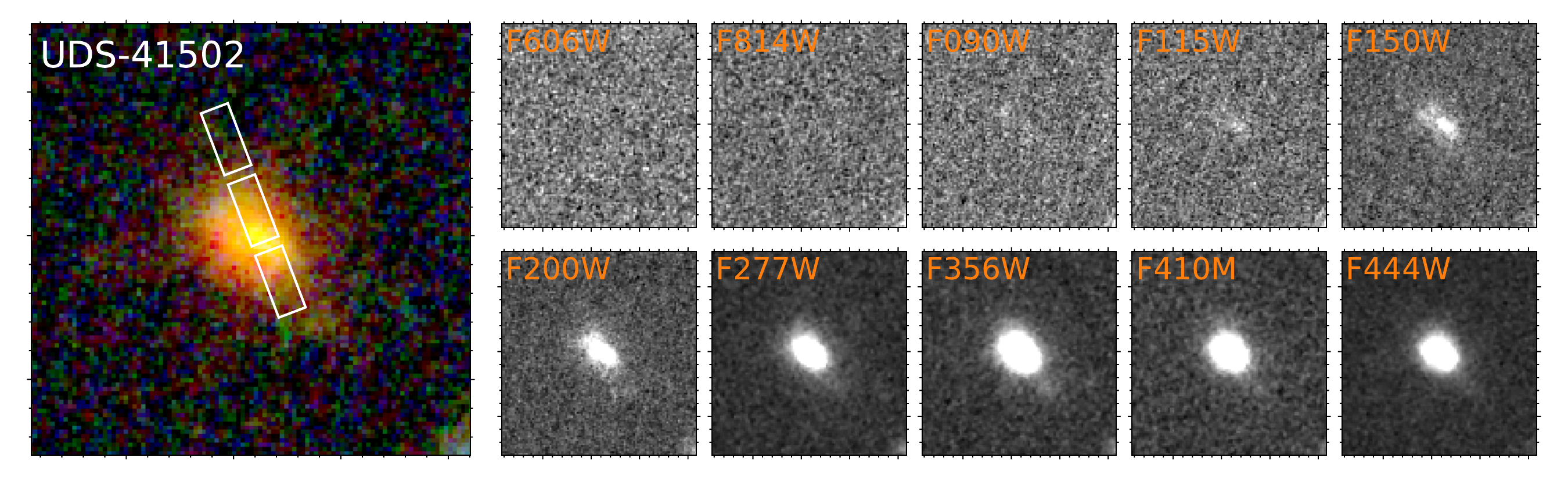} \\
\includegraphics[width=0.32\textwidth]{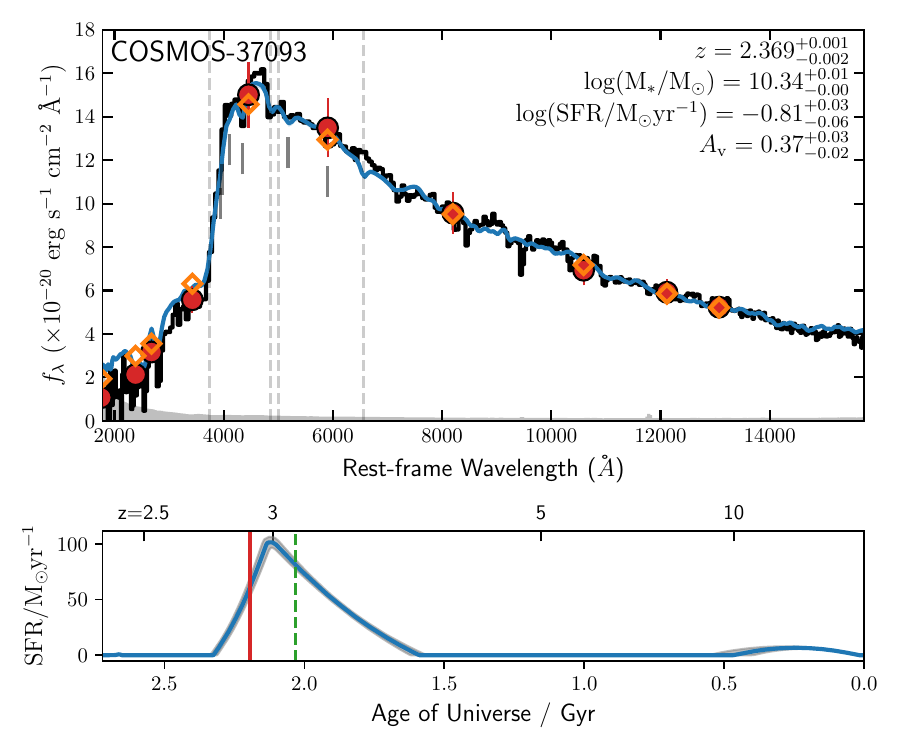} 
\includegraphics[width=0.32\textwidth]{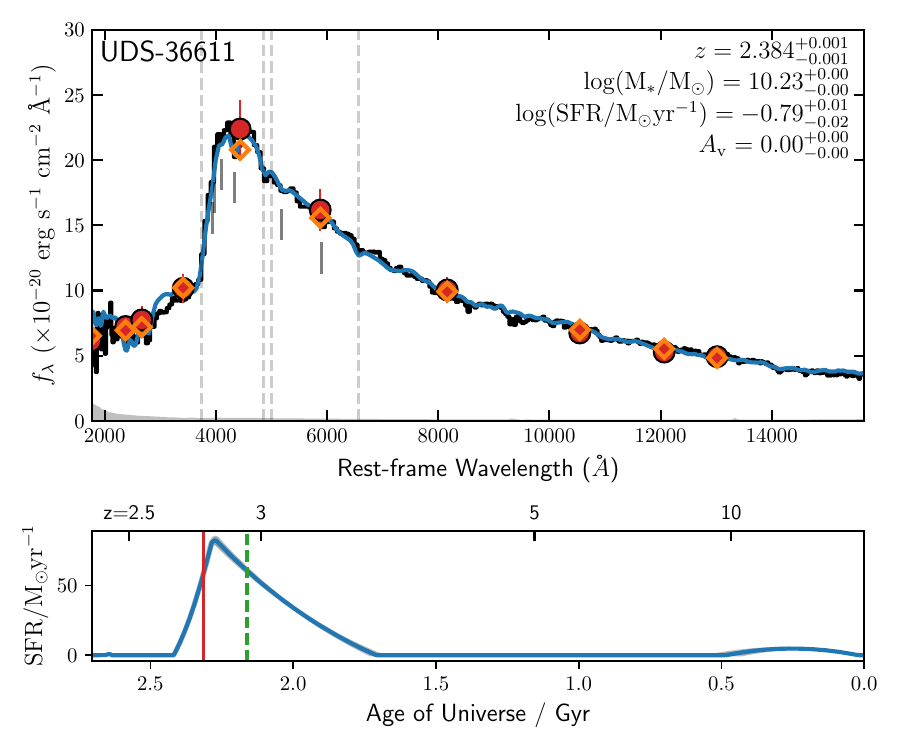} 
\includegraphics[width=0.32\textwidth]{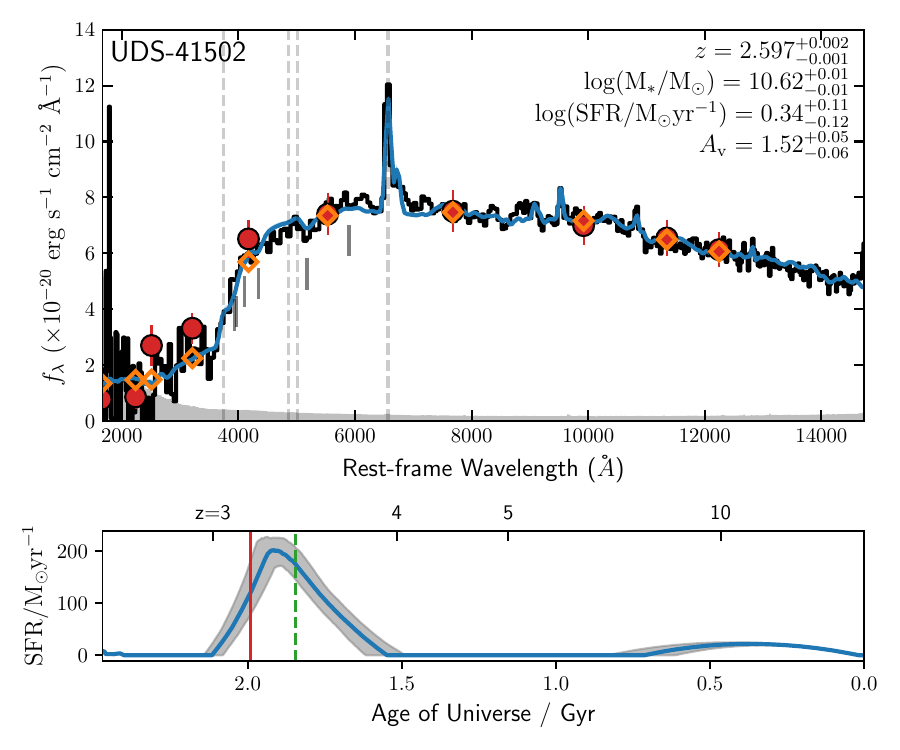} \\
\caption{Cutouts, SEDs, and star formation histories for spectroscopically-confirmed quiescent galaxies at $2 \leq z_\mathrm{spec} \leq 3$. The plotting convention follows Figure~\ref{fig:QGs_z3z4}. For UDS-6050, \Mgii\ $\lambda2798$ \AA\ line is marked with a short grey line above the spectrum. }
\label{fig:QGs_z2z3_1}
\end{figure*}

\begin{figure*}
\centering
\includegraphics[width=0.32\textwidth]{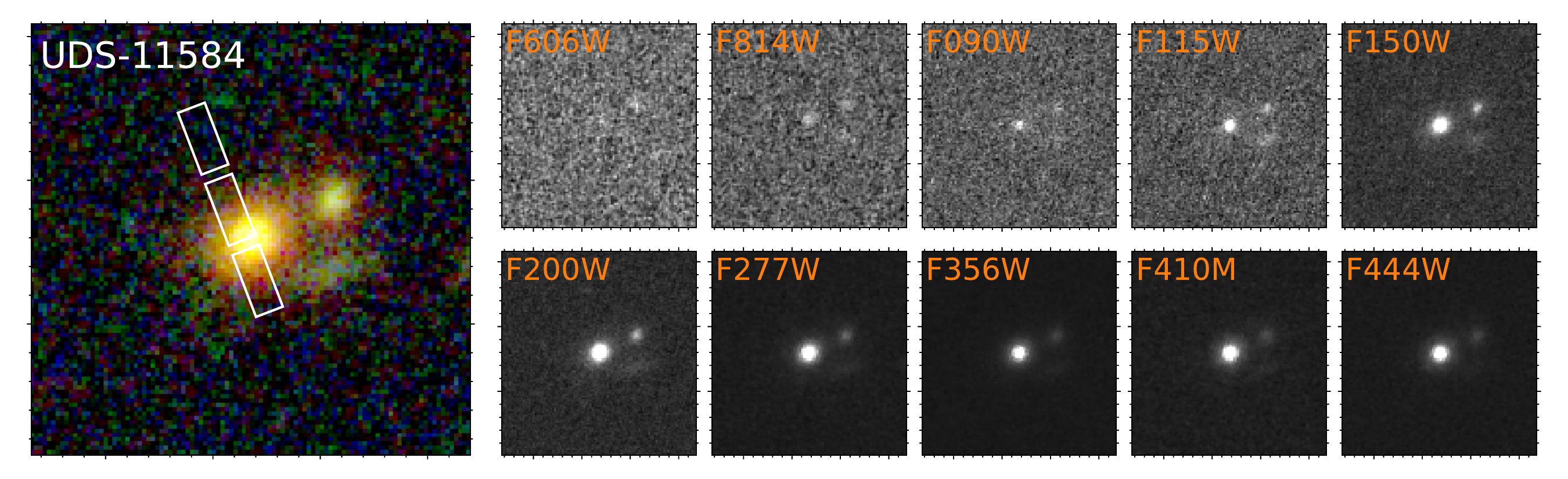} 
\includegraphics[width=0.32\textwidth]{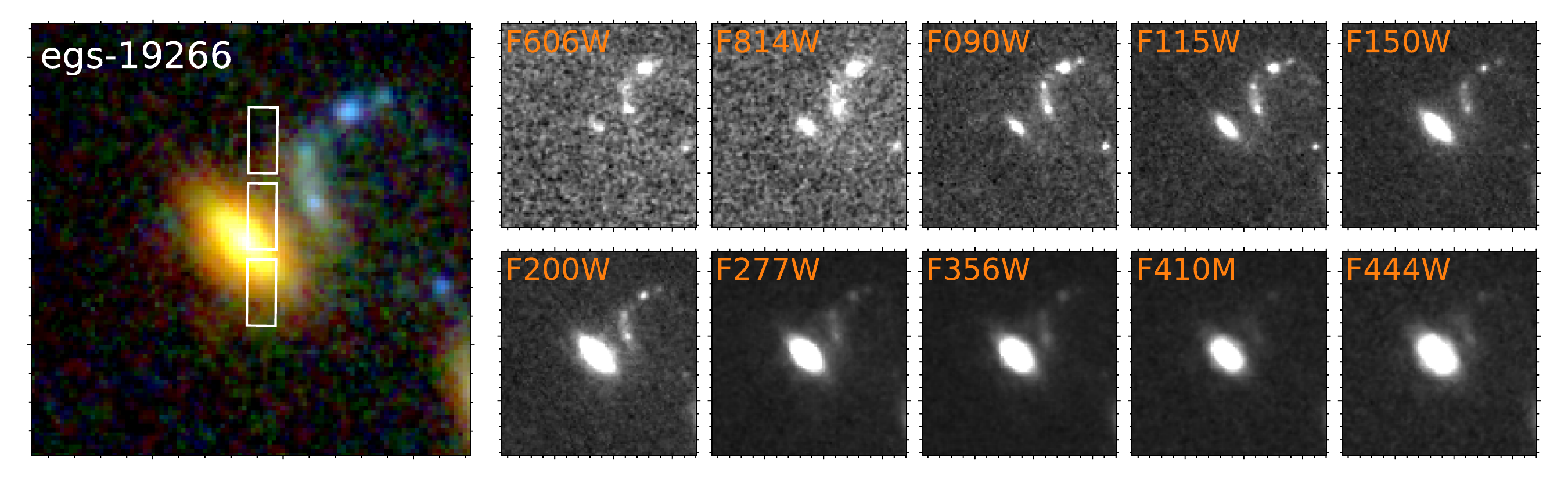}
\includegraphics[width=0.32\textwidth]{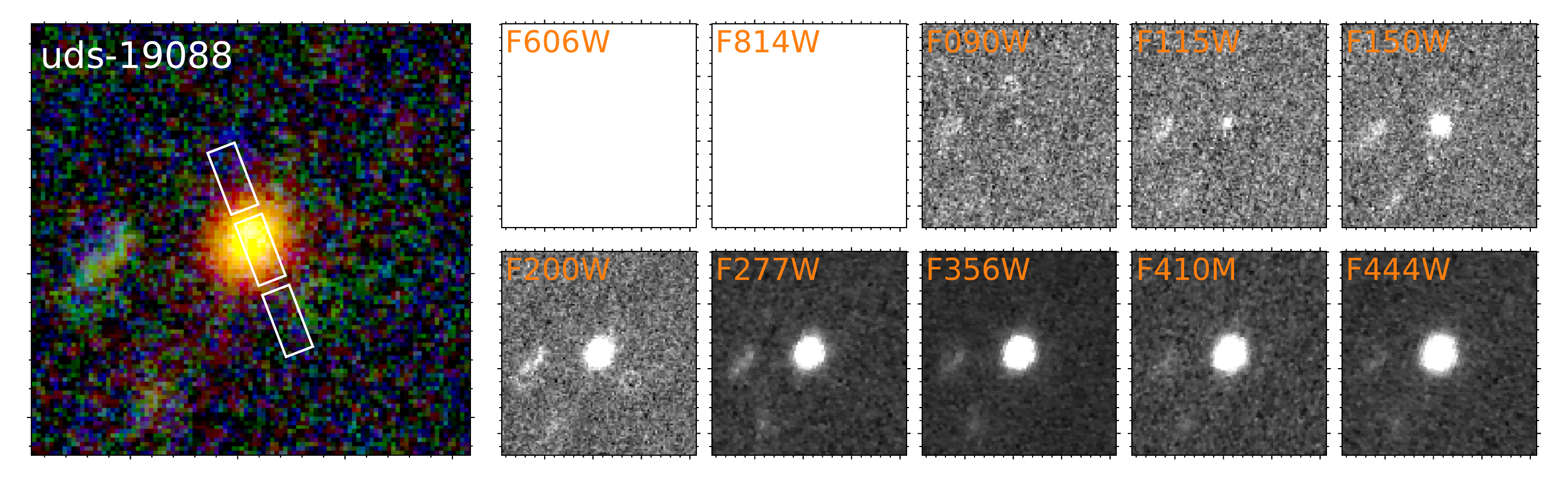}
\\
\includegraphics[width=0.32\textwidth]{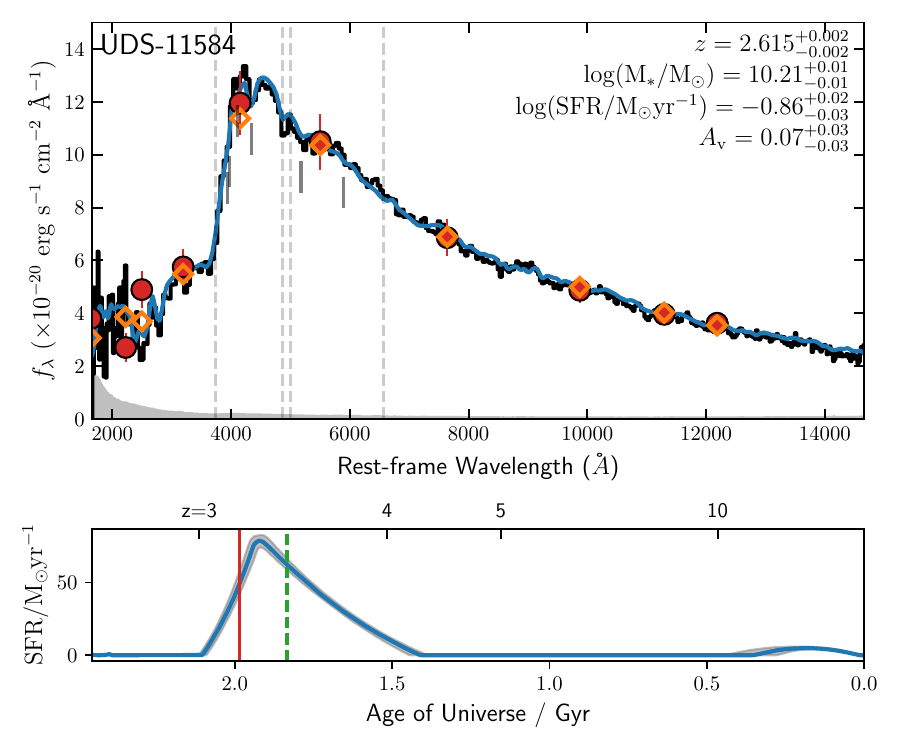} 
\includegraphics[width=0.32\textwidth]{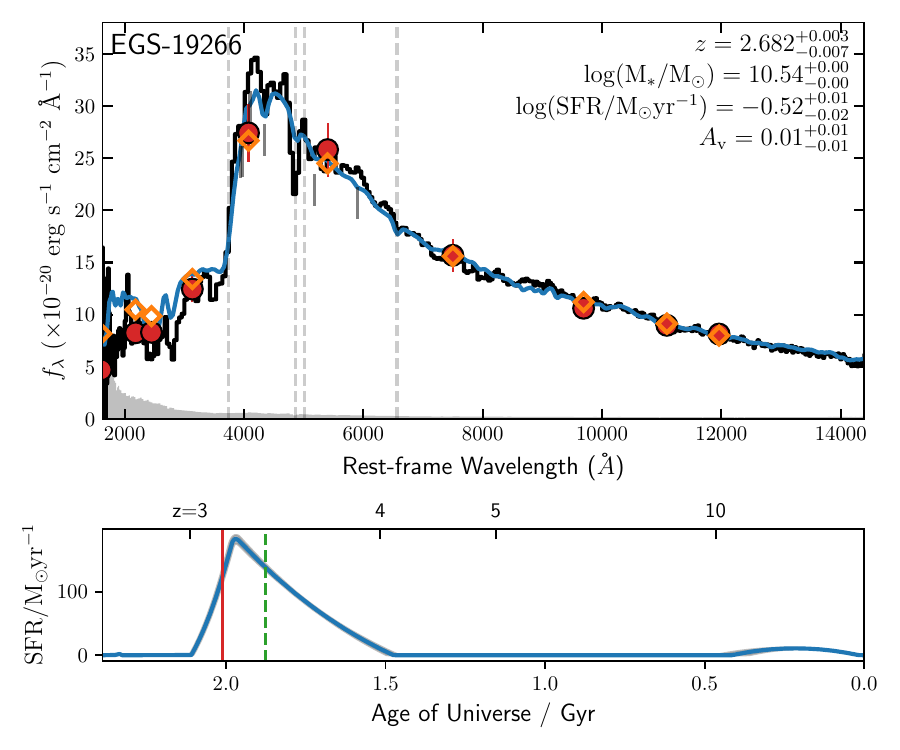}
\includegraphics[width=0.32\textwidth]{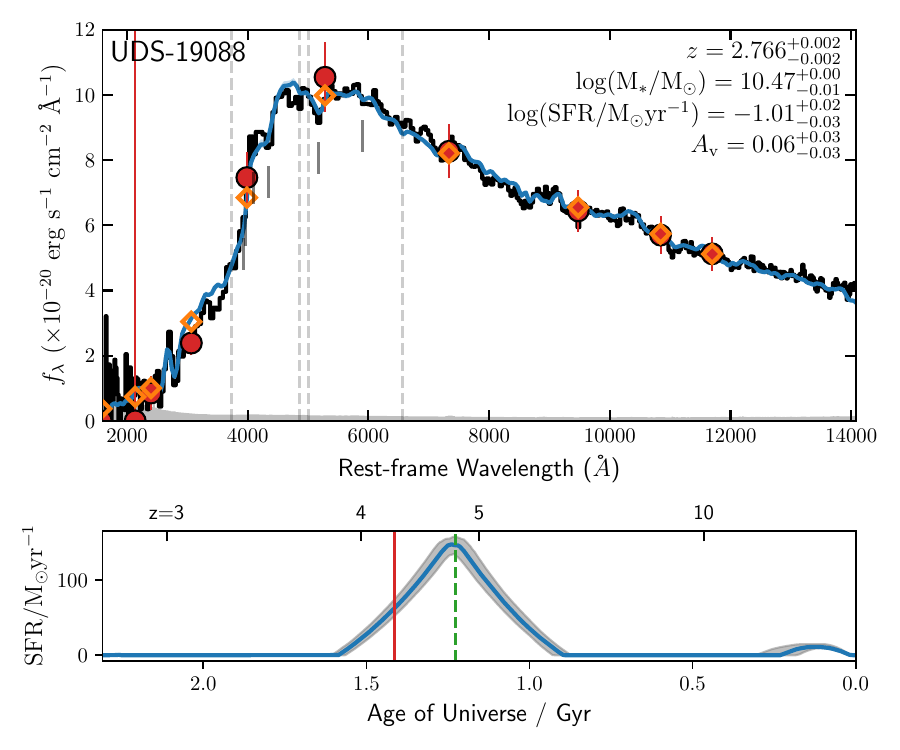} \\
\includegraphics[width=0.32\textwidth]{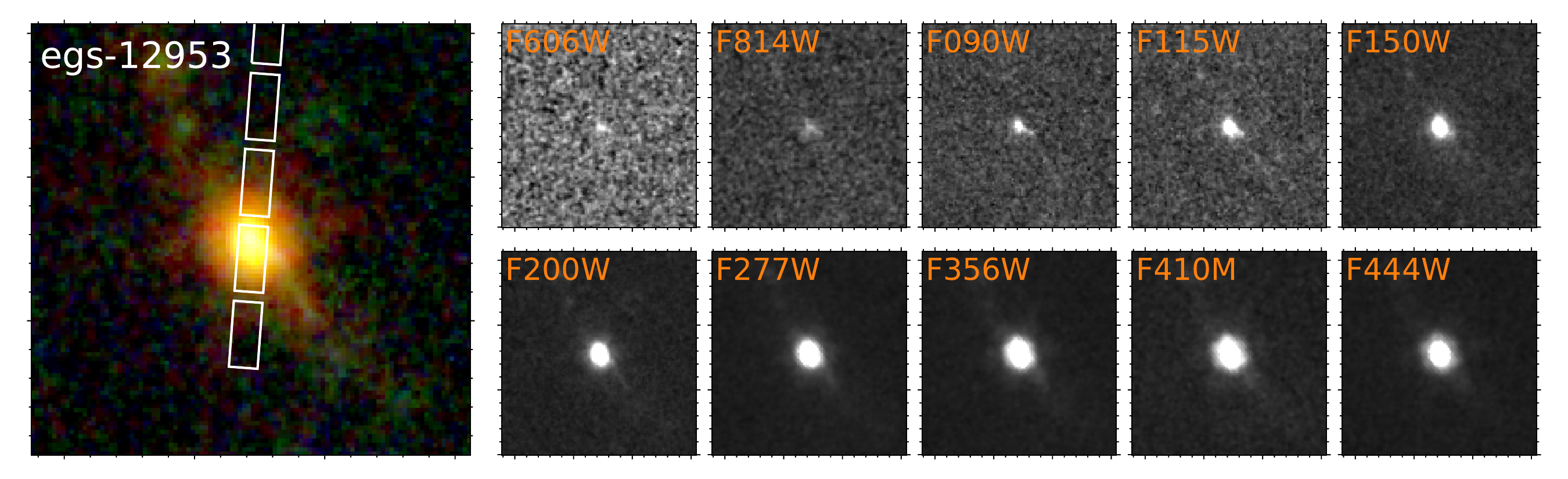} \\
\includegraphics[width=0.32\textwidth]{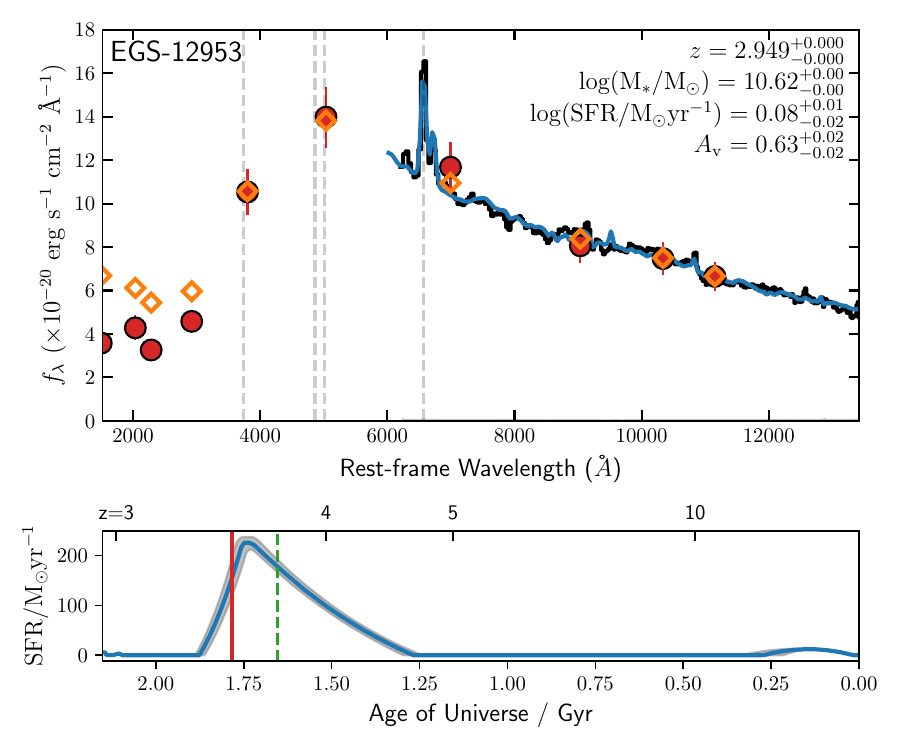} \\
\caption{Cutouts, SEDs, and star formation histories for spectroscopically-confirmed quiescent galaxies at $2 \leq z_\mathrm{spec} \leq 3$ that show evidence of AGN activity. The plotting convention follows Figure~\ref{fig:QGs_z3z4}.  }
\label{fig:QGs_z2z3_2}
\end{figure*}

Table \ref{tab:properties} shows the properties of spectroscopically confirmed quiescent galaxies.
Figure \ref{fig:QGs_z2z3_1} and Figure \ref{fig:QGs_z2z3_2} show the image, SED, and star formation history for spectroscopically confirmed quiescent galaxies at $2 < z < 3$ without and with AGN, respectively.

\begin{deluxetable*}{lccccccccc}
\tablecaption{Properties of Spectroscopically Confirmed Quiescent Galaxies. \label{tab:properties}}
\tablehead{\colhead{ID} & \colhead{R.A.} & \colhead{Dec.} & \colhead{Redshift} & \colhead{$\mathrm{log(M_* / M_\odot)}$} & \colhead{$\mathrm{log(SFR / M_\odot~yr^{-1})}$}  & \colhead{$t_{50}$} & \colhead{$t_{90}$} & \colhead{$D_n4000$} & \colhead{$D_B$} \\
\colhead{(1)} & \colhead{(2)} & \colhead{(3)} & \colhead{(4)} & \colhead{(5)} & \colhead{(6)}  & \colhead{(7)} & \colhead{(8)} & \colhead{(9)} & \colhead{(10)} }
\startdata
UDS-10310 & 34.41946 & -5.13378 & 2.18 & 10.470 $\pm$  0.005 & 0.02 $\pm$ 0.04 & 0.62 $\pm$  0.02 & 0.45 $\pm$ 0.02 & 1.14 $\pm$  0.09 & 2.27 $\pm$  0.34 \\
COSMOS-38203 & 150.14290 & 2.27850 & 2.20 & 10.819 $\pm$  0.004 & -0.42 $\pm$ 0.08 & 0.78 $\pm$  0.01 & 0.59 $\pm$  0.01 & 1.42 $\pm$  0.02 & 1.69 $\pm$  0.04 \\
EGS-20036${^*}$ & 214.92125 & 52.84573 & 2.23 & 10.541 $\pm$  0.002 & 0.01 $\pm$ 0.01 & 0.64 $\pm$  0.01 & 0.47 $\pm$  0.01 & 1.34 $\pm$  0.02 & 1.82 $\pm$  0.04 \\
UDS-6050${^*}$ & 34.43152 & -5.11436 & 2.29 & 11.008 $\pm$  0.002 & 0.77 $\pm$ 0.01 & 0.79 $\pm$ 0.01 & 0.61 $\pm$ 0.01 & 1.27 $\pm$ 0.02 & 1.57 $\pm$ 0.03 \\
EGS-12661 & 214.86648 & 52.85270 & 2.29 & 10.899 $\pm$ 0.002 & -1.60 $\pm$ 0.08 & 2.56 $\pm$ 0.01 & 0.86 $\pm$ 0.01 & 1.47 $\pm$ 0.03 & 1.81 $\pm$ 0.06 \\
COSMOS-43671 & 150.07654 & 2.24967 & 2.30 & 10.885 $\pm$ 0.003 & -0.15 $\pm$ 0.02 & 1.05 $\pm$ 0.02 & 0.86 $\pm$ 0.01 & 1.52 $\pm$ 0.04 & 1.68 $\pm$ 0.06 \\
COSMOS-37093 & 150.09387 & 2.28471 & 2.37 & 10.345 $\pm$ 0.005 & -0.81 $\pm$ 0.05 & 0.69 $\pm$ 0.01 & 0.53 $\pm$ 0.01 & 1.42 $\pm$ 0.04 & 2.18 $\pm$ 0.11 \\
UDS-36611 & 34.30934 & -5.25473 & 2.38 & 10.234 $\pm$ 0.003 & -0.79 $\pm$ 0.02 & 0.55 $\pm$ 0.00 & 0.39 $\pm$ 0.01 & 1.33 $\pm$ 0.02 & 1.98 $\pm$ 0.04 \\
UDS-41502 & 34.34351 & -5.27612 & 2.60 & 10.621 $\pm$ 0.007 & 0.34 $\pm$ 0.12 & 0.62 $\pm$ 0.04 & 0.48 $\pm$ 0.03 & 1.17 $\pm$ 0.08 & 2.50 $\pm$ 0.34 \\
UDS-11584 & 34.40638 & -5.13900 & 2.61 & 10.206 $\pm$ 0.006 & -0.86 $\pm$ 0.03 & 0.62 $\pm$ 0.01 & 0.47 $\pm$ 0.01 & 1.35 $\pm$ 0.03 & 1.94 $\pm$ 0.07 \\
EGS-19266 & 214.85385 & 52.80104 & 2.68 & 10.535 $\pm$ 0.003 & -0.52 $\pm$ 0.02 & 0.51 $\pm$ 0.01 & 0.38 $\pm$ 0.01 & 1.26 $\pm$ 0.03 & 2.51 $\pm$ 0.10 \\
UDS-19088 & 34.52518 & -5.17077 & 2.77 & 10.469 $\pm$ 0.005 & -1.01 $\pm$ 0.03 & 1.08 $\pm$ 0.02 & 0.90 $\pm$ 0.02 & 1.55 $\pm$ 0.05 & 1.84 $\pm$ 0.12 \\
EGS-12953 & 214.82774 & 52.82377 & 2.95 & 10.617 $\pm$ 0.004 & 0.08 $\pm$ 0.02 & 0.50 $\pm$ 0.02 & 0.37 $\pm$ 0.02 &  &  \\
COSMOS-24202 & 150.20902 & 2.34914 & 3.09 & 10.677 $\pm$ 0.002 & -0.15 $\pm$ 0.01 & 0.48 $\pm$ 0.01 & 0.36 $\pm$ 0.01 & 1.29 $\pm$ 0.01 & 2.20 $\pm$ 0.04 \\
UDS-16184 & 34.51366 & -5.15783 & 3.12 & 10.656 $\pm$ 0.010 & -0.59 $\pm$ 0.03 & 0.73 $\pm$ 0.02 & 0.59 $\pm$ 0.02 & 1.42 $\pm$ 0.04 & 1.79 $\pm$ 0.07 \\
COSMOS-26504 & 150.15180 & 2.33796 & 3.42 & 9.952 $\pm$ 0.008 & -1.24 $\pm$ 0.14 & 0.56 $\pm$ 0.03 & 0.43 $\pm$ 0.02 & 1.31 $\pm$ 0.06 & 1.74 $\pm$ 0.11 \\
EGS-7806 & 214.87909 & 52.88806 & 3.45 & 10.327 $\pm$ 0.005 & -0.33 $\pm$ 0.02 & 0.48 $\pm$ 0.01 & 0.37 $\pm$ 0.01 & 1.28 $\pm$ 0.04 & 2.06 $\pm$ 0.09 \\
UDS-35916 & 34.51197 & -5.25155 & 3.70 & 10.311 $\pm$ 0.007 & -1.05 $\pm$ 0.08 & 0.89 $\pm$ 0.06 & 0.70 $\pm$ 0.05 &  &  \\
UDS-38278 & 34.29044 & -5.26208 & 3.71 & 10.520 $\pm$ 0.004 & -0.65 $\pm$ 0.02 & 0.50 $\pm$ 0.01 & 0.40 $\pm$ 0.01 & 1.28 $\pm$ 0.02 & 2.37 $\pm$ 0.07
\enddata
\tablecomments{(1)-(3) ID and coordinates of galaxies. (4) Redshift from CAPERS spectra. (5) Stellar mass. (6) Star formation rate averaged over 100 Myrs. (7) and (8) the 
time since the galaxy formed 50 and 90 per cent of its total formed stellar mass. (9) Balmer break index strength. (10) 4000 \AA\ break strength. The reported uncertainties are half the 16th–84th percentile width of the corresponding distributions. QGs hosting AGN are marked with ${^*}$. }
\end{deluxetable*}

\begin{figure*}
\centering
\includegraphics[width=0.32\textwidth]{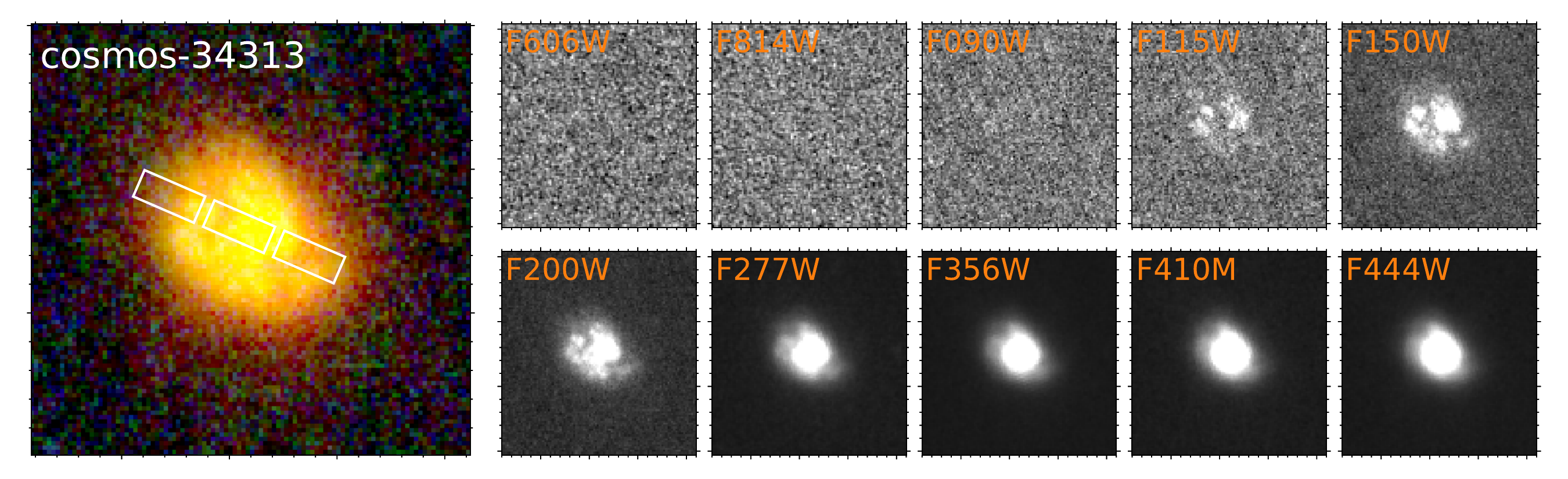} 
\includegraphics[width=0.32\textwidth]{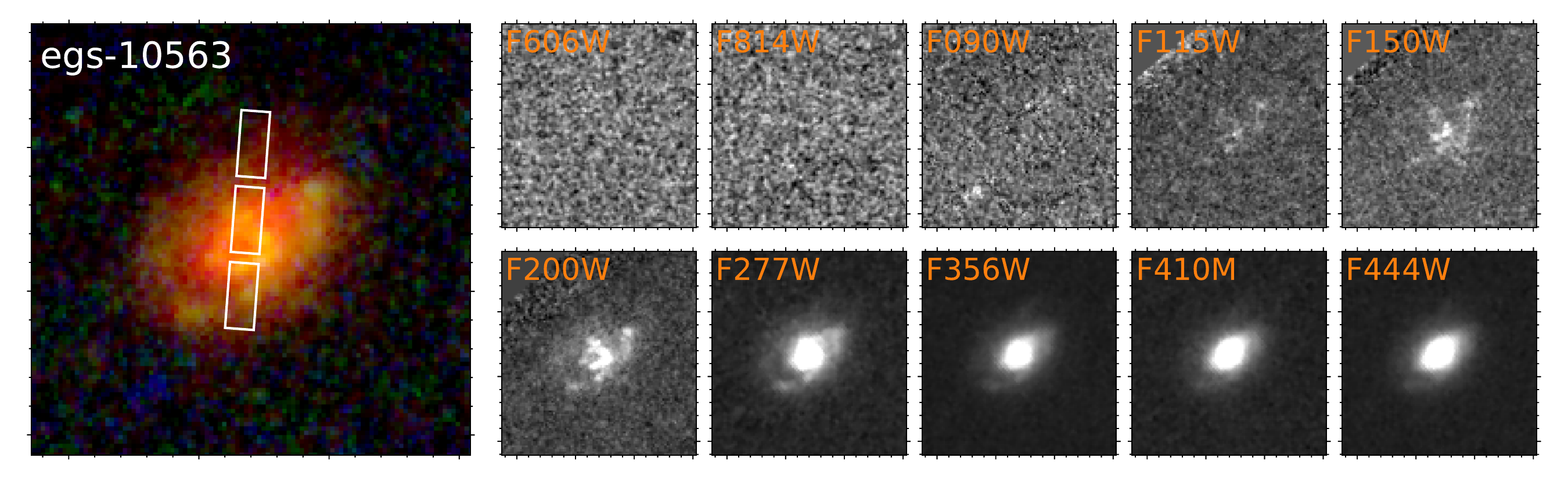}\\
\includegraphics[width=0.32\textwidth]{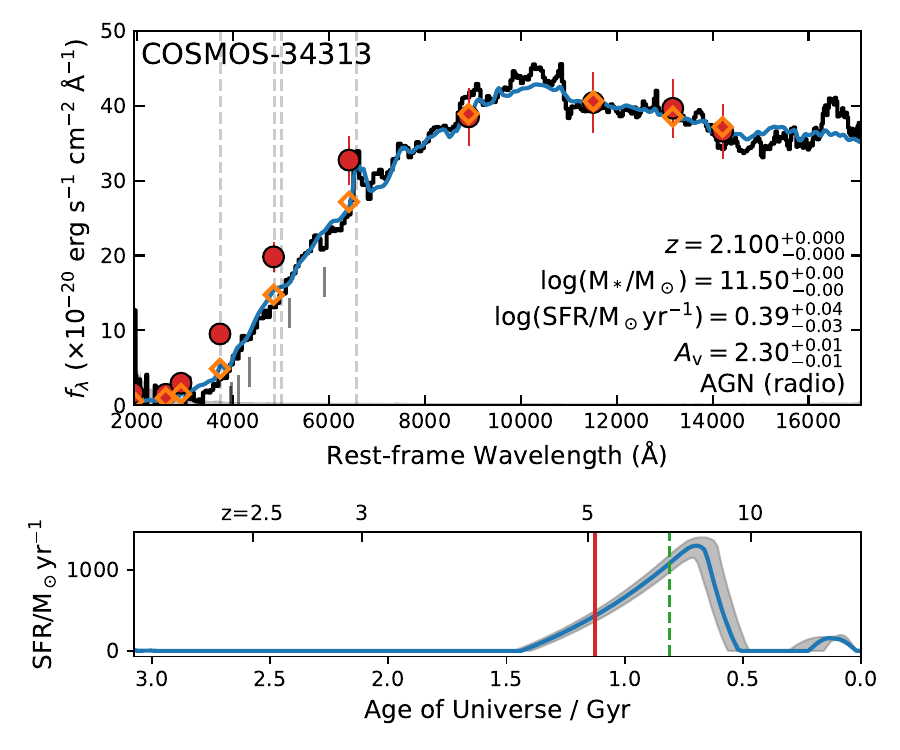} 
\includegraphics[width=0.32\textwidth]{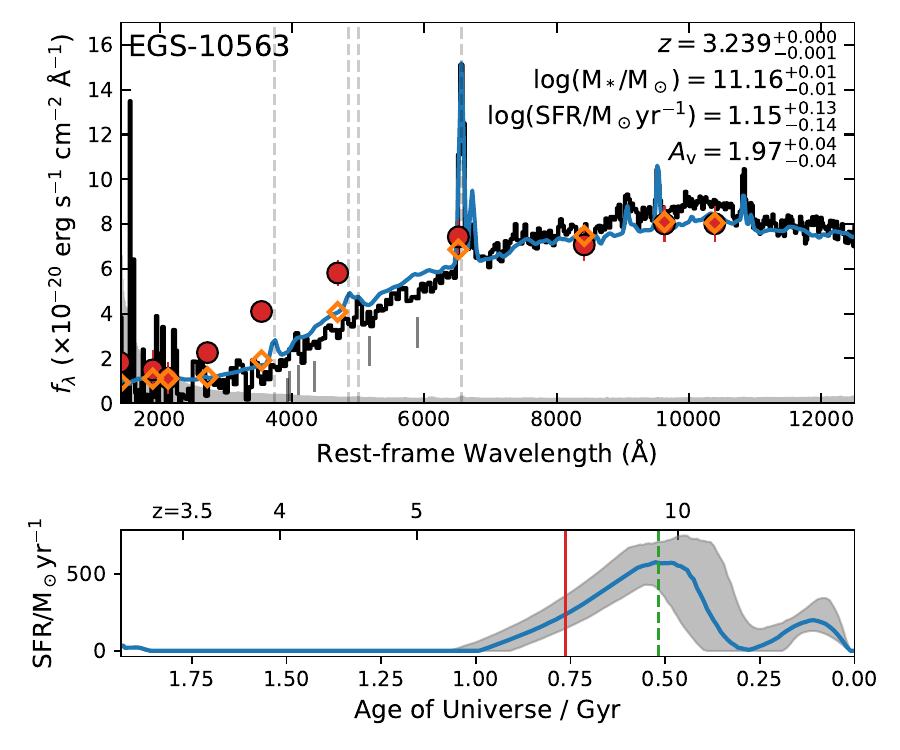} \\
\includegraphics[width=0.32\textwidth]{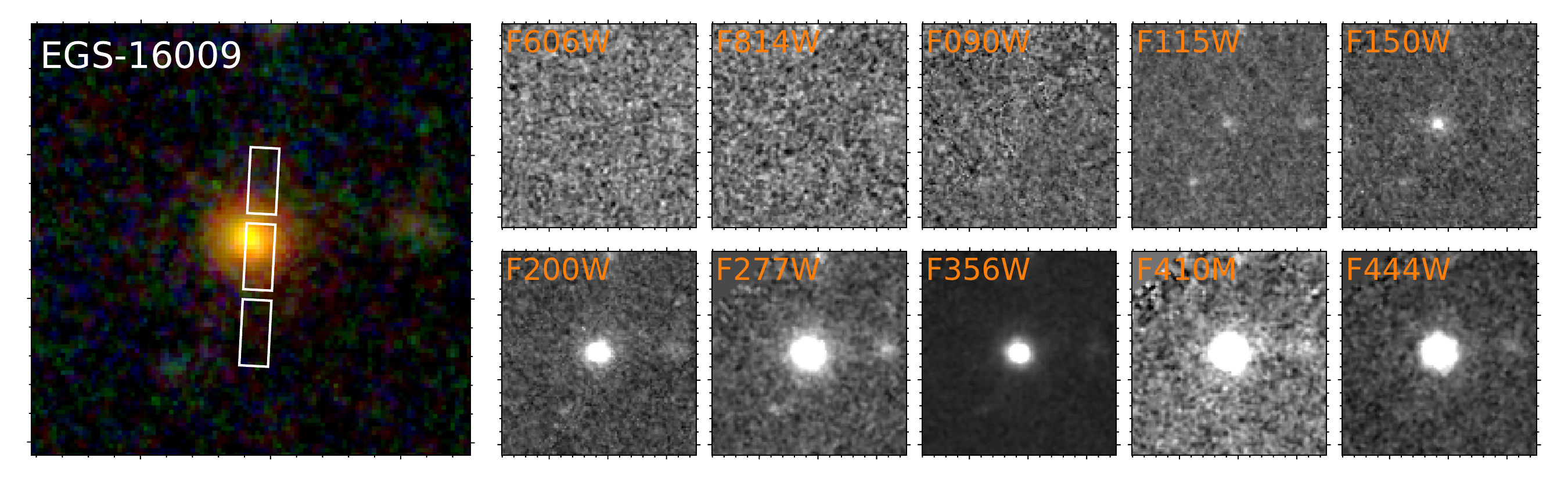} 
\includegraphics[width=0.32\textwidth]{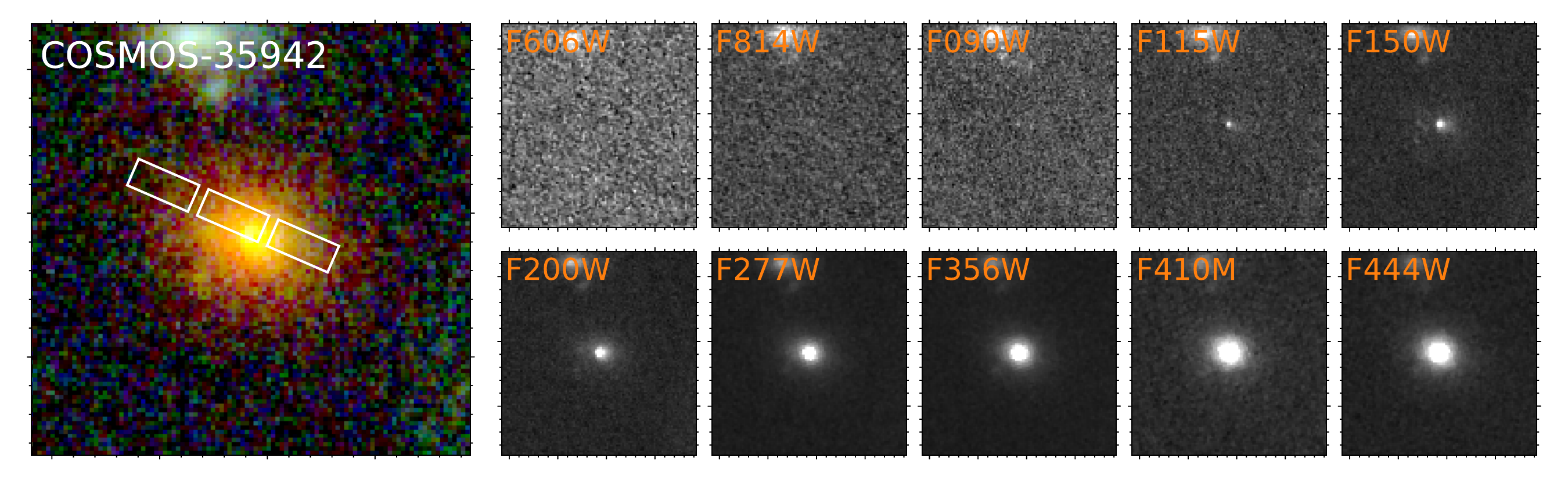} 
\includegraphics[width=0.32\textwidth]{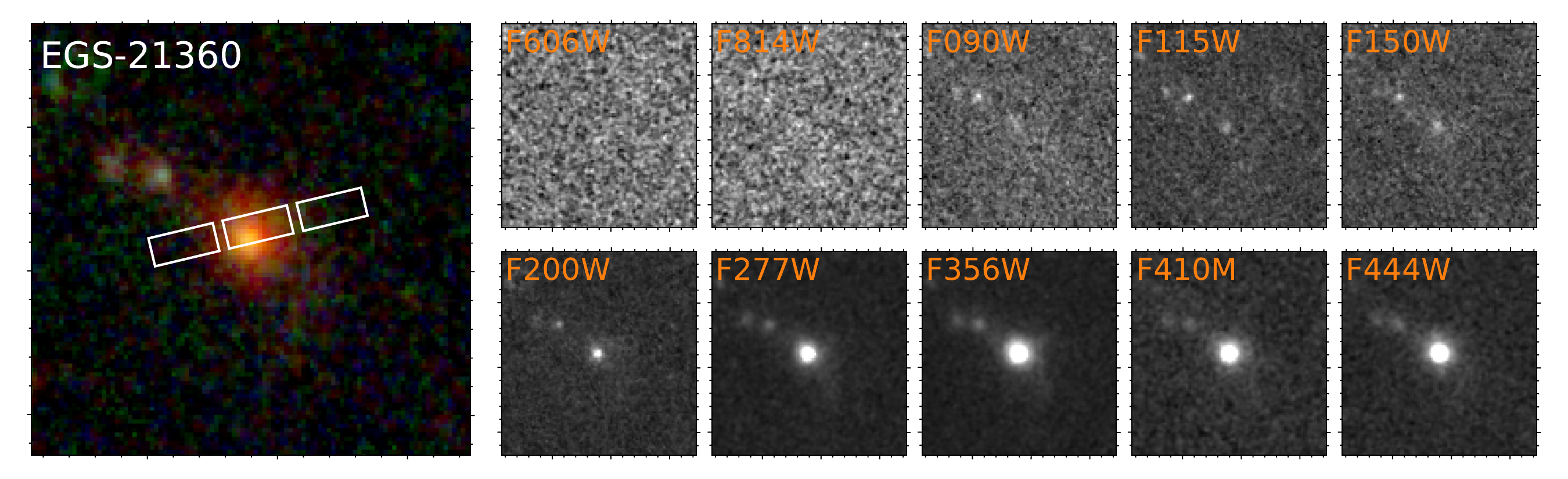} \\
\includegraphics[width=0.32\textwidth]{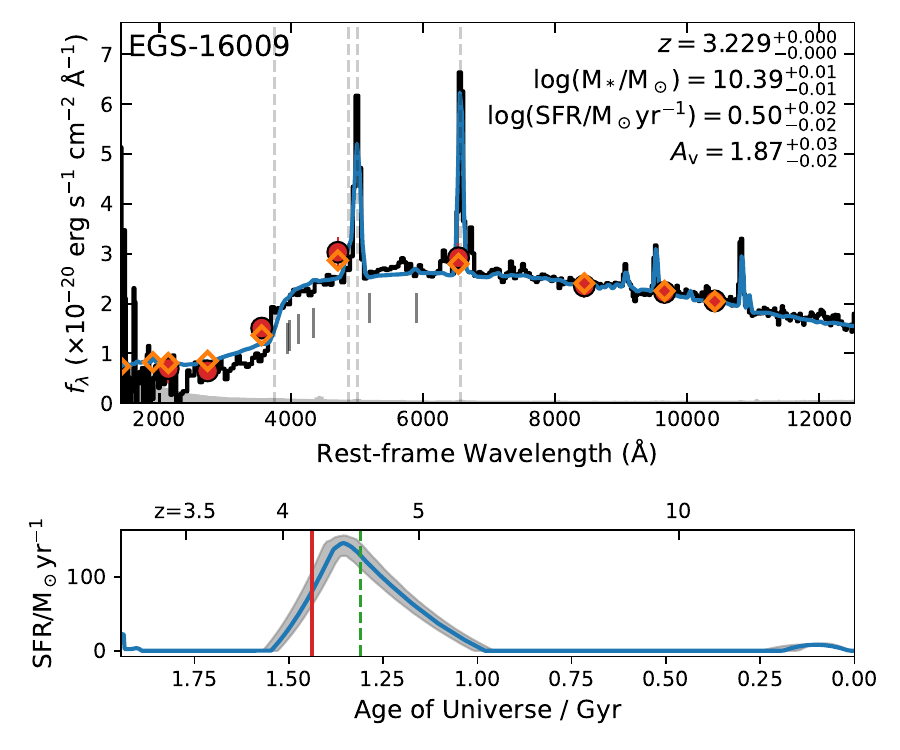} 
\includegraphics[width=0.32\textwidth]{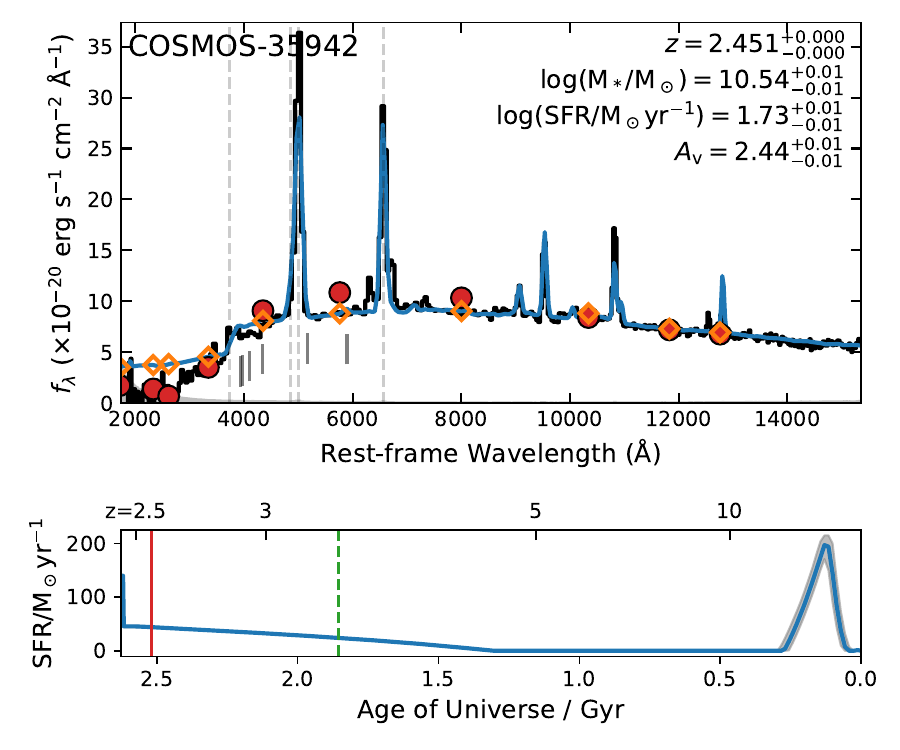}  
\includegraphics[width=0.32\textwidth]{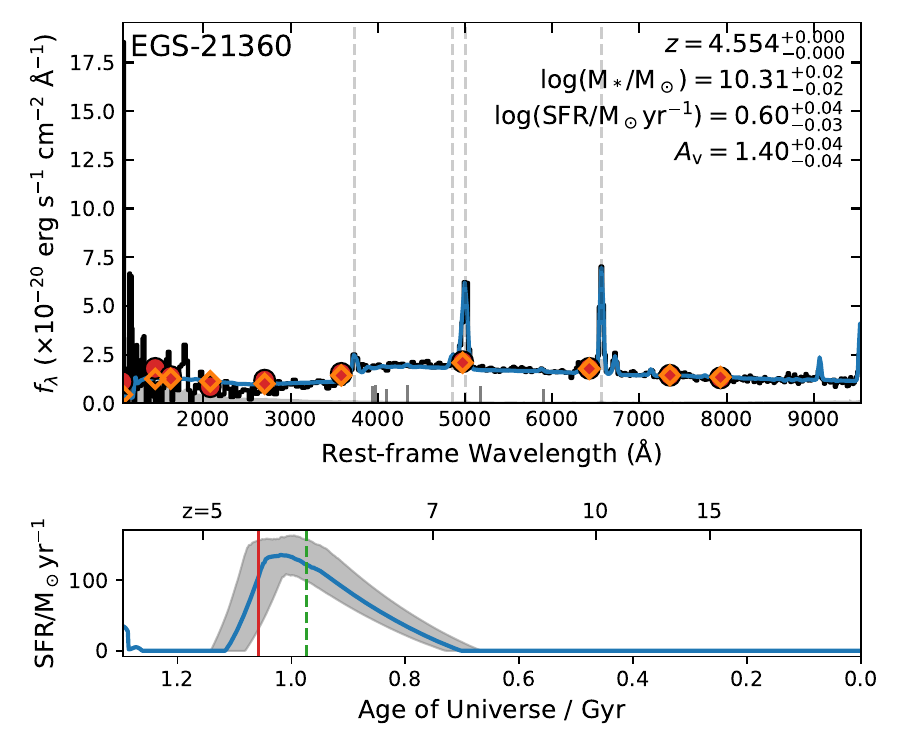}\\
\includegraphics[width=0.32\textwidth]{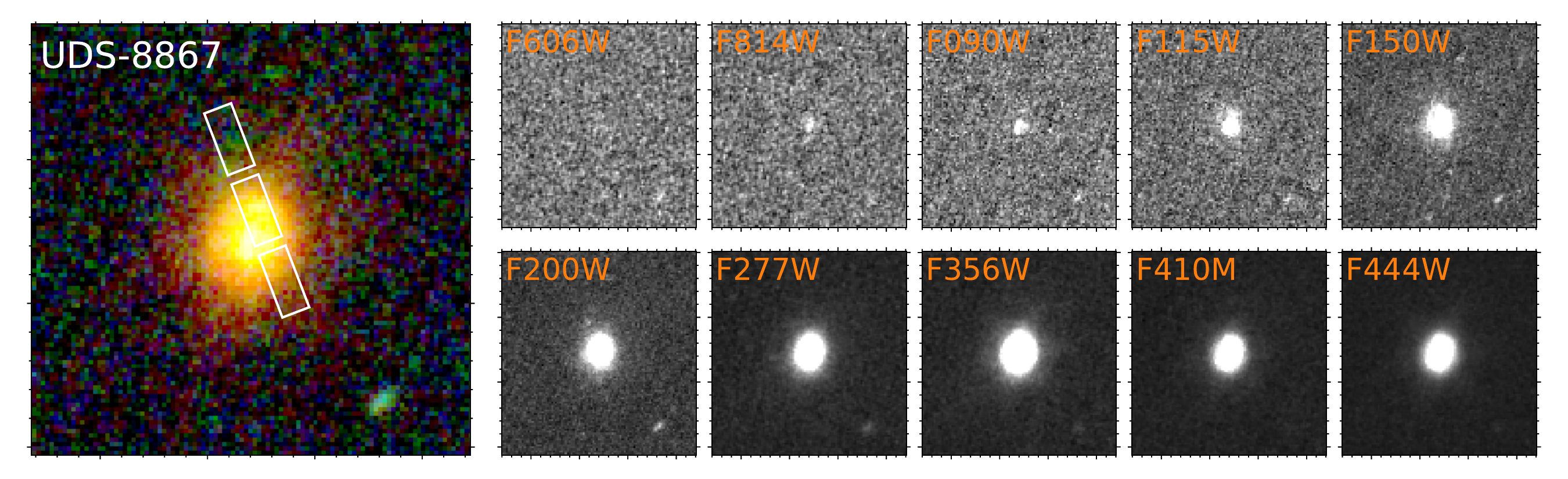} 
\includegraphics[width=0.32\textwidth]{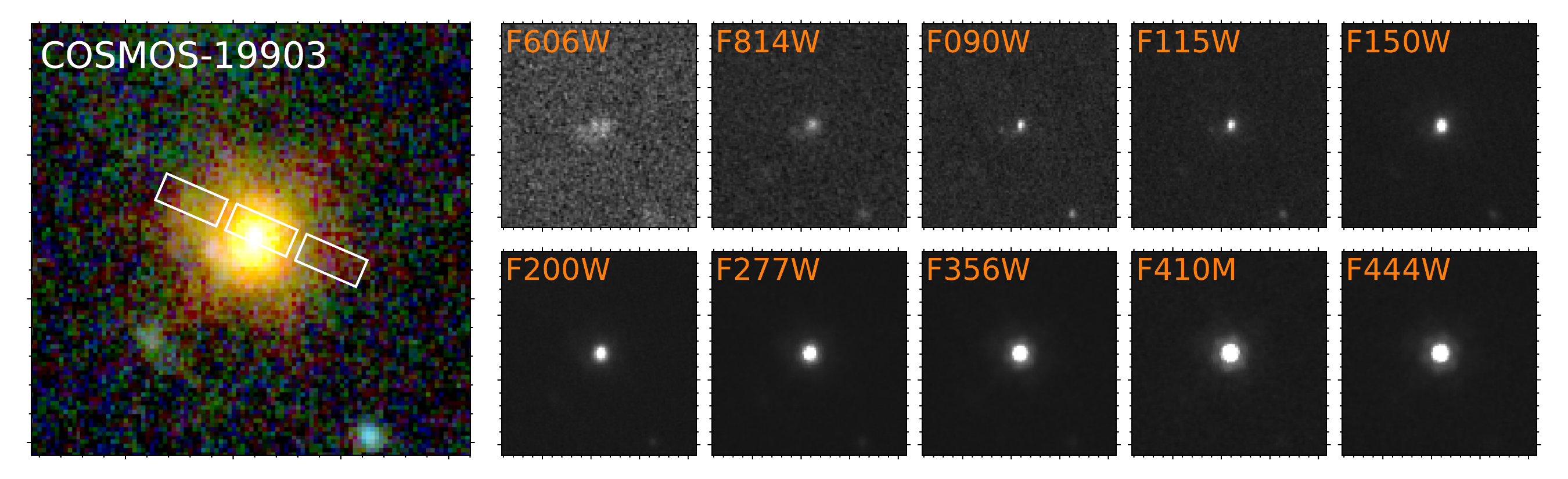} 
\includegraphics[width=0.32\textwidth]{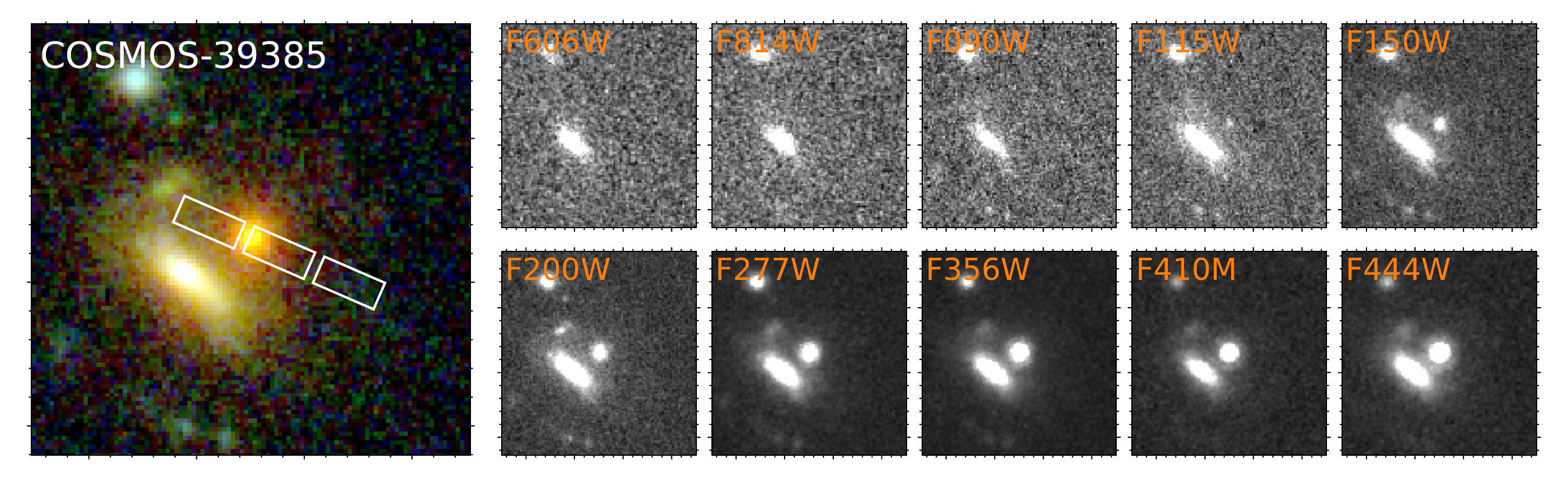}\\
\includegraphics[width=0.32\textwidth]{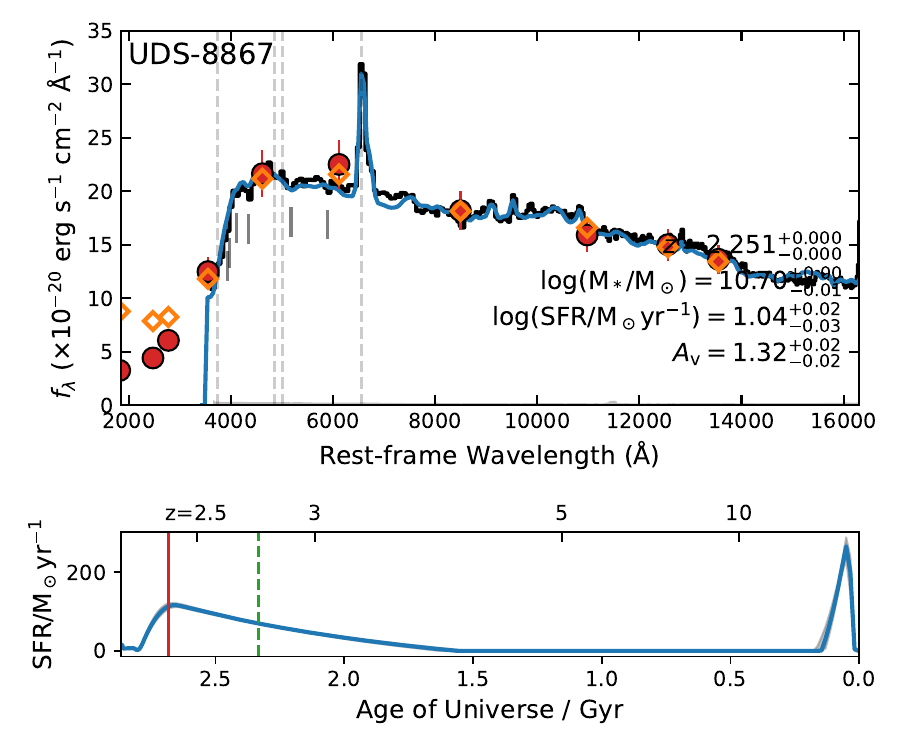} 
\includegraphics[width=0.32\textwidth]{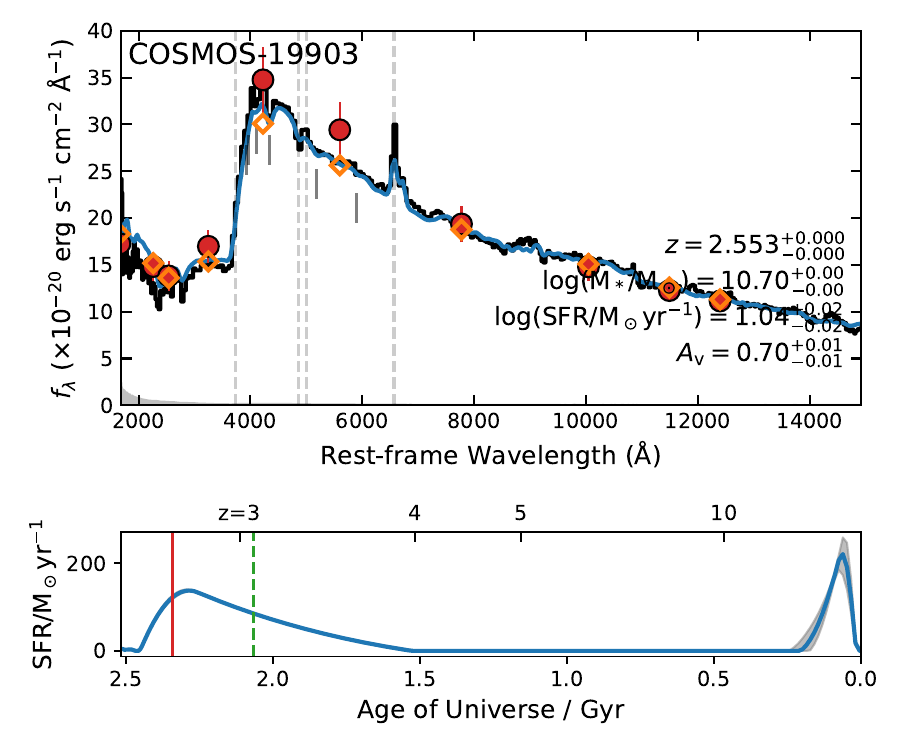} 
\includegraphics[width=0.32\textwidth]{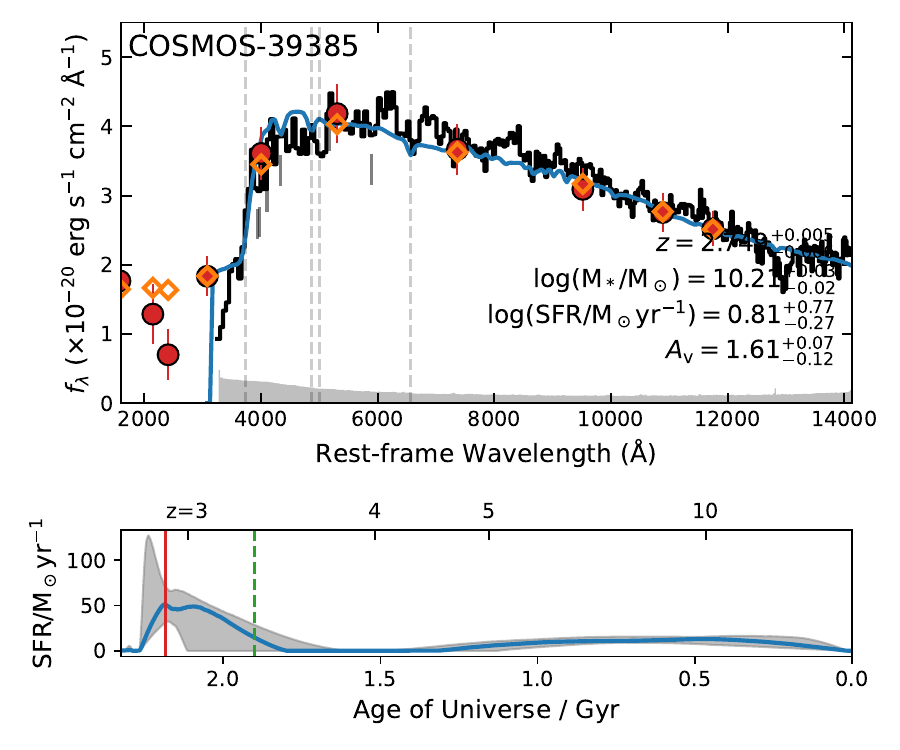} \\ 
\caption{Cutouts, SEDs, and star formation histories for galaxies selected by more than one quiescent selection criterion, but whose properties are more consistent with ongoing star formation and/or AGN activity. 
COSMOS-34313 and EGS-10563 selected by sSFR-$t(z)$ and sSFR--M$_\ast$; EGS-16009 is selected by the sSFR--M$_\ast$, sSFR$_\mathrm{p}$-$t(z)$, traditional $UVJ$, $ugi_s$; COSMOS-35942 is selected by the sSFR$_\mathrm{p}$-$t(z)$, traditional $UVJ$, $ugi_s$; EGS-21360 are selected by both sSFR--M$_\ast$, sSFR$_\mathrm{p}$-$t(z)$ and $ugi_s$. UDS-8867, COSMOS-19903 and COSMOS-39385 are selected sSFR$_\mathrm{p}$-$t(z)$ and $ugi_s$. These three galaxies show prominent Balmer/4000\AA\ breaks, and stellar absorption features, indicative of evolved stellar populations. }
\label{fig:sed_contaminators}
\end{figure*}

\section{Sensitivity of Derived Galaxy Properties and Quiescent Classification to the Assumed SFH} \label{app:diff_SFHs}

{Galaxy properties inferred through SED fitting can depend on the adopted SFH parameterization. We therefore test the sensitivity of the derived galaxy properties and quiescent classification by repeating the \bagpipes\ fits with two alternative parametric SFHs: a delayed-$\tau$ model and a double power-law model. For the double power-law model, the rising and falling slopes, $\beta$ and $\alpha$, are each allowed to vary from 0.1 to 1000 with log-uniform priors, while the turnover time, $\tau$, is assigned a uniform prior over $0.1$–$3.5$ Gyr. For the delayed-$\tau$ model, both the time since the onset of star formation and the $e$-folding timescale are assigned uniform priors over $0.1$–$3.5$ Gyr. In these tests, we retain the same spectra, photometry, and all other SED-fitting assumptions used in the fiducial analysis. }

\begin{figure*}
    \includegraphics[width=\textwidth]{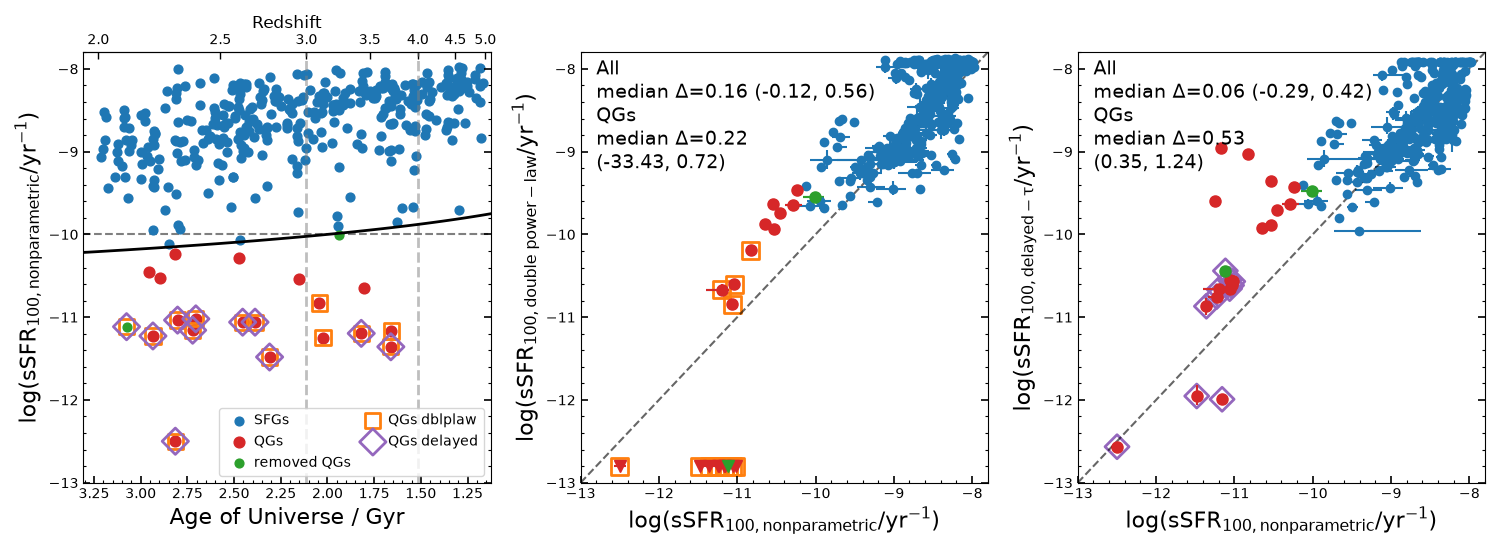}
    \caption{\textit{Left:} $\log(\mathrm{sSFR}/\mathrm{yr}^{-1})$ as a function of the age of the Universe, $t(z)$, for the CAPERS spectroscopic sample at $2\leq z\leq5$ and $\log(M_\ast/M_\odot)>9.5$, using the fiducial non-parametric SFH. This panel is the same as the right panel of Figure~\ref{fig:QGs_selection}. Galaxies satisfying $\mathrm{sSFR}<0.2/t(z)$ under the double-power-law and delayed-$\tau$ models are marked with open orange squares and open purple diamonds, respectively. \textit{Middle:} Comparison of the sSFRs inferred using the double-power-law and fiducial non-parametric SFHs. Ten galaxies have effectively zero SFRs under the double-power-law model, corresponding to $\log(\mathrm{sSFR}/\mathrm{yr}^{-1})<-13$, shown as downward-pointing triangles at the lower plotting boundary. \textit{Right:} Comparison of the sSFRs inferred using the delayed-$\tau$ and fiducial non-parametric SFHs. In the middle and right panels, the median offsets and their 16th–84th percentile ranges are reported separately for the full sample and the fiducial QGs.}
    \label{fig:ssfr_varingSFH}
\end{figure*}

{The stellar masses and SFRs estimated using the three SFH models are broadly consistent. Relative to the fiducial non-parametric SFH, the double power-law and delayed-$\tau$ models yield stellar masses that are lower by median values of $0.09$ and $0.04$ dex, respectively, and SFRs that are higher by $0.05$ and $0.02$ dex, respectively. Consequently, their inferred sSFRs are higher than the fiducial values by median offsets of $0.16$ and $0.06$ dex. The sSFR comparisons for the double-power-law and delayed-$\tau$ models are shown in the middle and right panels of Figure~\ref{fig:ssfr_varingSFH}, respectively. Ten galaxies have SFRs that are close to zero using the double power-law, corresponding to $\log(\mathrm{sSFR}/\mathrm{yr}^{-1})<-13$. Because these values fall below the plotted range, they are shown as downward-pointing triangles at the lower boundary of the panels.}

{The effect of the assumed SFH is more pronounced among the 19 fiducial QGs. For these galaxies, the double power-law and delayed-$\tau$ models yield sSFRs that are higher than the fiducial values by median offsets of $0.22$ and $0.53$ dex, respectively. Consequently, 13 of the 19 fiducial QGs remain classified as quiescent under the double power-law model, while 10 remain quiescent under the delayed-$\tau$ model. Neither alternative model identifies any additional QGs outside the fiducial QG sample. }

{The classification of individual galaxies is more sensitive to the adopted SFH parameterization. As shown in the left panel of Figure~\ref{fig:ssfr_varingSFH}, six galaxies with $\log(\mathrm{sSFR}_{100}/\mathrm{yr}^{-1})\sim-10$ to $-11$ in the fiducial fits have higher inferred sSFRs under one or both parametric models and consequently no longer satisfy the quiescence criterion. Visual inspection of the fitted spectra indicates that the non-parametric model generally provides a better description of the continuum and absorption features for the fiducial QGs whose classifications vary among the models. Moreover, the inferred SFHs presented in Figures~\ref{fig:QGs_z3z4}, \ref{fig:QGs_z2z3_1}, and \ref{fig:QGs_z2z3_2} reveal multiple episodes of star formation in some galaxies, such as EGS-12661 and UDS-16184. By contrast, the delayed-$\tau$ and double-power-law models each describe a single, smoothly varying episode of star formation and are therefore less flexible in representing such complex SFHs. This limitation can lead to either overestimation or underestimation of the recent SFR. }

\begin{figure*}
    \centering
    \includegraphics[width=0.8\textwidth]{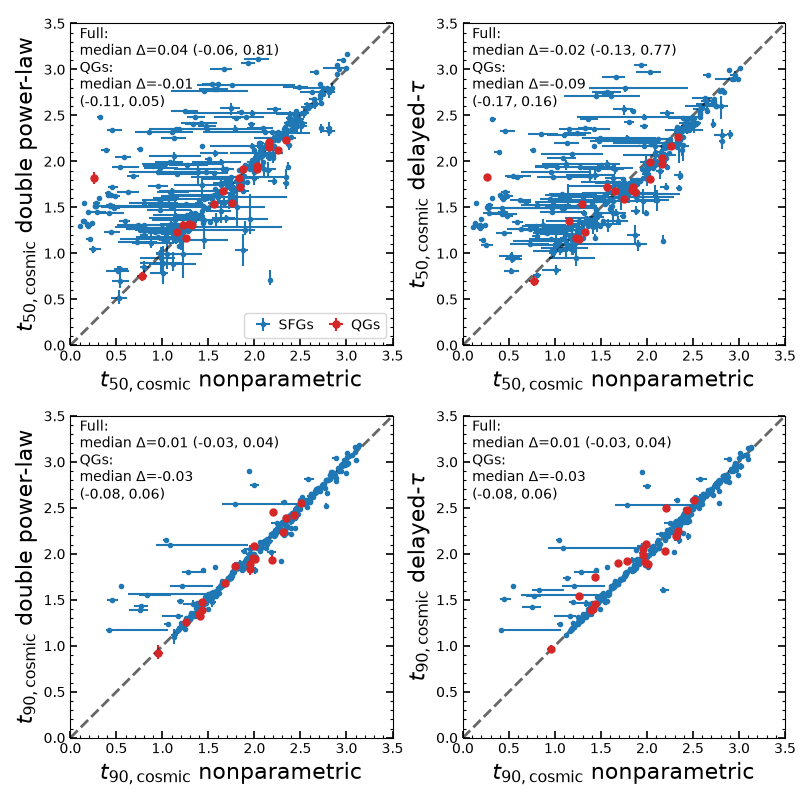}
    \caption{Comparison of the $t_{50, \mathrm{cosmic}}$ (tops) and $t_{50, \mathrm{cosmic}}$ (bottoms) obtained using the double-power-law and delayed-$\tau$ and fiducial non-parametric SFHs. The median offsets and their 16th–84th percentile ranges are reported separately for the full sample and the fiducial QGs.}
    \label{fig:time_varingSFH}
\end{figure*}

{In Figure~\ref{fig:time_varingSFH}, we also compare $t_{50,\mathrm{cosmic}}$ (top panels) and $t_{90,\mathrm{cosmic}}$ (bottom panels) across the different SFH models. Relative to the fiducial non-parametric model, the median $t_{50,\mathrm{cosmic}}$ offsets for the double-power-law and delayed-$\tau$ models are small, $0.04$ and $-0.02$ Gyr, respectively. However, the distributions contain a tail toward later $t_{50,\mathrm{cosmic}}$ values under the parametric models, dominated by SFGs. The corresponding median offsets in $t_{90,\mathrm{cosmic}}$ are smaller, $0.01$ Gyr for both models. For the fiducial QGs, the median offsets are $-0.01$ and $-0.09$ Gyr in $t_{50,\mathrm{cosmic}}$, and $-0.03$ Gyr for both models in $t_{90,\mathrm{cosmic}}$, respectively. 
These small offsets indicate that adopting different SFH models does not affect our conclusions regarding $t_{50,\mathrm{cosmic}}$ and $t_{90,\mathrm{cosmic}}$ for the QG sample. }

\section{sSFR--M$_{*}$ relations} \label{app:ssfr_m}

\begin{figure*}
    \centering
    \includegraphics[width=\textwidth]{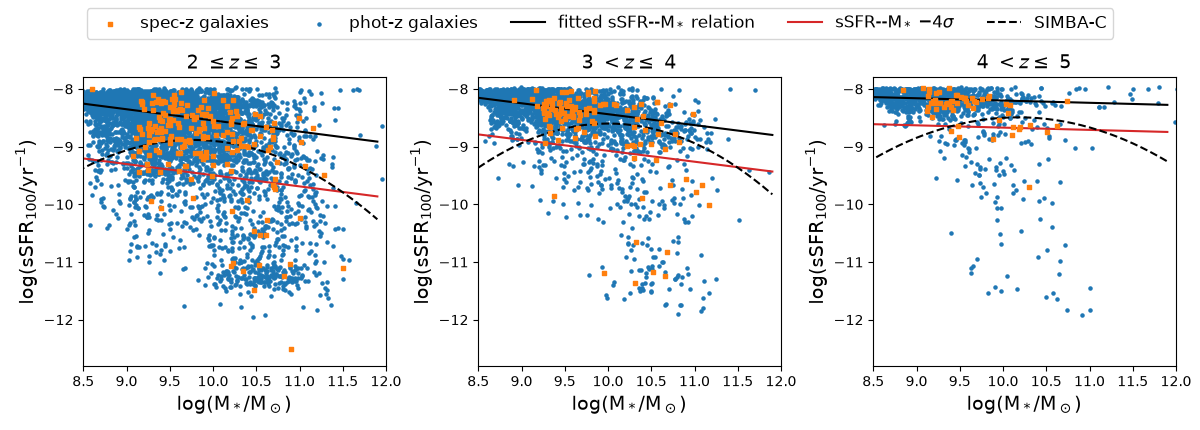}
    \caption{The sSFR--M$_{*}$ relations for photometric (phot-z, blue) and spectroscopic galaxies (spec-z, orange) with $\mathrm{M_*> 10^{8.5} M_\odot}$ at $2\leq z \leq 3$ (\textit{left}), $3 < z \leq 4$(\textit{middle}), and $4 < z \leq 5$ (\textit{right}). 
    The fitted sSFR--M$_{*}$ relations are shown in black lines, with the red lines indicating the 4$sigma$ below the relation. For comparison, the sSFR--M$_{*}$ relations from SIMBA-C simulation \citet{Szpila2025} are shown in dotted black lines. } 
    \label{fig:sSFR_M_fitting}
\end{figure*}

We derive a linear sSFR--M$_{*}$ relation for all photometric and spectroscopic galaxies in each of the three redshift ranges: $2\leq z \leq 3$, $3 < z \leq 4$, and $4 < z \leq 5$ with $\mathrm{M_*> 10^{8.5} M_\odot}$. In detail, we iteratively remove low sSFR outliers with $>3\sigma$ below the best fit to fit the relation with only SFGs, where $\sigma$ is the residual median standard deviation, and re-fit the relation until the sum of squared residuals does not decrease. The best-fitted sSFR--M$_{*}$ relations are: 
    \begin{align}\label{eq:ssfr-m}
    &2 \leq z \leq 3: \nonumber\\ 
    & \mathrm{log(sSFR_{100})} = -0.19 \times \mathrm{log(M_*)} -6.69, \\ 
    & 3< z \leq 4: \nonumber\\
    & \mathrm{log(sSFR_{100})} = -0.19 \times \mathrm{log(M_*)} -6.59, \\ 
    & 4 < z \leq 5: \nonumber\\
    & \mathrm{log(sSFR_{100})} = -0.03 \times \mathrm{log(M_*)} -7.87, 
    \end{align}
    where $\mathrm{sSFR_{100}}$ is the SFR averaged over 100 Myrs divided by the total formed stellar mass in units of $yr^{-1}$, and $\mathrm{log(M_*)}$ is stellar mass in units of $M_\odot$. 
    The $\mathrm{sSFR_{100}}$ and $\log(M_\ast)$ values are derived using \bagpipes, with photometry-only fitting for the photometric sample and joint spectroscopic–photometric fitting for the spectroscopic sample (see Section \ref{sec:bagpipes-phot} and Section \ref{sec:bagpipes-spec}). 
    The residual median standard deviations $\sigma$ are 0.23, 0.16, 0.13 for $2 \leq z \leq 3$, $3 < z \leq 4$ and $4 < z \leq 5$, respectively. 
    The best-fitted sSFR--M$_{*}$ relations are shown Figure \ref{fig:sSFR_M_fitting}. 
    We then select galaxies with 4$\sigma$ below the sSFR--M$_{*}$ relations as quiescent candidates. 


\end{document}